%% file: bernoulli.tex
\documentclass[11pt]{article}
\usepackage{amssymb,amsmath,epsf,graphicx}
\input epsf.sty
\newcommand{\be}{\begin{equation}}
\newcommand{\ee}{\end{equation}}
\newcommand{\bea}{\begin{eqnarray}\displaystyle}
\newcommand{\eea}{\end{eqnarray}}
\newcommand{\bdm}{\begin{displaymath}}
\newcommand{\edm}{\end{displaymath}}
\newcommand{\sectiono}[1]{\section{#1}\setcounter{equation}{0}}

\newcommand{\re}{\mathop{\rm Re}\nolimits}

\newcommand{\bpz}{\mathop{\rm bpz}\nolimits}

\newcommand{\arccot}{\mathop{\rm arccot}\nolimits}

\newcommand{\leftpartial}{\overleftarrow{\partial}}
\newcommand{\rightpartial}{\overrightarrow{\partial}}

\def\bra#1{\langle #1 |}
\def\ket#1{|#1 \rangle}
\def\kket#1{||#1 \rangle\!\rangle}
\def\aver#1{\langle\, #1 \,\rangle}

\def\res#1{\oint\! \frac{d#1}{2\pi i} \,}

\let\eps = \varepsilon

\def \rr {{\mathbb R}}

\def \ll {{\cal L}}

\def \hh {{\cal H}}

\def \bb {{\cal B}}

\def \oo {{\cal O}}

\def \sf  {string field }
\def \sft {string field theory }

\def \bpz {\dagger}
\def \lll {{\widehat{\cal L}}}
\def \bbb {{\widehat{\cal B}}}

\begin{document}
{}~ \hfill\vbox{\hbox{hep-th/0511286}\hbox{CERN-PH-TH/2005-220} }\break
\vskip 2.1cm

\centerline{\Large \bf Analytic solution for tachyon condensation } \vspace*{2.0ex}
\centerline{\Large \bf in open string field theory} \vspace*{8.0ex}

\centerline{\large \rm Martin Schnabl}

\vspace*{8.0ex}

\centerline{\large \it Department of Physics, Theory Division,}
\centerline{\large \it CERN, CH-1211, Geneva 23, Switzerland} \vspace*{2.0ex}
\centerline{E-mail: {\tt martin.schnabl@cern.ch}}

\vspace*{6.0ex}

\centerline{\bf Abstract}
\bigskip

We propose a new basis in Witten's open string field theory, in which the star product simplifies
considerably. For a convenient choice of gauge the classical string field equation of motion yields
straightforwardly an exact analytic solution that represents the nonperturbative tachyon vacuum.
The solution is given in terms of Bernoulli numbers and the equation of motion can be viewed as
novel Euler--Ramanujan-type identity. It turns out that the solution is the Euler--Maclaurin
asymptotic expansion of a sum over wedge states with certain insertions. This new form is fully
regular from the point of view of level truncation. By computing the energy difference between the
perturbative and nonperturbative vacua, we prove analytically Sen's first conjecture.

\vfill \eject

\baselineskip=16pt

\tableofcontents
\newpage

%%%%%%%%%%%%%%%%%%%%%%%%%%%%%%%%%%%%%%%%%%%%%%%%%%%%%%%%%%%%%%%%%%%%%%%%%%%%%%
\sectiono{Introduction}
\label{s_intro}
%%%%%%%%%%%%%%%%%%%%%%%%%%%%%%%%%%%%%%%%%%%%%%%%%%%%%%%%%%%%%%%%%%%%%%%%%%%%%%

Despite the beauty and simplicity of Witten's covariant field theory \cite{Witten-SFT} for open
bosonic string, only limited progress has been achieved over the years in practical applications
\cite{TZ,Sen-review}. The two main successes of the theory are computations of certain perturbative
string amplitudes and understanding the phenomenon of tachyon condensation. It is fair to say
nonetheless, that off-shell amplitudes in the Siegel gauge, the most popular covariant gauge,  are
rather unwieldy for practical purposes. To find explicitly even the simplest off-shell amplitudes,
one has to resort to numerical methods. For the tachyon condensation the situation is not much
better. Putting aside the interesting vacuum \sft proposal \cite{VSFT}, most of the results so far
were obtained by tedious numerical computations, following the seminal work of Sen and Zwiebach
\cite{SZ}, using the method of level truncation \cite{KS}.

The physics of tachyon condensation\footnote{Some early papers on this issue include
\cite{Bardakci1,Bardakci2,Bardakci3,Bardakci4}.} has made a major step forward when Ashoke Sen
identified the open-string tachyon with a physical instability of the D-brane on which the open
string ends. He made the following three conjectures \cite{senconj1, senconj2}. First, he related
the height of the tachyon potential at the true minimum to the tension of the D-brane on which the
tachyon lives. Second, he predicted existence of lump solutions with correct tensions which
describe lower dimensional D-branes popping out of the true vacuum. Finally he conjectured that
there are no physical excitations around the minimum and hence the cohomology of the BRST-like
kinetic operator there is empty. Sen's conjectures have been tested in variety of models, such as
noncommutative field theory, $p$-adic string, boundary \sf theory or vacuum \sf theory. Within
boundary \sft the first and second conjectures were proved in \cite{Gerasimov,KMM,GhSen}. The third
conjecture is true by construction in the vacuum \sft and the first two conjectures in this model
were proved in \cite{RSZ-BCFT,Okuyama,Okawa}.

The most accurate, beautiful and complete formulation of open bosonic \sft is Witten's cubic \sf
theory, but unfortunately due to the lack of exact analytic solutions, it allowed Sen's conjectures
to be tested only numerically. The height of the tachyon potential has been tested with ever
increasing accuracy in \cite{SZ,MT,GR}. The second conjecture was tested in a number of interesting
papers starting with \cite{HK,deMelloKoch,MSZ} and the third one in \cite{ET,EFHM,Giusto}. For more
references we refer to the reviews \cite{TZ,Sen-review} and \cite{Ohmori,DeSmet,Arefeva,Bonora}.

There was a large effort towards constructing analytic solutions. Various exact symmetries of the
Siegel-gauge solution have been identified \cite{HS,Trimming,constraints} and other were actively
looked for \cite{GRSZpatterns}. Exact solutions were sought in the pure-gauge-like or
partial-isometry form advocated in \cite{BfieldSFT}, but so far all such explicit solutions
\cite{Kluson,KO,TT} contained the identity state of the \sf algebra with some insertions and turned
out to be singular. There was another class of papers \cite{KP,Okawa-butt,haitang}, which attempted
to find systematic analytic approximations to the exact solutions. Unfortunately none of the above
papers succeeded in proving Sen's conjectures perhaps with the exception of the third conjecture
\cite{Kishimoto, Taka1, Taka2}. It is the goal of the present paper to provide the first
nonsingular solution and prove Sen's first conjecture.

The reason why most computations are hard in string field theory is twofold. First is that the
three-string vertex itself $\bra{V_{123}} \ket{\psi_1}\otimes \ket{\psi_2}\otimes \ket{\psi_3}$,
which defines the product in the string field algebra $\ket{\psi_1}* \ket{\psi_2} =
\bra{\psi_1}\otimes \bra{\psi_2}\ket{V_{123}}$, is quite complicated, especially when expressed in
the standard basis of $L_0$ eigenstates formed by matter and ghost oscillators. There is a basis in
which the star product simplifies \cite{Bars,spectroscopy,Douglas,Matsuo,Belov}, but manifest
background independence in the tachyon sector is lost and also conformal field theory techniques
become rather cumbersome. The second reason that makes all the computations even harder is the
choice of gauge fixing. Imposing the Siegel gauge $b_0 \psi = 0$ results in the propagator
$b_0/L_0$. Now every non-trivial \sft amplitude contains as part of its expression\footnote{ For
certain amplitudes one does not need the full information about the star product (\ref{star-ex}).
For example for the 4-point amplitude we need only the contraction $\bra{\psi_1}*\bra{\psi_2}
\frac{b_0}{L_0}\ket{\psi_3} * \ket{\psi_4} = \bra{I} \ket{\psi_1}*\ket{\psi_2}* \frac{b_0}{L_0}
\left(\ket{\psi_3} * \ket{\psi_4}\right)$. }
\be\label{star-ex}
\ket{\psi_1} * \frac{b_0}{L_0} \biggl(\ket{\psi_2} *  \ket{\psi_3}\biggr).
\ee
These building blocks of the string field theory Feynman diagrams have never been worked out
explicitly, but it is clear that they can be extracted from general off-shell amplitudes that have
been obtained in the past, and that they are going to depend on Schwarz--Christoffel maps of
polygons to the unit disk. Typically the parameters specifying the map depend on propagator lengths
(i.e. the Schwinger parameters) in a rather transcendental way
\cite{Giddings:1986,Sloan:1987,Samuel:1987,Taylor-pert}.

The string world-sheet is usually parameterized by a complex strip coordinate $w= \sigma + i \tau$,
$\sigma \in [0,\pi]$ or by $z = -e^{-i w} = -e^{-i \sigma + \tau}$, which takes values in the upper
half-plane. As has been shown in \cite{RSZ-BCFT}, the gluing conditions entering the geometrical
definition of the star product simplify if one uses another coordinate $\tilde z = \arctan z$, in
which the upper half-plane looks as a semi-infinite cylinder of circumference $\pi$. In fact, in
this coordinate we can write down simple closed form expression for arbitrary star products within
the subalgebra generated by Fock space states. Elements of this subalgebra are finite sums of the
so called wedge states with insertions \cite{RZ, wedge}, which we shall write in the form
\be\label{genstate}
U_r^\bpz U_r \, \tilde\phi_1(\tilde x_1) \tilde\phi_2(\tilde x_2)\ldots \tilde\phi_n(\tilde
x_n)\ket{0}.
\ee
By $\tilde\phi(\tilde x)$ we denote a local operator $\phi(z)$ expressed in the $\tilde z$
coordinate, which in the special case of a primary field of dimension $h$ is given by
\be\label{defphitilde}
\tilde\phi(\tilde z) = \left(\frac{dz}{d\tilde z}\right)^h \phi(z) = (\cos \tilde z)^{-2h}
\phi(\tan \tilde z).
\ee
The operator $U_r$ is a scaling operator in the $\tilde z$ coordinate, which can be written as $
U_r=\left( \frac{2}{r} \right)^{\ll_0}$, where
\be\label{LL0def1}
\ll_0 = \res{\tilde z} \tilde z T_{\tilde z \tilde z}(\tilde z)
\ee
is the zero mode of the worldsheet energy momentum tensor $T_{\tilde z \tilde z}$ in the $\tilde z$
coordinate. By a conformal transformation it can be expressed as
\be\label{LL0def2}
\ll_0 = \res{z} (1+z^2) \arctan z \, T_{zz}(z) = L_0 + \sum_{k=1}^\infty \frac{2(-1)^{k+1}}{4k^2-1}
L_{2k},
\ee
where the $L_n$'s are the ordinary Virasoro generators with zero central charge $c=0$ of the total
(i.e. matter and ghost) conformal field theory. The operator $U_r^\bpz$ in (\ref{genstate}) is
hermitian conjugate of $U_r$, which in our particular case coincides with the BPZ
conjugate.\footnote{Recall that the hermitian conjugate for a holomorphic field of dimension $h$ is
$\phi_n^\dagger = \phi_{-n}$, whereas the BPZ conjugate is $\mathrm{bpz}(\phi_n) = (-1)^{n+h}
\phi_{-n}$ .}

At first glance it might look surprising that we write (\ref{genstate}) with the factor
$U_r^\dagger U_r$ and not simply $U_r^\dagger$. After all, $U_r$ is just a scaling operator and its
action on conformal fields of dimension $h$ is particularly simple
\be\label{UphiU}
U_r \tilde\phi(\tilde z) U_r^{-1} = \left(\frac{2}{r}\right)^h \tilde\phi\left(\frac{2}{r}\tilde
z\right),
\ee
and it also keeps the vacuum invariant $U_r \ket{0} = \ket{0}$. There are at least two reasons why
we write (\ref{genstate}) the way we write it. The first reason is that the star product of two
such states takes a very simple form
\be\label{genstarprod}
U_r^\bpz U_r \, \tilde\phi(\tilde x) \ket{0} * U_s^\bpz U_s \, \tilde\psi(\tilde y) \ket{0} =
U_{r+s-1}^\bpz U_{r+s-1} \, \tilde\phi\!\left(\tilde x+\frac{\pi}{4}(s-1)\right)\,
\tilde\psi\!\left(\tilde y-\frac{\pi}{4}(r-1)\right) \ket{0},
\ee
where if there were more insertions, all insertions from the first string field would be shifted by
$\pi(s-1)/4$, whereas those from the second string field would move by $-\pi(r-1)/4$. We shall give
a detailed derivation of this formula in section~\ref{s_star}, although it follows easily from a
similar expression in \cite{wedge}. A nice feature of (\ref{genstarprod}) is, that it is valid for
any local operator insertions, not necessarily primary fields. Second reason for writing our states
in the form (\ref{genstate}) will become clear later, when we discuss expansion of the string field
in the $\ll_0$ eigenstates.

A well known special case of (\ref{genstate}) are the wedge states $\ket{r} \equiv U_r^\dagger
\ket{0}$ of Rastelli and Zwiebach \cite{RZ}. They have no operator insertions (one can view it as
an insertion of the operator identity) and by virtue of (\ref{genstarprod}) they obey the simple
algebra
\be
\ket{r}*\ket{s} = \ket{r+s-1}.
\ee
This family of states is pretty rich by itself, since it contains the identity string field
$\ket{I}=\ket{1}$ of the star algebra, the $SL(2,\rr)$ invariant vacuum $\ket{0}$ somewhat
confusingly being the wedge state $\ket{2}$, multiple products of the vacua
\be
\ket{n} = \underbrace{\ket{0} * \ket{0} * \ldots * \ket{0}}_{(n-1)\,\, \mbox{\small times}} =
U_n^\bpz \ket{0} = \left(\frac{2}{n}\right)^{\ll_0^\bpz} \ket{0},
\ee
and finally it contains a peculiar projector $\ket{\infty}$ called the sliver.

Given the simplicity of the star product (\ref{genstarprod}) in the $\tilde z$ coordinate one may
hope to be able to solve analytically the classical \sf equation of motion $Q_B\Psi+\Psi*\Psi=0$
coming from Witten's action
\be
S = -\frac{1}{g_o^2} \left[ \frac{1}{2} \aver{\Psi, Q_B \Psi} + \frac{1}{3} \aver{\Psi, \Psi*\Psi}
\right].
\ee
Beautiful aspect of this action is its enormous gauge invariance $\delta\Psi = Q_B\Lambda +
\Psi*\Lambda - \Lambda*\Psi$ which, however, has to be fixed in one way or another unless one wants
to deal with the full gauge orbit. The most popular choice for gauge fixing has been the Siegel
gauge $b_0 \Psi=0$. But alas, applying $b_0/L_0$ to the both sides of the equation of motion, one
finds
\be\label{bLeom}
\Psi + \frac{b_0}{L_0} \left(\Psi*\Psi\right) =0,
\ee
which cannot be solved easily within the states of the form (\ref{genstate}), since application of
the propagator $b_0/L_0 = b_0 \int_{t=0}^\infty e^{-t L_0}$ leaves the family of wedge states with
insertions. For this very reason also the off-shell amplitudes in Siegel gauge are doomed to be
rather complicated.

We are thus led to look for other gauge choices. Most natural one, and as far as we can tell, the
only one that works, is obtained by replacing the Siegel gauge $b_0 \Psi=0$ with $\bb_0\Psi =0$,
where $\bb_0$ is the zero mode of the $b$ ghost in the $\tilde z$ coordinate
\be\label{Bdef}
\bb_0 = \oint \frac{d\tilde z}{2\pi i} \tilde z b_{\tilde z \tilde z}(\tilde z) = \res{z} (1+z^2)
\arctan z \, b_{zz}(z) = \, b_0 + \sum_{k=1}^\infty \frac{2(-1)^{k+1}}{4k^2-1} b_{2k} \, .
\ee
Its anticommutator with the BRST charge $Q_B$ is $\left\{Q_B,\bb_0\right\} = \ll_0$ and hence,
multiplying the equation of motion with $\bb_0/\ll_0$, which itself is part of the
propagator,\footnote{Actually, as we shall discuss elsewhere, the propagator in our gauge is equal
to $\frac{\bb_0}{\ll_0} Q_B \frac{\bb_0^\bpz}{\ll_0^\bpz}$. Apparently the presence of two
Schwinger parameters for each propagator is the only disadvantage of our gauge. Note that the
propagator in the Siegel gauge can be written in a similar form since $\frac{b_0}{L_0} =
\frac{b_0}{L_0} Q_B \frac{b_0}{L_0}$.} we can write analogously to (\ref{bLeom}) the `projected'
equation of motion as
\be\label{BLLeom}
\Psi + \frac{\bb_0}{\ll_0} \left(\Psi*\Psi\right) =0.
\ee

It turns out that the operators $\ll_0$ and  $\ll_0^\bpz$ obey a very simple algebra
\be\label{IntroLLcom}
\left[\ll_0, \ll_0^\bpz\right]  =\ll_0+ \ll_0^\bpz,
\ee
and the algebra beautifully extends when generators $\bb_0$, $\bb_0^\dagger$, $B_1=b_1+b_{-1}$ and
$K_1=L_{1}+L_{-1}$ are added to it. The Lie algebra (\ref{IntroLLcom}) can be exponentiated and we
find a Lie group with the property
\be
x^{\ll_0} y^{\ll_0^\dagger} = \left(\frac{y}{x+y-x y}\right)^{\ll_0^\dagger}
\left(\frac{x}{x+y-xy}\right)^{\ll_0},
\ee
which has a natural interpretation in terms of gluing of surfaces \cite{wedge}. This relation
allows for easy application of $\bb_0/\ll_0$ to a product of several Fock states of the form
(\ref{genstate}). For the wedge states for example, we find
\be\label{bLaction}
\frac{\bb_0}{\ll_0} \ket{r} = - \bb_0^\bpz \int_2^r \frac{ds}{s} \ket{s},
\ee
which apart of the $\bb_0^\bpz$ factor is a superposition of states of the form (\ref{genstate}).
Enlarging our algebra of wedge states with insertions (\ref{genstate}) by allowing for the explicit
appearance of $\bb_0^\bpz$, we find a simple sector of the star algebra closed not only under the
star product, but also under the action of the BRST charge $Q_B$, the semi-propagator $\bb_0/\ll_0$
and many other operators.

The only method that has so far been used successfully for solving the \sft equation of motion in
Siegel gauge is the level truncation \cite{KS}. Essentially one expands the string field in the
eigenstates of the $L_0$ operator and truncates it to the first few levels, hoping that this
presents a good approximation for the physical problem in question. This method has been very
successful in finding the nonperturbative tachyon vacuum of the open strings \cite{SZ,MT,GR}. As we
have seen, the star algebra tremendously simplifies if one uses the $\tilde z$ coordinate. This
leads us immediately to the possibility that the $\ll_0$ level truncation might be the more natural
one for \sf theory. It turns out that states of the form (\ref{genstate}) have very simple
expansion in terms of the $\ll_0$ eigenstates. Unlike in the $L_0$ basis, where $U_r$ is rather
complicated, in the $\ll_0$ basis the combination $U_r^\dagger U_r$ is equal to
\be\label{UU}
U_r^\dagger U_r = \sum_{n=0}^\infty \frac{1}{n!} \left(\frac{2-r}{2}\right)^n {\lll}^{\; n},
\qquad\lll \equiv \ll_0 +\ll_0^\dagger.
\ee
By (\ref{IntroLLcom}) we see that the $n$-th term is an eigenstate (under the adjoint action) of
$\ll_0$ with eigenvalue $n$. Similarly, also the local operators in (\ref{genstate}) can be
naturally expanded in the basis of $\ll_0$ eigenstates. For example for the ghost field we have
\be
\tilde c(\tilde z) = \sum_{n=-\infty}^{\infty} \frac{\tilde c_n}{{\tilde z}^{n-1}},
\ee
where $\tilde c_n$ are $\ll_0$ eigenstates with eigenvalue $n$.

One rather unexpected feature arises when we combine the $\bb_0$ gauge with the $\ll_0$ level
truncation in certain sector of the theory (formed by the $\tilde c_n$ modes, and $\lll$ and $\bbb
\equiv \bb_0+\bb_0^\dagger$ operators acting on the vacuum). The entire set of equations of motion
for the individual components of $Q_B\Psi+\Psi*\Psi=0$ acquires such a simple structure, that they
can be solved exactly by a simple recursive procedure, level by level. The outcome of such a
calculation is surprisingly so simple, that a full all-levels form can be easily guessed to be
\bea\label{Introgn1solBern}
\Psi &=& \sum_{n=0}^\infty \sum_{\mbox{\scriptsize$\begin{array}{c} p = -1\\  p \ \mbox{odd}
\end{array}$}}^\infty \frac{\pi^p}{2^{n+2p+1}n!} (-1)^n B_{n+p+1}
{\lll}^{\; n} \,\tilde c_{-p} \ket{0} +
\\\nonumber
&& +\sum_{n=0}^\infty \sum_{\mbox{\scriptsize$\begin{array}{c} p,q = -1\\  p+q \ \mbox{odd}
\end{array}$}}^\infty \frac{\pi^{p+q}}{2^{n+2(p+q)+3} n!} (-1)^{n+q} B_{n+p+q+2} \,\bbb\, {\lll}^{\; n}\, \tilde c_{-p} \tilde
c_{-q}\ket{0},
\eea
where $B_n$ are the Bernoulli numbers; see appendix~\ref{a_Bernoulli} for the definition and few
basic properties.

Although a direct attempt to express the solution (\ref{Introgn1solBern}) in the conventional $L_0$
basis gives rise to a divergent series, it turns out that (\ref{Introgn1solBern}) is the
Euler--Maclaurin asymptotic expansion of the following sum over wedge states with insertions
\bea\label{Introgn1sol}
\Psi &=& \lim_{N \to \infty} \left[\psi_N - \sum_{n=0}^N \partial_n \psi_n\right],
\\\label{Introgn1sol2}
\psi_n &=& \frac{2}{\pi^2} U_{n+2}^\bpz U_{n+2}\left[ \bbb \tilde c\left(-\frac{\pi}{4} n\right)
\tilde c\left(\frac{\pi}{4} n\right) + \frac{\pi}{2} \left(\tilde c\left(-\frac{\pi}{4} n\right) +
\tilde c\left(\frac{\pi}{4} n\right)\right) \right] \ket{0}.
\eea
As is well known, in most cases the Euler--Maclaurin series are badly divergent (although they are
often Borel summable as is the case here), so we should not be surprised by the divergence.

For the re-summed form (\ref{Introgn1sol}), we prove that the solution is a true solution of the
equation of motion, and we give a fully analytic proof of Sen's first conjecture. This new form is
also suitable for the decomposition into the $L_0$ eigenstates. We find numerically that the
coefficients are well behaved, higher level coefficients seem to decay quite rapidly, and the
solution resembles many features of the Siegel gauge solution \cite{SZ,MT,GR}. This is in fact a
rather pleasing feature of our gauge. Just as $\tan x \simeq x$ for small $x$, we have $\bb_0 =
b_0+ \frac{2}{3} b_2 + \cdots$ and it seems that the dominant effect of the gauge fixing comes from
the $b_0$ part.\footnote{This proximity to the Siegel gauge distinguishes our $\bb_0$ gauge from
another interesting old proposal \cite{PTY} which uses the star algebra derivative
$B_1=b_1+b_{-1}$. The $B_1$ gauge shares some of the nice algebraic properties with the $\bb_0$
gauge, but it seems to fail in describing the tachyon condensation.} Also truncating our exact
solution to finite $L_0$ levels, gives us a good approximation to the energy. Third way at arriving
at the right energy is to start with the solution in the $\ll_0$ basis and use Pad\'e approximants.
By this method one confirms Sen's first conjecture with accuracy about $10^{-6}$ at level 18.

The paper is organized as follows. In section~\ref{s_star} we will review and further develop
properties of the star product using the $\tilde z$ coordinate. We will also prove a simple but
powerful lemma, which will later allow direct construction of the tachyon vacuum. In
section~\ref{s_gn0} we will solve a simple toy model equation $\left( \ll_0 -1\right)\Phi +
\Phi*\Phi =0$ whose solution will be given in terms of Bernoulli numbers. The equation of motion
will become rather elegant and novel identity for the Bernoulli numbers, somewhat akin to the
Euler--Ramanujan identities. This example will serve a useful lesson for the true \sf theory with
ghost number one string field in section~\ref{s_gn1}. Here we shall describe how to find the
solution and provide an alternative form useful for proving Sen's first conjecture, which we
explicitly prove. Apart of the analytic proof we provide two other rather distinct numerical
confirmations, one using the Pad\'e approximants and another one using ordinary level truncation.
Some details are left for the appendices.

%\newpage
%%%%%%%%%%%%%%%%%%%%%%%%%%%%%%%%%%%%%%%%%%%%%%%%%%%%%%%%%%%%%%%%%%%%%%%%%%%%%%
\sectiono{Star algebra}
\label{s_star}
%%%%%%%%%%%%%%%%%%%%%%%%%%%%%%%%%%%%%%%%%%%%%%%%%%%%%%%%%%%%%%%%%%%%%%%%%%%%%%

\subsection{The Fock space and the two-vertex}

\begin{figure}[ht]
\begin{center}
\input{two_vertex.pstex_t}
\caption{\small String worldsheet in three different coordinate systems related by $z = -e^{-i w}$
and $\tilde z = \arctan z$. In the $\tilde z$ coordinate the lines marked with an arrow are
identified, so that the worldsheet forms semi-infinite cylinder $C_\pi$. Fock states are given by
the insertion of local operators at the puncture $P$. Inserting operators also at $\tau = +\infty$,
i.e. $z=\infty$ or $\tilde z = -\pi/2 = \pi/2 \mod \pi$ would correspond to taking the BPZ inner
product. We have also marked the left and right (looking backwards in time) parts of the string at
$\tau =0$ separated by the midpoint $M$. }
\label{Fig2vertex}
\end{center}
\end{figure}
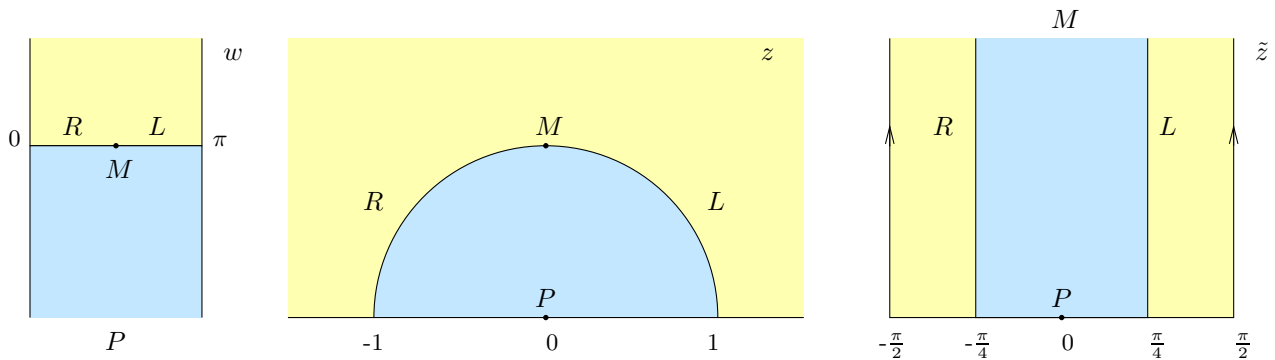

The \sft star algebra is an algebra built on the Hilbert space of the first quantized string.
Postponing questions about its completeness, such a space must contain the Fock space, which we
define here as the set of states created from the vacuum by the action of finitely many creation
operators, or equivalently by the insertion of local operators in the far past being represented by
the puncture $P$ on the worldsheet, see Fig.~\ref{Fig2vertex}.

Traditionally two coordinate systems have been used most often. The first, the more intuitive one,
uses worldsheet time $\tau \in (-\infty,\infty)$ and coordinate $\sigma \in [0,\pi]$ which are
often combined to form a new complex coordinate $w = \sigma + i \tau$ defined on a strip. Second
coordinate system obtained by the map $z=-e^{-iw}$ is the most practical one for conformal field
theory computations, since correlation functions on the upper half-plane are easily found by the
method of images.

For the purposes of \sft a third coordinate system is the most useful one. It is obtained by the
map $\tilde z = \arctan z$ which takes the upper half-plane (UHP) into the semi-infinite cylinder
$C_\pi$ with circumference $\pi$. The conformal field theory in this coordinate remains easy. As in
the case of the upper half-plane, one can also employ the doubling trick to restrict our attention
to a single holomorphic sector only.  General $n$-point functions on $C_\pi$ can be readily found
in terms of correlators on the upper half-plane by conformal mapping \footnote{In the modern
language of \sft \cite{RZ,RSZ-BCFT} one uses a global coordinate $z$ defined on the upper
half-plane and local coordinates defined around punctures. In that approach one never needs to
discuss explicitly correlators anywhere else than in the upper half-plane. It helps our intuition
however, to introduce at intermediate stages correlators of local operators on `real' cylinders,
even though in practise they are evaluated by mapping them to the upper half-plane. Care must be
taken when translating formulas from one formalism to another. We thank Barton Zwiebach for a
discussion that helped clarify this issue.}
%\footnote{When we talk about conformal field theory on the cylinder $C_\pi$ we adopt an {\it
%active} viewpoint on the conformal transformation. In the passive viewpoint the analog of equation
%(\ref{CpiUHPcorr}) for primary fields would read
%\bdm
%\aver{\phi_1( x_1)\ldots \phi_n(x_n)}_{UHP} = \left(\left.\frac{d\tilde z}{d z}\right|_{z=x_1}\right)^{h_1}\ldots
%\left(\left.\frac{d\tilde z}{d z}\right|_{z=x_n}\right)^{h_n} \aver{\tilde\phi_1(\tilde x_1)\ldots
%\tilde\phi_n(\tilde x_n)}_{UHP}.
%\edm
%Both correlators are evaluated on the upper half-plane, since under a passive conformal
%transformation only coordinates change and not the space itself. We thank Barton Zwiebach for a
%discussion that helped clarify this issue.}
\be\label{CpiUHPcorr}
\aver{\phi_1(\tilde x_1)\ldots \phi_n(\tilde x_n)}_{C_\pi} = \aver{\tilde\phi_1(\tilde x_1)\ldots
\tilde\phi_n(\tilde x_n)}_{UHP}.
\ee
The fields $\tilde \phi_i (\tilde x_i)$ were defined in (\ref{defphitilde}) as a coordinate change
(i.e. a {\it passive} conformal transformation) of $\phi_i(x_i)$. Alternatively they can be
expressed as an {\it active} conformal transformation $\tilde \phi_i (\tilde x_i) = \tan \circ
\,\phi_i(\tilde x_i)$, where in general $f \circ \oo$ denotes an active conformal transformation of
the operator $\oo$. If, for example, $\oo$ is a primary field $\phi(x)$ of dimension $h$, then $f
\circ \,\phi (x) = \left(f'(x)\right)^h \phi\left(f(x)\right)$. As we shall discuss below, the
active conformal transformation can be represented by a similarity transformation on the string
Hilbert space $f \circ \oo = U_f \oo U_f^{-1}$.

Consider for example the two and three-point functions. Let $\phi_i(z)$ be appropriately normalized
holomorphic primary fields of dimension $h_i$. Then the standard correlators in the upper
half-plane
\bea
\aver{\phi_i(x) \phi_j(y)}_{UHP} &=& \frac{\delta_{ij}}{(x-y)^{2h_i}},
\\
\aver{\phi_i(x) \phi_j(y) \phi_k(z)}_{UHP} &=&
\frac{C_{ijk}}{(x-y)^{h_i+h_j-h_k}(x-z)^{h_i+h_k-h_j}(y-z)^{h_j+h_k-h_i}}
\eea
readily imply
\bea\label{Cpicorr2}
\aver{\phi_i(\tilde x) \phi_j(\tilde y)}_{C_\pi} &=& \frac{\delta_{ij}}{\sin(\tilde x- \tilde
y)^{2h_i}},
\\\label{Cpicorr3}
\aver{\phi_i(\tilde x) \phi_j(\tilde y) \phi_k(\tilde z)}_{C_\pi} &=& \frac{C_{ijk}}{\sin(\tilde
x-\tilde y)^{h_i+h_j-h_k}\sin(\tilde x-\tilde z)^{h_i+h_k-h_j}\sin(\tilde y-\tilde
z)^{h_j+h_k-h_i}}
\eea
on the semi-infinite cylinder $C_\pi$.  The correlators are indeed well defined on $C_\pi$, as they
are invariant under a shift of any of the coordinates by $\pi$, e.g. $x \to x+\pi$, provided that
all dimensions $h_i$ are integer valued. Also note that the leading short distance behavior is the
same for $C_\pi$ and $UHP$, as it should be.

As we have already mentioned, the Fock states are defined by insertions of local operators in the
far past on the world-sheet
\be
\ket{\phi} = \phi(0) \ket{0}.
\ee
But unless we are considering states corresponding to insertions of primary operators (on-shell
states for example), the states depend on the coordinate system used to insert the local operators.
From the \sft point of view, it is more natural to work with states
\be
\ket{\tilde \phi} =  \tilde \phi(0) \ket{0}
\ee
created from the vacuum by the insertion of $\phi(0)$ in the $\tilde z$ coordinate. By conformal
transformation this state can be expressed as
\be\label{cttotilde}
\ket{\tilde \phi} = U_{\tan} \ket{\phi} = e^{\frac{1}{3} L_2 - \frac{1}{30} L_4 + \frac{11}{1890}
L_6 - \frac{1}{1260} L_8 -  \frac{34}{467775} L_{10} + \cdots} \ket{\phi},
\ee
where $U_{\tan}$ is an operator which represents the action of conformal transformation $\tilde z
\to z=\tan \tilde z$ and can be explicitly constructed following \cite{LPP2,RZ}, see also
\cite{wedge}. Note that $U_{\tan}$ is the inverse of $U_{\arctan}$ which is used to define the
sliver state \cite{RZ}

In general, for any conformal map $f(z)$ holomorphic at $z=0$ one can construct the operator $U_f$
as an exponential $\exp(\sum v_n L_n )$, where $n \ge 0$ and
$v_n$ are Laurent coefficients of a vector field $v(z) = \sum v_n z^{n+1}$ related to the
map $f(z)$ by the Julia equation $v(z)\partial_z f(z) = v(f(z))$. We should mention however, that
the vector field $v(z)$ often exists only as a formal power series, i.e. with zero radius of
convergence. This is the case for $f(z)=\tan z$  and $f(z)=\arctan z$, (whose generating vector
fields differ by an overall minus sign) as was shown in \cite{wedge}.

One of the key ingredients of \sf theory is the two-vertex, which is the familiar BPZ inner product
of conformal field theory, see Fig.~\ref{Fig2vertex}. It is defined as a map
$\hh \otimes \hh \to \rr$
\be
\aver{\phi_1, \phi_2} = \aver{I \circ \phi_1(0) \, \phi_2(0)}_{UHP},
\ee
where $I: z \to -1/z$ is the inversion symmetry. For the states $\ket{\tilde \phi_i}$ the
two-vertex can be written as
\be\label{2vert}
\aver{\tilde\phi_1, \tilde\phi_2} = \aver{I \circ \tilde\phi_1(0) \, \tilde\phi_2(0)}_{UHP} =
\aver{\phi_1\!\left(\frac{\pi}{2}\right) \phi_2(0)}_{C_\pi}.
\ee
Note that in the $\tilde z$ coordinate the inversion symmetry $I: z \to -1/z$ becomes just a
translation (i.e. a rotation) along the circumference  $I: \tilde z \to \tilde z \pm \pi/2$. The
correlators (\ref{Cpicorr2}) and (\ref{Cpicorr3}) on $C_\pi$ are manifestly invariant under it.

\subsection{The three-vertex and the star product}

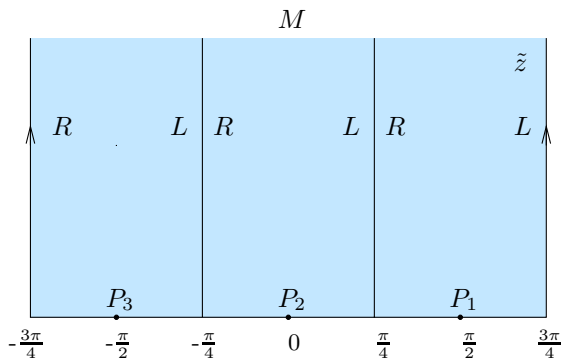
\begin{figure}[ht]
\begin{center}
\input{three_vertex.pstex_t}
\caption{\small Worldsheet of disk topology (after adding the midpoint $M$ at infinity) glued
together out of three semi-infinite strips. The lines marked with an arrow are identified. The
three vertex $\aver{\phi_1, \phi_2, \phi_3}$ is defined as a correlator of three local operators
$\phi_i$ inserted in the punctures $P_i$. } \label{Fig3vertex}
\end{center}
\end{figure}

Unlike in closed string field theory \cite{closed}, in open \sft there is a single vertex which
determines all the interactions. The three-vertex is a map $\hh \otimes \hh \otimes \hh \to \rr$
and is defined as a correlator on a surface formed by gluing together three strips representing
three open string worldsheets, see Fig.~\ref{Fig3vertex}.

Traditionally \cite{LPP1,RZ}, for states $\ket{\phi_i}$ defined using the $z$ coordinate, the
three-vertex has been written~as
\be
\aver{\phi_1, \phi_2, \phi_3} = \aver{f_1 \circ \phi_1(0) \, f_2 \circ \phi_2(0) \, f_3 \circ
\phi_3(0)}_{UHP},
\ee
where $f_n(z)=\tan\left(\frac{(2-n)\pi}{3}+\frac{2}{3}\arctan z \right)$. For states defined using
the $\tilde z$ coordinate it can be expressed directly as
\be\label{3vert1}
\aver{\tilde\phi_1, \tilde\phi_2, \tilde\phi_3} = \aver{\phi_1\!\left(\frac{\pi}{2}\right) \,
\phi_2(0) \, \phi_3\!\left(-\frac{\pi}{2}\right)}_{C_\frac{3\pi}{2}},
\ee
without the need of any conformal map. Here the correlator is taken on a semi-infinite cylinder
$C_{\frac{3\pi}{2}}$ of circumference $3\pi/2$, see Fig.~\ref{Fig3vertex}.

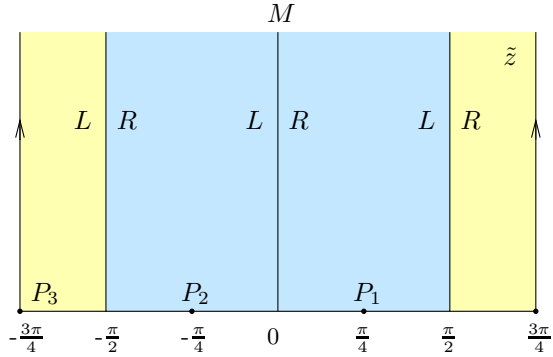
\begin{figure}[ht]
\begin{center}
\input{star_product.pstex_t}
\caption{\small Star product of two states $\ket{\tilde\phi_1} * \ket{\tilde\phi_2}$ represented by
local operator insertions at punctures $P_1$ and $P_2$. A local operator $\chi$ corresponding to
the `test state' $\ket{\tilde\chi}$ can be inserted at the puncture $P_3$. The correlator is
evaluated on a semi-infinite cylinder of circumference $3\pi/2$. } \label{FigStarProd}
\end{center}
\end{figure}

The three-vertex allows us to introduce the star product $*: \hh \otimes \hh \to \hh$.  Given two
states $\ket{\phi_1}$ and $\ket{\phi_2}$ the star product is defined by matching the three-vertex
with an additional `test state' $\ket{\chi}$ to the two-vertex
\be\label{32match}
\aver{\tilde \chi \, ,\, \tilde\phi_1 \, , \, \tilde\phi_2} = \aver{\tilde \chi \, , \,
\tilde\phi_1
* \tilde\phi_2}, \qquad \forall \chi.
\ee
Graphically the star product of two Fock states can be represented by the surface in
Fig.~\ref{FigStarProd}.
To find an explicit formula for the star product it is useful to rewrite the left hand side of
(\ref{32match}) as a correlator on a semi-infinite cylinder $C_\pi$ of circumference~$\pi$
\be\label{3vert3}
\aver{\tilde \chi \, ,\, \tilde\phi_1 \, , \, \tilde\phi_2} = \aver{s \circ \chi \!\left(\pm
\frac{3\pi}{4}\right) \, s \circ \phi_1\left(\frac{\pi}{4}\right) \, s \circ
\phi_2\!\left(-\frac{\pi}{4}\right)}_{C_\pi}.
\ee
%\be\label{3vert2}
%\aver{\tilde\phi_1, \tilde\phi_2, \tilde\phi_3} = \aver{s \circ \phi_1\!\left(\frac{\pi}{2}\right) \, s
%\circ \phi_2(0) \, s \circ \phi_3\!\left(-\frac{\pi}{2}\right)}_{C_\pi}
%\ee
using a simple conformal map $s: \tilde z \to \frac{2}{3} \tilde z$. Note that the scaling
transformation $s$ is implemented by $U_3 \equiv (2/3)^{\ll_0}$, where $\ll_0$ was introduced in
(\ref{LL0def1}) and (\ref{LL0def2}). Thinking of $s \circ \phi_1\left(\frac{\pi}{4}\right) \, s
\circ \phi_2\!\left(-\frac{\pi}{4}\right)$ in terms of its local operator product expansion around
$\tilde z =0$, the right hand side of (\ref{3vert3}) has the form of the two-vertex (\ref{2vert}).
To see it more clearly, let us restrict to the set of test states $\chi$ with definite scaling
dimension $h$. Then indeed $s \circ \chi(\pm3\pi/4) = (2/3)^h \chi(\pm \pi/2)$. Writing thus
(\ref{3vert3}) as the two-vertex, the factor $(2/3)^h$ can be traded for an operator $U_3^\dagger$
acting on the second entry $s \circ \phi_1\left(\frac{\pi}{4}\right) \, s \circ
\phi_2\!\left(-\frac{\pi}{4}\right)$, so that we have
\be
\aver{\tilde \chi \, ,\, \tilde\phi_1 \, , \, \tilde\phi_2} = \aver{\tilde\chi, U_3^\dagger\left(
\,s \circ \tilde\phi_1\left(\frac{\pi}{4}\right) \, s \circ
\tilde\phi_2\!\left(-\frac{\pi}{4}\right)\right)}
\ee
and hence we find
\be\label{sp22}
\tilde\phi_1(0)\ket{0} * \tilde\phi_2(0)\ket{0} =  U_3^\bpz U_3 \,
{\tilde\phi}_1\left(\frac{\pi}{4}\right) {\tilde\phi}_2 \left(-\frac{\pi}{4}\right) \ket{0}.
\ee

When the local fields $\phi_{1,2}$ are, for example, primary fields of conformal dimensions
$h_{1,2}$ we can use the fact that $U_r$ has a simple action (\ref{UphiU}) on them, and we can
re-express (\ref{sp22}) in the standard form
\be\label{sp22hh}
\phi_1(0)\ket{0} * \phi_2(0)\ket{0} = \left(\frac{8}{9}\right)^{h_1+h_2} U_3^\bpz \,
{\phi}_1\left(\tan\frac{\pi}{6}\right) {\phi}_2 \left(-\tan\frac{\pi}{6}\right) \ket{0}.
\ee
This formula agrees with few explicit examples given in \cite{RZ} and generalized in \cite{wedge}.
It will be however formula (\ref{sp22}), and its generalizations given in the next subsection, that
will be most useful for the rest of the paper.

Let us now explain how to translate the expression (\ref{sp22}) to the ordinary Virasoro basis
based on the coordinate $z$. By Virasoro basis we essentially mean the basis in the Verma module
formed by the action of matter or total Virasoro generators on the highest weight states. In
general the operator $U_r \equiv (2/r)^{\ll_0}$ represents the scaling $\tilde z \to \frac{2}{r}
\tilde z$, which in the $z$ coordinate becomes
$z \to f_r(z)$, where
\be
f_r(z) = \tan\left(\frac{2}{r} \arctan z\right).
\ee
The operators $U_r$ can be written as $\exp(\sum v_n L_n )$, by solving recursively the Julia
equation $v(z)\partial_z f_r(z) = v(f_r(z))$ following \cite{RZ}. One finds
\be\label{wedger}
U_r = \left(\frac{2}{r}\right)^{L_0} e^{-\frac{r^2 -4}{3\,r^2} \,L_{2} +\frac{r^4-16}{30\,r^4}\,
L_{4} -\frac{\left(r^2-4\right) \, \left( 176 + 128\,r^2 + 11\,r^4\right)}{1890\,r^6} \, L_{6}
+\frac{\left(r^2-4 \right) \,\left( r^2 +4\right) \, \left( 16 + 32\,r^2 + r^4 \right) }{1260\,r^8}
\, L_{8} + \cdots }.
\ee
Using the composition rule $U_{f \circ g} = U_f U_g$ which reflects the fact that $U_f$ form a
representation of the conformal group, one can arrive to a more convenient canonically ordered
form\footnote{It is easy to write a simple recursive algorithm similar to the one of \cite{RZ} to
find out the coefficients in front of $L_{n}$ for almost arbitrarily high $n$. We provide more
details in appendix~\ref{a_surface}.}
\be\label{nowedge}
U_r = \left(\frac{2}{r}\right)^{L_0} e^{ -\frac{r^2-4}{3r^2}\,L_{2}} e^{\frac{r^4-16}{30r^4} \,
L_{4}} e^{-\frac{16 (r^2-4)(r^2-1)(r^2+5)}{945 r^6}\,L_{6}} e^{\frac{(r^2-4)(109 r^6 + 436 r^4 -
944r^2+1344)}{11340 r^8}\,L_{8}} \ldots,
\ee
which is advantageous in level truncation computations. The least ordered, but most beautiful form
of $U_r$ is of course the one already mentioned
\be
U_r = \left(\frac{2}{r}\right)^{\ll_0} = e^{\log\left(\frac{2}{r}\right) \left(L_0 + \frac{2}{3}
L_2 - \frac{2}{15} L_4 + \frac{2}{35} L_6 - \frac{2}{63} L_8 + \cdots \right)}.
\ee

\subsection{Wedge states with insertions}

\begin{figure}[ht]
\begin{center}
\input{wedge.pstex_t}
\caption{\small Multiple star product $\ket{\tilde\phi_1} * \ket{\tilde\phi_2}* \cdots *
\ket{\tilde\phi_n}$, the so called wedge state with insertions. Without insertions it would be
denoted as $\ket{n+1}$. The correlator is evaluated on a semi-infinite cylinder of circumference
$(n+1)\pi/2$. }
\label{Figwedge}
\end{center}
\end{figure}
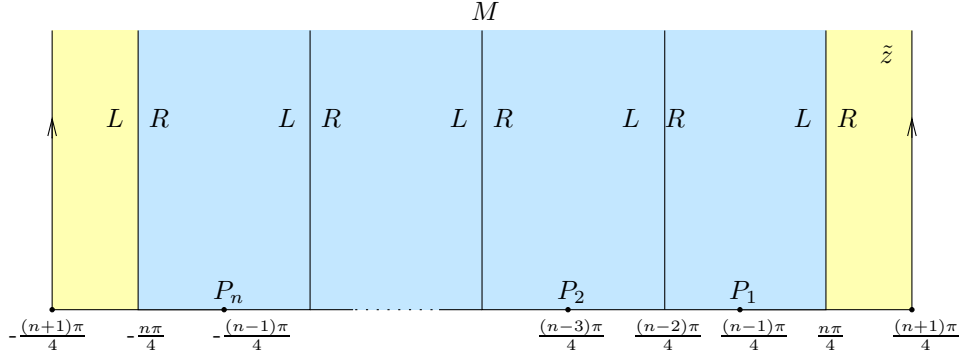

So far we have considered only a star product of two Fock states. Generalization to the multiple
star product $\ket{\tilde\phi_1} * \ket{\tilde\phi_2}* \cdots * \ket{\tilde\phi_n}$, where
$\ket{\tilde\phi_j} \equiv \tilde\phi_j(0) \ket{0}$, is rather straightforward and is obtained by gluing
together $n+1$ strips as in Fig.~\ref{Figwedge}.
The analog of (\ref{sp22}) is
\be\label{sp222}
\ket{\tilde\phi_1} * \ket{\tilde\phi_2}* \cdots * \ket{\tilde\phi_n} = U_{n+1}^\bpz U_{n+1} \,
{\tilde\phi}_1\left(\frac{(n-1)\pi}{4}\right) {\tilde\phi}_2 \left(\frac{(n-3)\pi}{4}\right) \ldots
{\tilde\phi}_n \left(-\frac{(n-1)\pi}{4}\right) \ket{0}.
\ee
In more generality we could consider a family of states
\be\label{wedgewi}
U_r^\bpz U_r \, \tilde\phi_1(\tilde x_1) \tilde\phi_2(\tilde x_2)\ldots \tilde\phi_n(\tilde
x_n)\ket{0},
\ee
for arbitrary real $r \ge 1$ and arbitrary insertion points $\tilde x_i$, $|\re \tilde x_i| \le
(r-1)\pi/4$. How do such states star multiply? States of the form (\ref{wedgewi}) are represented
by cylinders of circumference $r\pi/2$ and punctures at points $\tilde x_i$ as in
Fig.~\ref{Figwedge}, regardless of whether they can be constructed by gluing Fock states or not.
The star multiplication proceeds as for Fock states by simply gluing together the parts of the two
or more  cylinders with strips of length $\pi/2$ cut out (in light yellow on Fig.~\ref{Figwedge}),
and then gluing back one such strip to form a new bigger cylinder.\footnote{Note that under star
multiplication the circumference can only grow, or in a limiting case with $r=1$ can remain the
same. Having $r<1$ would formally correspond to deleting a part of surface, it is hard to make
sense of it in case there are some punctures, and it is also ill behaved in level truncation
\cite{wedge}.} Mathematically we can write it as
\bea
\label{wwistar}
&& U_r^\bpz U_r \, \tilde\phi_1(\tilde x_1)\ldots \tilde\phi_n(\tilde x_n)\ket{0} * U_s^\bpz U_s \,
\tilde\psi_1(\tilde y_1)\ldots \tilde\psi_m(\tilde y_m)\ket{0}=
\\\nonumber
&& \quad = U_t^\bpz U_t \,  \tilde\phi_1(\tilde x_1 + \frac{\pi}{4}(s-1) )\ldots
\tilde\phi_n(\tilde x_n + \frac{\pi}{4}(s-1)) \, \tilde\psi_1(\tilde y_1 - \frac{\pi}{4}(r-1)
)\ldots \tilde\psi_m(\tilde y_m - \frac{\pi}{4}(r-1)) \ket{0},
\eea
where $t=r+s-1$. We leave it as an exercise to the reader to check the associativity.

Before we end this discussion let us look in more detail on the simplest case with no insertions,
i.e. when all operators $\phi_i$ are taken to be the identity operator. These are the original
wedge states
\be
\ket{r} =  U_r^\bpz U_r \ket{0} = U_r^\bpz \ket{0}
\ee
introduced by Rastelli and Zwiebach in \cite{RZ}. They obey a simple algebra
\be
\ket{r} * \ket{s} = \ket{r+s-1},
\ee
which is a special case of (\ref{wwistar}). Note that the $SL(2,R)$ invariant vacuum $\ket{0}$ is
the wedge state $\ket{2}$. The wedge state $\ket{r}$ with lowest allowed $r=1$ is the identity of
the star algebra. For the limiting value $r \to \infty$ one finds a projector, so called sliver
state which has attracted much attention in the literature, especially in the context of the vacuum
\sf theory \cite{VSFT}.

\subsection{Operator algebra in the $\tilde z$ coordinate}

To tackle such a complicated task such as solving the string field equations of motion,
we found it very useful to use an operator formalism and to algebraize the problem.
In fact our formula (\ref{wwistar}) was a first step in this programm.
Let us now take few steps further.

We have already noted that the wedge states can be naturally written in terms of the (hermitian or
BPZ conjugate) of the scaling operator $U_r = (2/r)^{\ll_0}$. The infinitesimal generator of the
scaling is given by the zero mode $\ll_0$ of the total energy momentum tensor $T_{\tilde z \tilde
z}(\tilde z)$ with zero central charge. Let us now look at other modes. We define
\be\label{Tarctan}
\ll_n =  \oint \frac{d\tilde z}{2\pi i} {\tilde z}^{n+1} T_{\tilde z \tilde z}(\tilde z) = \res{z}
(1+z^2) \left(\arctan z\right)^{n+1} T_{zz}(z).
\ee
Note that there would be a central charge contribution in the last equation if $c$ were nonzero.
The hermitian conjugate is then given by
\be\label{Tarccot}
\ll_m^\dagger = \oint \frac{dz}{2\pi i} (1+z^2) \left(\arccot z\right)^{m+1} T_{zz}(z).
\ee
Both sets of operators obey standard Virasoro algebra with zero central charge.
\bea
\left[\ll_n,\ll_m\right] &=& (n-m) \ll_{n+m}, \\
\left[\ll_n^\dagger,\ll_m^\dagger\right] &=& -(n-m) \ll_{n+m}^\dagger.
\eea
What about the mixed commutators? It turns out that three operators $\ll_0$, $\ll_0^\dagger$ and
$\ll_{-1}=K_1 \equiv L_1 + L_{-1}$, which will be of particular importance, form an interesting
closed algebra \cite{wedge, Cargese}
\bea\label{LLda}
\left[\ll_0,\ll_0^\dagger\right] &=& \ll_0 + \ll_0^\dagger, \\
\left[\ll_0, K_1\right] &=&K_1, \\
\left[\ll_0^\dagger, K_1\right] &=& -K_1.
\eea
There are three different ways of deriving it. The first, the most straightforward way, is to use
the explicit form (\ref{LL0def2})
\bdm
\ll_0 = L_0 + \sum_{k=1}^\infty \frac{2(-1)^{k+1}}{4k^2-1} L_{2k}
\edm
and simply calculate the commutators as we did in \cite{wedge}. The second, rather indirect way is to use
the gluing theorem to argue \cite{wedge} that
\be
U_r U_s^\dagger  = U_{2+\frac{2}{r}(s-2)}^\dagger U_{2+\frac{2}{s}(r-2)}.
\ee
Differentiating with respect to $r$ and $s$ and setting $r=s=2$ one recovers (\ref{LLda}).
The third method, which is also applicable for general modes $\ll_n$ is to use standard contour
arguments\footnote{I thank Ian Ellwood for suggesting the method.}
to find
\be
\left[\ll_n,\ll_m^\dagger\right] = \res{z} (1+z^2) (\arctan z)^{n}(\arccot z)^{m}
\left((m+1)\arctan z + (n+1)\arccot z\right) T(z)
\ee
There is an important subtlety however, in that the contours must pass precisely through the points
$\pm i$; one can take the unit circle for example. The reason is that because of the cuts in
$\arctan z$, the contour in (\ref{Tarctan}) must cross the imaginary axis within the segment
$[-i,i]$, whereas the contour in (\ref{Tarccot}) must cross it outside this range. The choice of
operator ordering is determined by the time ordering (i.e. the $|z|$-ordering in the radial
quantization) in the path integral formalism, and therefore to make sense of the operator product
$\ll_0 \ll_0^\dagger$ in the path integral, the contour defining  $\ll_0^\dagger$ must lie inside
the one defining $\ll_0$. To satisfy these two conflicting requirements the contours must pass
through points $\pm i$, which are fortunately integrable singularities. This would not be the case
for commutators $\left[\ll_r,\oo_s^\dagger\right]$, if $\oo$ were an operator of dimension $h \le
0$.

Some other nontrivial examples, which can be obtained by this method are
\bea
\left[\ll_1,\ll_1^\dagger\right] &=& \frac{\pi^2}{6}( \ll_0 + \ll_0^\dagger) - \frac{2}{3}(\ll_2 +
\ll_2^\dagger),
\\
\left[\ll_0,\ll_1^\dagger\right] &=& \frac{\pi^2}{4} \ll_{-1} - \ll_1 \,.
\eea
It is interesting to note that in general $\left[\ll_n,\ll_m^\dagger\right]$ are given as finite
linear combinations of the generators $\ll_k$ and $\ll_k^\dagger$ as long as $n,m \ge -1$.

In terms of ordinary Virasoro operators our new Virasoro operators are given explicitly by
\footnote{Had we worked with nonzero central charge, the only modification would be an additional
term $c/6$ in $\ll_{-2}$. Nevertheless, as shown in \cite{wedge} some commutators such as
$\left[\ll_0,\ll_0^\dagger\right]$ would become divergent.}
\bea\label{lln}
\ll_2 &=& L_2 - \frac{1}{15} L_6 + \frac{64}{945} L_8  + \cdots \nonumber\\
\ll_1 &=& L_1 + \frac{1}{3} L_3 - \frac{7}{45} L_5 + \frac{29}{315} L_7 + \cdots  \nonumber\\
\ll_0 &=& L_0 + \frac{2}{3} L_2 - \frac{2}{15} L_4 + \frac{2}{35} L_6 + \cdots  \nonumber\\
\ll_{-1} &=& L_{-1} + L_1 \\
\ll_{-2} &=& L_{-2} + \frac{4}{3} L_0 + \frac{11}{45} L_2 - \frac{8}{189} L_4  \cdots  \nonumber \\
\ll_{-3} &=& L_{-3} + \frac{5}{3} L_{-1} + \frac{3}{5} L_1 - \frac{31}{945} L_3  \cdots  \nonumber \\
\ll_{-4} &=& L_{-4}+ 2L_{-2} + \frac{16}{15} L_0 + \frac{62}{945} L_2 - \frac{1}{225} L_4  \cdots. \nonumber
\eea
Note that the operators $\ll_{1}$, $\ll_{0}$, $\ll_{-1}$, $\ll_{-2}\ldots$ are conservation laws
for the sliver \cite{RSZ-classical}. To see that, we note that $\ll_n$ defined in (\ref{Tarctan})
can be alternatively written as a conformal transformation of $L_n$
\bea
\ll_n &=& U_{\tan} L_n U_{\tan}^{-1} = \tan \circ \, L_n \, = \res{z} z^{n+1} \, \tan \circ
\,T_{zz}(z)
\\\nonumber
&=& \res{z} z^{n+1} \, \cos^{-4}z T_{zz}(\tan z) = \res{z} (1+z^2)
\left(\arctan z\right)^{n+1} T_{zz}(z)
\eea
and hence
\be\label{sliverconslaw}
\bra{\infty} \ll_{-n} = \bra{0} U_{\arctan z} U_{\tan} L_{-n} U_{\tan}^{-1} =  \bra{0} L_{-n}
U_{\arctan z} = 0
\ee
for $n \ge -1$. We shall say more on the conservation laws for wedge states in
appendix~\ref{a_surface}.

Attentive reader might have noticed from (\ref{LLda}) that the combination $\ll_0+\ll_0^\dagger$
commutes with $K_1$. In fact there is a deeper reason for that.
Note that from (\ref{Tarctan}) and (\ref{Tarccot})
\bea\label{LLres1}
\ll_0+\ll_0^\dagger &=& \oint \frac{dz}{2\pi i} (1+z^2) \left(\arctan z+ \arccot z \right) T(z)
\\\label{LLres2}
&=& \frac{\pi}{2} \oint \frac{dz}{2\pi i} (1+z^2) \eps\left(\re z\right) T(z),
\eea
where $\eps(x)$ is the step function equal to $\pm 1$ for positive or negative values respectively.
(We also abbreviate $T_{zz}(z)$ to $T(z)$.) In order to be able to write expression (\ref{LLres1})
for both terms using a single contour integral, we have used the unit circle in both
(\ref{Tarctan}) and (\ref{Tarccot}). Splitting the integration contour into two halves in
(\ref{LLres2}), one in the $\re z>0$ half-plane and the other in $\re z<0$, we observe that these
two semi-circle contour integrals are in fact the definition of $K_1^L$ and $K_1^R$ respectively.
We thus find
\be
\ll_0+\ll_0^\dagger =  \frac{\pi}{2} \left( K_1^L - K_1^R \right),
\ee
and since $K_1^L+K_1^R=K_1$, we also have
\bea
K_1^L &=&  \frac{1}{2} K_1 + \frac{1}{\pi} \left( \ll_0 + \ll_0^\dagger \right),
\\
K_1^R &=&  \frac{1}{2} K_1 - \frac{1}{\pi} \left( \ll_0 + \ll_0^\dagger \right).
\eea
Now we see that the relation $\left[\ll_0+\ll_0^\dagger, K_1 \right]=0$ is responsible for
$\left[K_1^L, K_1^R\right]=0$. Here we are quite lucky, since such commutators between the left and
right string operators are often anomalous.

The operators $K_1^L$,  $K_1^R$ and $K_1$ also have rather simple properties with regard to the
star product
\bea
K_1^L \left( \phi_1 * \phi_2 \right) &=&  \left(K_1^L \phi_1 \right) * \phi_2, \\
K_1^R \left( \phi_1 * \phi_2 \right) &=&  \phi_1 * \left(K_1^R \phi_2\right), \\
\label{K1der}
K_1   \left( \phi_1 * \phi_2 \right) &=& \left(K_1 \phi_1 \right) * \phi_2 +  \phi_1 * \left(K_1
\phi_2\right).
\eea
The first two relations reflect the geometry of the Witten vertex, in that the left part of the
first string becomes the left part of the product and the right part of the right string becomes
the right part of the star product. The last relation is the well known fact that $K_1$ is a
derivation of the star product. Sometimes an analogous relation might also be useful
\be D  \left(
\phi_1 * \phi_2 \right) = \left(D \phi_1 \right) * \phi_2 +  \phi_1 * \left(D \phi_2\right),
\ee
where $D=\ll_0-\ll_0^\dagger$ is another star algebra derivative. The operators $K_1^L$ and $K_1^R$
play a further role, in that their operator action generates star multiplication by the family of
wedge states for the full Hilbert space. Explicitly, as follows readily from the results in
\cite{wedge}, we find
\bea
\ket{n} * \ket{\psi} &=& e^{-(n-1)\frac{\pi}{2} K_1^L} \ket{\psi}, \\
\ket{\psi} * \ket{n} &=& e^{(n-1)\frac{\pi}{2} K_1^R} \ket{\psi}.
\eea
One can thus alternatively write the wedge states as\footnote{One could also write the wedge states
as $\ket{n}=e^{-(n-1)\frac{\pi}{2} K_1^L} \ket{I}$, which is reminiscent of the formal
considerations in \cite{Kluson}.}
\be
\ket{n} = e^{-(n-2)\frac{\pi}{2} K_1^L} \ket{0} =  e^{(n-2)\frac{\pi}{2} K_1^R} \ket{0} =
e^{-(n-2)\frac{\pi}{4} \left(K_1^L-K_1^R \right)} \ket{0} = e^{-\frac{n-2}{2}
\left(\ll_0+\ll_0^\dagger \right)} \ket{0}.
\ee

Number of interesting relations can be obtained by exponentiating the Lie algebra (\ref{LLda}). The
most important ones are
\bea
U_r U_s &=& U_{\frac{rs}{2}},
\\
U_r U_s^\dagger  &=& U_{2+\frac{2}{r}(s-2)}^\dagger U_{2+\frac{2}{s}(r-2)},
\\
U_r e^{\alpha X} &=& e^{\frac{2\alpha}{r} X} U_r, \qquad \mbox{valid for $X=K_1$, $K_1^{L,R}$,
$\ll_0+\ll_0^\dagger$},
\\\label{eLL}
e^{\beta\left(\ll_0+\ll_0^\dagger\right)} &=& U_{2-2\beta}^\dagger U_{2-2\beta}.
\eea
The first two were derived in \cite{wedge}, the latter two can be obtained by similar methods. Let
us illustrate such a derivation on (\ref{eLL}) which plays a central role in this paper. Let us
denote $f(x) = x^{\ll_0^\dagger} x^{\ll_0}$. Clearly $f(1)=1$. The derivative $f'(x)$ can be easily
computed with the help of (\ref{ULU}) given below
\bdm
f'(x)=\frac{1}{x} x^{\ll_0^\dagger}\left(\ll_0^\dagger+\ll_0\right) x^{\ll_0} = \frac{1}{x^2} f(x)
\left(\ll_0^\dagger+\ll_0\right) =\frac{1}{x^2} \left(\ll_0^\dagger+\ll_0\right) f(x).
\edm
Integrating this as a differential equation we find
$f(x)=\exp\left[\left(1-\frac{1}{x}\right)\left(\ll_0^\dagger+\ll_0\right)\right]$ and (\ref{eLL})
readily follows.\footnote{Barton Zwiebach has suggested alternative derivation based on embedding
the two-dimensional algebra $\left[\ll_0,\ll_0^\dagger\right] =\ll_0+\ll_0^\dagger$ inside $gl(2)$
and using its explicit representation in terms of two dimensional matrices.} Other useful
identities are
\bea\label{ULU} U_{r}\, \ll_0^\dagger \,U_{r}^{-1} &=& \frac{2-r}{r}
\,\ll_0 + \frac{2}{r} \,\ll_0^\dagger,
\nonumber\\
U_{r}^{\dagger\,-1}\, \ll_0\, U_{r}^{\dagger} &=& \frac{2}{r} \, \ll_0 + \frac{2-r}{r} \,
\ll_0^\dagger,
\nonumber\\
U_{r}^{-1}\, \ll_0^\dagger \,U_{r} &=& \frac{r-2}{2}  \,\ll_0 + \frac{r}{2} \,\ll_0^\dagger,
\nonumber\\
U_{r}^{\dagger}\, \ll_0\, U_{r}^{\dagger\,-1} &=& \frac{r}{2} \, \ll_0 + \frac{r-2}{2} \,
\ll_0^\dagger.
\eea
Finally for completeness we remind the reader that on a primary field $\tilde\phi$ of dimension $h$
the exponentiated generators $\ll_0$ and $K_1$ act as scaling and translation
\bea
\lambda^{\ll_0} \tilde\phi\left(\tilde z\right) \lambda^{-\ll_0} &=& \lambda^h
\tilde\phi\left(\lambda \tilde z\right),
\\
e^{\alpha K_1} \tilde\phi\left(\tilde z\right) e^{-\alpha K_1} &=& \tilde\phi\left(\tilde z +
\alpha \right).
\eea

\subsection{Star product in the $\ll_0$-basis}
\label{ss_lemma}

It turns out that the most general \sf algebra elements (\ref{wedgewi}) we have considered so far
can be very naturally expressed in the basis of $\ll_0$ eigenstates. To start with, consider first
pure wedge states with no insertions. They can be written using (\ref{eLL}) as
\be\label{rinLL}
\ket{r} = U_r^\dagger U_r \ket{0} = e^{\frac{2-r}{2} (\ll_0+\ll_0^\dagger)}\ket{0} =
\sum_{n=0}^\infty \frac{1}{n!} \left(\frac{2-r}{2}\right)^n \left(\ll_0+\ll_0^\dagger\right)^n
\ket{0}.
\ee
Note that by (\ref{LLda}) the states $\left(\ll_0+\ll_0^\dagger\right)^n \ket{0}$ are eigenstates
of $\ll_0$ with eigenvalue $n$. Although these states are far from being normal ordered, they are
quite convenient. Almost normal ordered expression (normal ordered up to some $L_0$'s hidden inside
$\ll_0^\dagger$) can be written as
\bea
\left(\ll_0+\ll_0^\dagger\right)^n \ket{0} &=& \left(n-1+\ll_0^\dagger\right)
\left(n-2+\ll_0^\dagger\right) \ldots \left(1+\ll_0^\dagger\right) \ll_0^\dagger \ket{0}  =
\frac{\Gamma\left(\ll_0^\dagger+n\right)}{\Gamma\left(\ll_0^\dagger\right)} \ket{0}
\nonumber\\
&=& \sum_{k=1}^n (-1)^{n-k} S_n^{(k)} \left(\ll_0^\dagger\right)^k \ket{0},
\eea
where $S_n^{(k)}$ are the (signed) Stirling numbers of the first kind. They are defined in such a
way that $(-1)^{n-k} S_n^{(k)}$ is the number of permutations of $n$ symbols which have precisely
$k$ cycles. This expression might be useful for deriving various startling mathematical identities,
but for our purposes it will be the form $\left(\ll_0+\ll_0^\dagger\right)^n \ket{0}$ which will
prove to be most useful.

There is a second kind of $\ll_0$ eigenstates, which are perhaps more obvious, which are obtained
simply by conformal transformation (\ref{cttotilde}) of the $L_0$ eigenstates. As an example
consider modes of the $\tilde c$ ghost
\be
\tilde c(\tilde z) = \sum_{n=-\infty}^{\infty} \frac{\tilde c_n}{{\tilde z}^{n-1}},
\ee
given by
\be
\tilde c_n = \tan \circ \, c_n = \sum_{m=n}^\infty c_m \res{\tilde z} {\tilde z}^{n-2} \cos^2
\tilde z \,(\tan \tilde z)^{-m+1},
\ee
since $\tilde c(\tilde z) = \tan \circ \, c(\tilde z) = \cos^2 \tilde z \, c(\tan \tilde z)$.
Equivalently, using the more conventional passive viewpoint these modes can be expressed as
\be
\tilde c_n =  \res{\tilde z} {\tilde z}^{n-2} \tilde c(\tilde z) =\sum_{m=n}^\infty c_m \res{z}
\frac{1}{(1+z^2)^2} (\arctan z)^{n-2} z^{-m+1}.
\ee
First few $\ll_0$ eigenstates are explicitly given by
\bea
\tilde c_1 \ket{0} &=& c_1 \ket{0}
\nonumber\\
\tilde c_0 \ket{0} &=& c_0 \ket{0}
\nonumber\\
\tilde c_{-1} \ket{0} &=& \left(c_{-1}-c_{1} \right) \ket{0}
\nonumber\\
\tilde c_{-2} \ket{0} &=& \left(c_{-2}-\frac{2}{3} c_{0} \right) \ket{0}
\nonumber\\
\tilde c_{-3} \ket{0} &=& \left(c_{-3}-\frac{1}{3}c_{-1} + \frac{1}{3} c_{1} \right) \ket{0}.
\eea
More complicated examples are given by products of several $\tilde\phi_n$ modes of any number of
primary fields. Just to give an example of a case where there are contractions between two mode
operators we use (\ref{lln}) to write a weight $5$ $\ll_0$-eigenstate
\be
\ll_{-3} \ll_{-2} \ket{0} = \left( L_{-3} L_{-2} + \frac{5}{3} L_{-3} \right) \ket{0}.
\ee

We have seen that there are basically two types of $\ll_0$ eigenstates. Ones which use a $n$-th
power of $\ll_0+\ll_0^\dagger$ (or a factor of $\bb_0+\bb_0^\dagger$) and ones which use modes of
primary operators $\tilde \phi_n$. The former ones contain infinite sum of terms in the ordinary
$L_0$ basis, whereas the second ones only finite number of them. Looking at (\ref{wedgewi}) we see
that we really should combine and use these two kinds of states together. One might be worried
about overcounting if we include both kinds of states, but note that for instance the state
$\ll_0^\dagger \ket{0}$ with $\ll_0$ eigenvalue equal to one, is truly impossible to write as a
linear combination of states like $\tilde \phi_n \ket{0}$. In fact the only viable candidate
$\ll_{-1} \ket{0}$ is identically equal to zero.

The star product rules for the above states of the $\ll_0$ basis can be readily worked out using
(\ref{wwistar}). This leads to the following trivial but powerful lemma which belongs to the main
results of the paper:

%\newpage
\noindent
{\bf Lemma:}\\
\emph{Let $\psi_1$ and  $\psi_2$ be two eigenstates of $\ll_0$ with eigenvalues $h_1$ and $h_2$
respectively. Let us further assume that they are linear combinations of states of the form
(\ref{wedgewi}) with the only operator insertions allowed being $\bb_0^\dagger$, arbitrary power of
$\ll_0^\dagger$ and any number of the $\tilde c$ ghosts. Then the star product $\psi_1*\psi_2$ is
an infinite linear combination of $\ll_0$ eigenstates with eigenvalues $h \ge h_1+h_2$.}
\\
%\newpage
\noindent
{\bf Proof:}\\
Let us write a basis of states with a definite $\ll_0$ eigenvalue $h$ in the form
\bea\label{Belem1}
\left(\ll_0+\ll_0^\dagger\right)^n \tilde c_{-p_1} \tilde c_{-p_2} \ldots \tilde c_{-p_k} \ket{0},
&
\\\label{Belem2}
\left(\bb_0+\bb_0^\dagger\right) \left(\ll_0+\ll_0^\dagger\right)^m \tilde c_{-q_1} \tilde c_{-q_2}
\ldots \tilde c_{-q_l} \ket{0}, &
\eea
where $h = n+p_1+\cdots + p_k = 1 + m+ q_1+ \cdots + q_l$. The first basis element (\ref{Belem1})
can be rewritten up to a numerical factor as
\be
\left.\frac{d^{n+(p_1+1)+\cdots+(p_k+1)}}{dr^n d\tilde x_1^{(p_1+1)} \ldots d\tilde x_k^{p_k+1}}
U_r^\bpz U_r \, \tilde c(\tilde x_1)\ldots \tilde c(\tilde x_k)
\ket{0}\right|_{\!\!\mbox{\scriptsize$\begin{array}{c} r=2\\ \tilde x_i = 0\end{array}$}}.
\ee
Multiplying two states of this form using the formula (\ref{genstarprod}) or (\ref{wwistar})
\be
U_r^\bpz U_r \, \tilde\phi_1(\tilde x) \ket{0} * U_s^\bpz U_s \, \tilde\phi_2(\tilde y) \ket{0} =
U_{r+s-1}^\bpz U_{r+s-1} \, \tilde\phi_1\left(\tilde
x+\frac{\pi}{4}(s-1)\right)\tilde\phi_2\left(\tilde y-\frac{\pi}{4}(r-1)\right) \ket{0},
\ee
we see that the total number of derivatives acting on the right hand side will be equal to the sum
of the number of derivatives acting on the two factors on the left hand side. Some of the
derivatives on the right hand side can act both on $U_t^\dagger U_t$ and the $\tilde c$ ghosts, but
regardless of where they act they always increase the $\ll_0$ eigenvalue by 1. Since setting
$r=s=2$ at the end leaves us with $U_3^\dagger U_3$ apart of powers of $\ll_0+\ll_0^\dagger$ and
modes of the $\tilde c$ ghosts, we have proven only $h \ge h_1+h_2$ and not the equality.

For the states (\ref{Belem2})) we may use the identities (see appendix~\ref{a_Bgauge})
\bea
\left(\left(\bb_0 + \bb_0^\bpz\right) \phi_1 \right) * \phi_2 &=&  \left(\bb_0 +
\bb_0^\bpz\right)\left(\phi_1
* \phi_2 \right) +(-1)^{{\rm gh}(\phi_1)} \frac{\pi}{2} \, \phi_1 * B_1 \phi_2,
\\\nonumber
\phi_1 * \left(\left(\bb_0 + \bb_0^\bpz\right) \phi_2 \right) &=&  (-1)^{{\rm gh}(\phi_1)}
\left(\bb_0 + \bb_0^\bpz\right)\left(\phi_1 * \phi_2 \right) -(-1)^{{\rm gh}(\phi_1)} \frac{\pi}{2}
\left(B_1\phi_1\right)
* \phi_2,
\\
\left(\left(\bb_0 + \bb_0^\bpz\right) \phi_1 \right) * \left(\left(\bb_0 + \bb_0^\bpz\right) \phi_2
\right) &=& -(-1)^{{\rm gh}(\phi_1)} \, \frac{\pi}{2}\left(\bb_0 + \bb_0^\bpz\right) B_1
\left(\phi_1 * \phi_2 \right) + \left(\frac{\pi}{2}\right)^2  \left(B_1\phi_1\right) *
\left(B_1\phi_2\right),
\nonumber
\eea
and thanks to the fact that  $B_1$ and $\left(\bb_0 + \bb_0^\dagger\right)$ raise the $\ll_0$
eigenvalue by one, we can reduce this case to the previous one.

Let us note that the lemma in its simplest form holds only for the assumed subsector of the string
field theory, which fortunately is big enough for the goals of the present paper.  As soon as one
starts to introduce other operator insertions such as  matter operators $\partial X$, $e^{ikX}$
etc., operator contractions seem to spoil the nice property that $h \ge h_1+h_2$. It would be nice
to find a way out, in order to be able to look efficiently for space-time dependent solutions. For
the case of Wilson line marginal deformations generated by $i \partial X$ one possibility might be
to replace in the above lemma the operator $\ll_0$ with $\ll_0+N$, where $N$ is an $\alpha'$
counting operator. This could work, since each contraction of $i \partial X$'s is accompanied by an
explicit factor of $\alpha'$.

%\newpage
%%%%%%%%%%%%%%%%%%%%%%%%%%%%%%%%%%%%%%%%%%%%%%%%%%%%%%%%%%%%%%%%%%%%%%%%%%%%%%
\sectiono{Ghost number zero toy model}
\label{s_gn0}
%%%%%%%%%%%%%%%%%%%%%%%%%%%%%%%%%%%%%%%%%%%%%%%%%%%%%%%%%%%%%%%%%%%%%%%%%%%%%%

It has been suggested \cite{GRSZpatterns}, that since $L_0$ has eigenvalue $-1$ on
$c_1 \ket{0}$, solving an equation \mbox{$(L_0-1) \ket{\Phi} + \ket{\Phi}*\ket{\Phi}=0$}
for ghost number zero field $\Phi$ could teach us something about the true ghost number one
solution. Indeed it was found that some of the coefficients of the tachyon solution in the matter
sector of the ghost number one theory were strikingly close to the corresponding coefficients in
the ghost number zero solutions. Although the precise relationship has never been discovered, and
if it exists it is very likely not a simple one, we shall start with an analogous equation
\be\label{gn0eom}
\left( \ll_0 -1\right)\Phi + \Phi*\Phi =0,
\ee
replacing $L_0$ with $\ll_0$ and hoping to find some clues for the ghost number one case.  Let us
start with an ansatz in the form
\be\label{gn0ansatz}
\Phi = \sum_{n=0}^\infty f_n \kket{n},
\ee
where we have introduced states
\be
\kket{n} =  \frac{(-1)^n}{2^n n!} \left(\ll_0 + \ll_0^\bpz\right)^n \ket{0}.
\ee
These states appear in the expansion of the wedge states
\be
\ket{r} = e^{\frac{2-r}{2}\left(\ll_0 + \ll_0^\bpz\right)} \ket{0} = \sum_{n=0}^\infty (r-2)^n
\kket{n}
\ee
and can be formally written as
\be
\kket{n} = \oint \frac{dr}{2\pi i} \frac{1}{(r-2)^{n+1}} \ket{r}.
\ee
We do not pretend here to give meaning to wedge states $\ket{r}$ with complex $r$, we use the
residue integral merely as a shorthand for taking derivatives and setting $r=2$. Thanks to the
commutation relation $\left[\ll_0, \ll_0+\ll_0^\bpz\right]= \ll_0+\ll_0^\bpz$ the states $\kket{n}$
are eigenstates of $\ll_0$
\be
\ll_0  \kket{n} = n \kket{n},
\ee
and using $\ket{r} * \ket{s} = \ket{r+s-1}$ one can derive easily their star products
\be
\kket{n} *  \kket{m} = \sum_{k=n+m}^\infty \frac{k!}{n! m! (k-n-m)!} \kket{k}.
\ee
The fact that the star product of two states with $\ll_0$ weights $n$ and $m$ contains weights only
greater or equal to $n+m$ is one of the key observations of the present paper that allowed much of
the subsequent progress. Strictly speaking, as we have mentioned earlier, the statement is correct
only in certain subsector of the string field theory. We will see in the next section that it is
fortunately large enough for the physics of tachyon condensation.

Plugging our ansatz to the equation (\ref{gn0eom}) we find a simple set of equations
\be\label{bernoulliId}
(n-1) f_n \, = \, -\!\! \sum_{\mbox{\scriptsize$\begin{array}{c} 0 \le p,q \le n \\ p+q \le n \end{array}$}}
\frac{n!}{p! q! (n-p-q)!} f_p f_q.
\ee
First equation for $n=0$ is simply $-f_0=-f_0^2$ and requires us to set $f_0 =1$ or $f_0=0$. In the
first case the rest of the coefficients $f_1, f_2, \ldots$ can be successively and uniquely
determined and will be discussed in the next subsection. In the second case one has the freedom to
set $f_1$ to an arbitrary value. These solutions resemble one parameter pure gauge solutions and we
shall comment on them in subsection \ref{ss_puregauge}.

\subsection{`Tachyon' solutions}

Let us focus on the case $f_0 =1$ first. Calculating recursively first few coefficients from the
equation (\ref{bernoulliId}) we find $f_0 =1, f_1 = -\frac{1}{2}, f_2 = \frac{1}{6}, f_3=0, f_4 =
-\frac{1}{30}, \dots $. Surprisingly these are nothing but the Bernoulli numbers, so that our
solution becomes
\be
\Phi = \sum_{n=0}^\infty B_n \kket{n}.
\ee
The Bernoulli numbers $B_n$ are one of the most important number sequences in mathematics, with
many properties, the most basic ones are for the readers convenience collected in
appendix~\ref{a_Bernoulli}. The equation (\ref{bernoulliId}) appears to be a novel identity for the
Bernoulli numbers, somewhat similar to the Euler--Ramanujan identity. We present an elementary
proof in the appendix~\ref{a_Bernoulli}.

Having found the solution to (\ref{gn0eom}) in the $\kket{n}$ basis we can express it in other
forms as well. Using the generating function for the Bernoulli numbers (\ref{BernDef}), geometric
series expansion,  wedge state conservation laws and definition of the Riemann zeta function, we
can write it in various forms
\bea\label{gn0manip}
\Phi &=& \sum_{n=0}^\infty B_n \kket{n} = \frac{1}{2}
\frac{(\ll_0+\ll_0^\bpz)}{1-e^{-\frac{1}{2}(\ll_0+\ll_0^\bpz)}} \ket{0}
\\\label{gn0manip1}
&=&  \frac{1}{2} \sum_{n=0}^\infty  (\ll_0+\ll_0^\bpz) e^{-\frac{n}{2}(\ll_0+\ll_0^\bpz)} \ket{0}
=    \frac{1}{2} \sum_{n=0}^\infty (\ll_0+\ll_0^\bpz) \ket{n+2}
\\\label{gn0sol}
&=&  \sum_{n=0}^\infty \frac{1}{n+2} \ll_0^\bpz \ket{n+2} =
\sum_{n=0}^\infty \frac{1}{2} \ll_0^\bpz \left(\frac{2}{n+2}\right)^{\ll_0^\bpz+1} \ket{0} \\
\label{gn0solzeta}
&=& \ll_0^\bpz \, 2^{\ll_0^\bpz} \left(\zeta(\ll_0^\bpz+1)-1\right) \ket{0}
\eea
demonstrating the richness (or perhaps redundancy) of our formalism\footnote{ As a side remark let
us note that by expanding $\sum B_n \kket{n}$ in the powers of $\ll_0^\bpz$ Stirling numbers of the
first kind appear naturally. Comparing this to the same expansion of $\ll_0^\bpz \, 2^{\ll_0^\bpz}
\left(\zeta(\ll_0^\bpz+1)-1\right)$ we find rather curious relation between the Stieltjes constants
$\gamma_n$ and products of Bernoulli and Stirling numbers. The sums which appear are only
asymptotic series, but they can be summed to arbitrary precision using Pad\'e approximants or
exactly via Borel summation.}. From (\ref{gn0manip}) and (\ref{gn0solzeta}) we can see that there
is formally a term of the form $0/0$ or $0\times \infty$ for $\ll_0+\ll_0^\dagger$ or
$\ll_0^\bpz=0$. One has to be therefore a bit careful. In fact (\ref{gn0manip1}) and (\ref{gn0sol})
cannot be correct, since the expressions are missing the $\ket{0}$ component. The step from
(\ref{gn0manip}) to (\ref{gn0manip1}) allows for writing the $\ll_0+\ll_0^\bpz$ factor inside or
outside the sum. If we write it outside the sum, we immediately find using (\ref{UrNO},
\ref{ucoeff}) that it acts on
\bea \sum_{n=0}^\infty \ket{n+2} &=& \sum_{n=0}^\infty \left(\ket{0}
-\frac{1}{3} L_{-2} \ket{0}+ \frac{1}{30} L_{-4} \ket{0} + \frac{1}{18} L_{-2} L_{-2} \ket{0} +
\cdots \right) +
\nonumber\\
&& +\sum_{n=0}^\infty \frac{1}{(n+2)^2} \left(\frac{4}{3} L_{-2} \ket{0} - \frac{4}{9} L_{-2}
L_{-2} \ket{0} + \cdots \right) +
\nonumber\\
&& + \sum_{n=0}^\infty \frac{1}{(n+2)^4} \left(-\frac{8}{15} L_{-4} \ket{0} + \frac{8}{9} L_{-2}
L_{-2} \ket{0} + \cdots \right) + \cdots.
\eea
All terms here are regular except the first term which is just the sliver state $\ket{\infty}$
found in \cite{RZ} multiplied by a divergent factor. Acting with $\ll_0+\ll_0^\bpz$ on
$\left(\sum_{n=0}^\infty 1\right)\ket{\infty}$ produces an ambiguous answer which has to be fixed
to be the sliver state $\ket{\infty}$ itself with unit coefficient. First it has to be in the
kernel of $\ll_0+\ll_0^\dagger$ and hence proportional to the sliver, second it has to contain the
vacuum $\ket{0}$ with unit coefficient. Adding the sliver to the (\ref{gn0sol}) and trading the
$\ll_0^\bpz$ for a derivative with respect to the wedge angle, we can rewrite it in a simple form
\be\label{gn0solfinal}
\Phi = \ket{\infty} - \sum_{n=2}^\infty \left. \frac{d}{d\alpha} \ket{n+\alpha}\right|_{\alpha=0}.
\ee
There is actually a much direct connection between the form $\Phi=\sum B_n \kket{n}$ and wedge
state representation (\ref{gn0solfinal}) which will be useful in the ghost one case. It follows
from (\ref{rinLL}) that
\bea
\sum_{n=0}^\infty  B_n \kket{n} &=& \sum_{n=0}^\infty  \frac{B_n}{n!} \left. \frac{d^n}{dr^n}
U_r^\dagger U_r \ket{0} \right|_{r=2} = \ket{\infty} - \sum_{n=0}^\infty  \frac{B_n}{n!} \left(
\left. \frac{d^n}{dr^n} U_r^\dagger U_r \ket{0} \right|_{r=\infty} -\left. \frac{d^n}{dr^n}
U_r^\dagger U_r \ket{0} \right|_{r=2}\right)
\nonumber\\
&=& \ket{\infty} - \sum_{n=2}^\infty \frac{d}{d n} \ket{n},
\eea
where we used the fact that $U_r^\dagger \ket{0} = \ket{\infty} + O(1/r^2)$ and hence all the
derivatives $\frac{d^n}{dr^n} U_r^\dagger \ket{0}$ vanish at $r=\infty$, except for $n=0$. In the
last line we have used the Euler--Maclaurin sum formula
\be\label{EMcseriesgn0}
\sum_{n=0}^\infty \frac{B_n}{n!} \left[ f^{(n)}(b) - f^{(n)}(a)\right] = \sum_{k=a}^{b-1} f'(k).
\ee
In practise, the sum formula is most often used in a form (\ref{EMc}) as a finite sum
$n=0,\ldots,N$ with a remainder $R_N$. As a series, it is usually rapidly divergent, although there
are important exceptions such as polynomials and exponentials. The Euler--Maclaurin series is often
Borel summable though, as was shown by Hardy \cite{Hardy}. In the ghost number one case we will
demonstrate for the tachyon coefficient, that the Borel summation indeed bridges the solution in
the $\ll_0$ basis written in terms of Bernoulli numbers and a corresponding sum over wedge states.

The form (\ref{gn0solfinal}) is particularly useful for showing that it is indeed a solution of the
equations of motion. For the kinetic term we find
\bea
(\ll_0-1) \Phi &=& - \ket{\infty} -\sum_{n=2}^\infty \left.\frac{d}{d\alpha}
\left((n+\alpha-2)\frac{d}{d\alpha} -1 \right) \ket{n+\alpha}\right|_{\alpha=0}
\nonumber\\
&=& -\ket{\infty}  -\sum_{n=2}^\infty (n-2)\left. \left(\frac{d}{d\alpha}\right)^2
\ket{n+\alpha}\right|_{\alpha=0},
\eea
and similarly for the interaction term
\bea
\Phi * \Phi &=& \ket{\infty} + \sum_{n,m=2}^\infty \left.\frac{d}{d\alpha} \frac{d}{d\beta}
\ket{n+m+\alpha+\beta-1}\right|_{\alpha=\beta=0}
\nonumber\\
&=& \ket{\infty} + \sum_{k=3}^\infty (k-2)\left. \left(\frac{d}{d\alpha}\right)^2
\ket{k+\alpha}\right|_{\alpha=0},
\eea
which completes our proof. To calculate the term $(\ll_0-1) \Phi$ we had to use $\ll_0
\ket{\infty}=0$, which as we discuss in the appendix \ref{a_surface}, is true only if we first
regulate the sliver by replacing it with $\ket{r}$ for large $r$, act with $\ll_0$, do all the
normal ordering and take the limit $r \to \infty$ at the end. Without the regularization one would
encounter divergent sums in the course of normal ordering.

It could seem therefore, that our solution is not that well behaved after all.
Fortunately, this is not the case, as closer inspection of (\ref{gn0solfinal}) reveals.
In fact there is a large cancellation between the two terms in (\ref{gn0solfinal}) at large levels.
A simple way to see that is to replace the sum with the integral
\be
\Phi \sim \ket{\infty} - \int_2^\infty dn \left. \frac{d}{d\alpha} \ket{n+\alpha}\right|_{\alpha=0}
\, = \ket{0},
\ee
which is quite good, albeit trivial approximation to the exact solution.

We can confirm the cancellation between the sliver and the sum parts by a more direct computation
in the standard Virasoro basis of $L_0$ eigenstates. For example the coefficient of
$\left(L_{-2}\right)^m \ket{0}$ for the sliver is $\frac{(-1)^m}{3^m m!}$, whereas for the sum part
$\sum_{n=2}^\infty \left. \frac{d}{d\alpha} \ket{n+\alpha}\right|_{\alpha=0}$ it is
\bdm
\frac{(-1)^m}{3^m m!} \sum_{n=2}^\infty \left. \frac{d}{d\alpha}
\left(1-\frac{4}{(n+\alpha)^2}\right)^m \right|_{\alpha=0}
\edm
For finite $m$ the sum
can be expressed readily in terms of Riemann zeta function and one does not see much signs of
cancellation. For $m$ very large however, this infinite sum of Riemann zeta functions can be
exactly evaluated up to small corrections
\be\label{bound2}
\sum_{n=2}^\infty \left. \frac{d}{d\alpha} \left(1-\frac{4}{(n+\alpha)^2}\right)^m
\right|_{\alpha=0} = 1 + O\left(e^{-A m^{1/3}}\right), \qquad A \gtrsim 1.21.
\ee
The error can be rigorously bounded from above by the use of Euler--Maclaurin formula, the proof is
relegated to the appendix~\ref{a_sumsliver}. We thus see almost perfect cancellation between the
two terms in (\ref{gn0solfinal}). One could look for similar cancellations for other coefficients,
we have done it also for $\left(L_{-4}\right)^m \ket{0}$. This time we find the relevant sum to be
\be\label{bound4}
\sum_{n=2}^\infty \left. \frac{d}{d\alpha} \left(1-\frac{16}{(n+\alpha)^4}\right)^m
\right|_{\alpha=0} = 1 + O\left(e^{-B m^{1/5}}\right), \qquad B \gtrsim 0.80.
\ee
In both cases there is thus a large cancellation between the sliver and the sum parts of the solution and
that is the reason why the solution has good properties in level truncation.

Let us finish this section by giving an explicit expression for the coefficients in the standard Virasoro basis.
From the form (\ref{gn0solfinal}) it is easy to find
\bea
\Phi &=& \ket{0} + \frac{8\zeta(3)-9}{3}  L_{-2} \ket{0} +\frac{-64\zeta(5)+65}{30}  L_{-4} \ket{0}
+ \frac{64 \zeta(5) -16 \zeta(3) -47}{18} L_{-2} L_{-2} \ket{0} + \cdots
\nonumber\\
&=& \ket{0} + 0.2054  L_{-2} \ket{0} - 0.04544 L_{-4} \ket{0} + 0.007248 L_{-2} L_{-2} \ket{0} +
\cdots \,.
%\nonumber\\
%&& \quad + 0.01862 L_{-6} \ket{0} - 0.002514 L_{-4} L_{-2} \ket{0} -0.0002138 L_{-2} L_{-2} L_{-2}
%\ket{0} +\cdots
\eea
Just for comparison, the sliver is
\bea
\ket{\infty} &=& \ket{0} - \frac{1}{3}  L_{-2} \ket{0} +\frac{1}{30}  L_{-4} \ket{0}
+ \frac{1}{18} L_{-2} L_{-2} \ket{0} + \cdots
\nonumber\\
&=& \ket{0} - 0.3333  L_{-2} \ket{0} + 0.03333 L_{-4} \ket{0} + 0.005556 L_{-2} L_{-2} \ket{0} +
\cdots\,.
%\nonumber\\
%&& \quad - 0.01693 L_{-6} \ket{0} - 0.01111 L_{-4} L_{-2} \ket{0} -0.006173 L_{-2} L_{-2} L_{-2} \ket{0}
%+\cdots
\eea
Although not very obvious from the first four levels, one can easily go to much higher levels, to
see clearly that the coefficients of $\Phi$ decay much faster than those of the sliver
$\ket{\infty}$.

Finally let us compare our exact solution with a solution obtained by level truncation whose first
few coefficients we give in Table~\ref{Tabgn0}.
\begin{table}[t]
\footnotesize
\begin{tabular}{|r|l|l|l|l|l|l|l|l|}
\hline
                                 & 2 & 4 & 6 & 8 & 10 & 12 & 14 &  exact \\
\hline
$\ket{0}\!\!$                      & 1 & 1 & 1 & 1 &  1 &  1 &  1 &  1  \\
\hline
$L_{-2} \ket{0}\!\!$               & 0.1060 & 0.1302 & 0.1417 & 0.1488 & 0.1537 & 0.1574 & 0.1602 & 0.2054 \\
\hline
$L_{-4} \ket{0}\!\!$               &   & -0.01002 & -0.01422 & -0.01683 & -0.01868 & -0.02008 & -0.02121 & -0.04544 \\
\hline
$L_{-2} L_{-2} \ket{0}\!\!$       &   & -0.0003507 & -0.0001659 & 5.487 $10^{-5\!\!\!}$ & 0.0002593 & 0.0004410 & 0.0006018 & 0.007248 \\
\hline
$L_{-6} \ket{0}\!\!$               &   &   & 0.002826 & 0.004204 & 0.005134 & 0.005830 & 0.006380 & 0.01862 \\
\hline
$L_{-4} L_{-2} \ket{0} \!\!$        &   &   &  0.0004060 & 0.0005317 & 0.0005840 & 0.0006052 & 0.0006107 & -0.002514 \\
\hline
$L_{-2} L_{-2} L_{-2} \ket{0}\!\!$ &   &   & -6.3985$\,10^{-5}\!\!\!$ & -9.887 $10^{-5}\!\!\!$ & -0.0001231 & -0.0001413 & -0.0001556
& -0.0002138 \\
\hline
$L_{-3} L_{-3} \ket{0}\!\!$        &   &   &  2.068  $10^{-5}\!\!\!$ & 2.565 $10^{-5\!\!\!}$& 2.736 $10^{-5}\!\!\!$& 2.788 $10^{-5}\!\!\!$
& 2.790 $10^{-5}\!\!\!$& 0\\
\hline
\end{tabular}
\caption[Ghost number zero solution in level truncation]{\small Solution of the ghost number zero
equations of motion in ordinary level truncation. The lowest level coefficients converge best to
the exact answer. The convergence is slower than in the toy model \cite{GRSZpatterns}, presumably
because the level truncation truncates not only the field but also the kinetic term.}
\label{Tabgn0}
\end{table}
The convergence of the level truncation computation is rather slow, presumably due to the fact that
in contrast to the equation $(L_0-1)\Phi + \Phi*\Phi=0$, here the level truncation affects the
equation itself by approximating $\ll_0$.

\subsection{`Pure gauge' solutions}
\label{ss_puregauge}

As we mentioned earlier there is another class of solutions to (\ref{bernoulliId}) which starts as
\be
f_0 =0, \qquad f_1 = \beta, \qquad f_2 = -2\beta^2, \qquad f_3 = 3\beta^2(2\beta-1), \qquad
f_4 = -4\beta^2(6\beta^2-6\beta +1), \ldots,
\ee
where $\beta$ is an arbitrary parameter determining the solution. It looks quite impossible to guess the form of
a general term, but the reader may check that it is given by
\be
f_n = n! \oint \frac{dz}{2\pi i} \frac{1}{z^{n+1}} \frac{ \lambda z}{\lambda e^z -1}
= - n \lambda\,  {\rm Li}_{-n+1}(\lambda) - \delta_{n,1} \lambda,
\ee
where $\lambda = \frac{\beta}{\beta-1}$ and ${\rm Li}_n(z)$ is the polylogarithm function.
The solution can be recast in a form similar to (\ref{gn0solfinal})
\bea
\label{gn0pure1} \Phi_\lambda &=& \frac{1}{2} \frac{(\ll_0+\ll_0^\bpz) \lambda }{1-\lambda
e^{-\frac{1}{2}(\ll_0+\ll_0^\bpz)}} \ket{0}
\\
\label{gn0pure2}
&=& - \sum_{n=2}^\infty \left. \lambda^{n-1} \frac{d}{d\alpha} \ket{n+\alpha}\right|_{\alpha=0},
\eea
and now it is a rather trivial task to show that (\ref{gn0pure2}) is a solution, just as we did for (\ref{gn0solfinal}).

Note that since the coefficients of the wedge states are polynomials in $1/n^2$, the solution
(\ref{gn0pure2}) is convergent for $|\lambda|\le 1$ and hence makes most sense for $\beta \in
(-\infty, \frac{1}{2})$. For $|\lambda| > 1$ one attempt could be to use (\ref{gn0pure1}) to expand
around $\lambda=\infty$, but this would generate wedge states with $n=1, 0, -1, -2, \ldots$.
Although one might think that in some sense $\ket{n}=\ket{-n}$ as is true for the coefficients, the
presence of $n=0$ seems to invalidate the expansion. In fact it follows from the empirical study in
\cite{wedge}, that wedge states $\ket{n}$ with $-1<n<1$ are not well behaved in level truncation.
In spite of that, it seems that at least some of the values for  $|\lambda| > 1$ could be
meaningful.

For example for $\lambda =+\infty$, (i.e. $\beta=1$) the series truncates after the first term and
one finds $\Phi_{\lambda=\infty} =
-\frac{1}{2}\left(\ll_0+\ll_0^\bpz\right)\ket{I}=-\frac{1}{\pi}\left(K_1^L-K_1^R\right)\ket{I}$,
where $\ket{I} = \ket{1}$ is the identity in the star algebra.\footnote{This state has been
discussed previously in \cite{HorMartin}. It was used to show that global symmetries generated by
$K_n$ can be viewed as part of the gauge symmetry of string field theory.} Although this state does
not look as ill behaved as wedge states $\ket{n}$ for $n \sim 0$, similar states have been shown to
possess anomalous properties \cite{HorStrom}.

Finally, let us note that for the special value  $\lambda =1$, i.e. $\beta =\infty$ we find
$\Phi_{\lambda=1}=- \sum_{n=2}^\infty \partial_n \ket{n}$ which looks just as (\ref{gn0solfinal})
except for the sliver part. It reminds us of Yang-Mills theory, where the instantons can be viewed
as singular limits of pure gauge configurations. In fact, one could regard our toy model 'tachyon
solution' as a $\lambda \to 1$ limit of a pure gauge solution if one defines $\Phi_\lambda= \lim_{M
\to \infty} \left[ \lambda^{M-1} \ket{M}  - \sum_{n=2}^M \lambda^{n-1} \partial_n \ket{n} \right]$
and takes the limit $\lambda \to 1$ first.

%\newpage
%%%%%%%%%%%%%%%%%%%%%%%%%%%%%%%%%%%%%%%%%%%%%%%%%%%%%%%%%%%%%%%%%%%%%%%%%%%%%%
\sectiono{Ghost number one -- the real thing}
\label{s_gn1}
%%%%%%%%%%%%%%%%%%%%%%%%%%%%%%%%%%%%%%%%%%%%%%%%%%%%%%%%%%%%%%%%%%%%%%%%%%%%%%

Let us now face the real challenge, to solve the ghost-number-one equation of motion
$Q_B\Psi+\Psi*\Psi=0$ of string field theory. As we have anticipated we will look for solutions in
the $\bb_0\Psi=0$ gauge, where $\bb_0$ was introduced in (\ref{Bdef}). We shall start by
constructing the true vacuum solution in the basis of $\ll_0$ eigenstates discussed in
section~\ref{ss_lemma}.

One of the methods used to solve \sf equations of motion which worked in the Siegel gauge was to
use a recursive approach \cite{GR}
\be
\psi \to - \frac{b_0}{L_0} \left(\Psi*\Psi\right)
\ee
starting with $\Psi^{(0)} \propto c_1 \ket{0}$.\footnote{There are some subtleties to this method,
such as the need for adjusting the overall normalization at every step and also tricks to break
some peculiar limit cycles. These issues do not affect the present discussion, the interested
reader is referred to \cite{GR}.} One could hope that the same strategy would work in our new
gauge. Starting with $\Psi^{(0)} \propto \tilde c_1\ket{0}$ and repeatedly star multiplying and
acting with $\bb_0/\ll_0$ (see (\ref{bLaction})) leaves us in a simple invariant space which we can
parameterize as
\be\label{gn1ansatz}
\Psi = \sum_{n,p} f_{n,p} \left(\ll_0 + \ll_0^\bpz\right)^n \tilde c_p \ket{0} + \sum_{n,p,q}
f_{n,p,q} \left(\bb_0 + \bb_0^\bpz\right) \left(\ll_0 + \ll_0^\bpz\right)^n \tilde c_p \tilde
c_q\ket{0},
\ee
where $n = 0,1,2,\dots$, and $p,q = 1, 0, -1, -2,\dots$. This will be thus our ansatz for finding
the exact solution. Let us now plug the ansatz into the equations of motion $Q_B\Psi+ \Psi*\Psi=0$.
For the coefficient of lowest level ghost number two state $\tilde c_1 \tilde c_0 \ket{0}$ which
appears in the equation of motion we find
\be
f_{0,1} + \pi \left[ -\frac{1}{2} f_{0,1}^2 +  f_{0,1} \left( f_{1,1} + 2 f_{0,1,0} \right) \right]
= 0.
\ee
Somewhat unexpectedly we see that imposing $\bb_0$ gauge sets $f_{1,1} + 2 f_{0,1,0}=0$ and
therefore the equation can be solved easily, giving two solutions $f_{0,1} = 0$ or  $f_{0,1} =
\frac{2}{\pi}$. As we discuss further in the appendix~\ref{a_details}, the first one corresponds to
the pure gauge transformation of the vacuum.

Going further to the next level (i.e. the $\ll_0$ eigenvalue $h=0$) we have two states. For $\tilde
c_1 \tilde c_{-1} \ket{0}$ the equation can be trivially satisfied by the usual requirement of
twist invariance which works in our gauge and basis as usually. For  $ \left(\ll_0 +
\ll_0^\bpz\right) \tilde c_1 \tilde c_0 \ket{0}$ we find
\be
f_{1,1}+ 2f_{0,1,0} + \pi \left[ \frac{1}{4} f_{0,1}^2 - \frac{3}{2} f_{0,1} f_{1,1} - f_{0,1}
f_{0,1,0} + f_{1,1}^2 + 2 f_{1,1} f_{0,1,0} + 2 f_{0,1} \left(f_{2,1}+f_{1,1,0}\right) \right] = 0
\ee
Imposing the $\bb_0$ gauge the equation reduces to
\be
\frac{1}{4} f_{0,1}^2 - f_{0,1} f_{1,1}=0
\ee
which uniquely determines $f_{1,1}= \frac{1}{4} f_{0,1} = \frac{1}{2\pi}$. Had we chosen in the
previous step $f_{0,1} = 0$ , then $f_{1,1}$ would be a free gauge parameter. It might be
surprising that we find pure gauge solutions in our $\bb_0$ gauge, the reason is that the state
$\left[\left(\ll_0 + \ll_0^\bpz\right) \tilde c_1 -  \left(\bb_0 + \bb_0^\bpz\right) \tilde c_1
\tilde c_0 \right]\ket{0}$ is annihilated by all three operators $\bb_0, Q_B$ and $\ll_0$. There
are no other states like this, as one can easily check, since the kernel of $\ll_0$ at ghost number
one is spanned by just three states. \footnote{This state is a bit reminiscent of the ghost dilaton
$\left(c_{1} c_{-1} - \bar c_{1} \bar c_{-1}\right) \ket{0}$ in closed \sft which is also a $Q_B$
exact state annihilated by $b_0$, but cannot be written as $Q_B\Lambda$ with $b_0 \Lambda =0$. It
would be therefore interesting to study spectrum and interactions of \sft around this solution.}

Using the formulas in appendix~\ref{a_Bgauge} one can show that for any states $\psi_1, \psi_2$
which satisfy $\bb_0\psi_1=\bb_0\psi_2=0$ their star product obeys
\be
\bb_0 \left(\psi_1 * \psi_2 \right) = \frac{\pi}{4} \left( B_1^R \psi_1 * \psi_2 - (-1)^{{\rm
gh}(\psi_1)} \psi_1 *  B_1^L\psi_2 \right).
\ee
Since both $B_1^L$ and  $B_1^R$ increase $\ll_0$ eigenvalue by one, the coefficient in front of a
state with $\ll_0=h$ in $\bb_0 \left(\psi_1 * \psi_2 \right)$ can receive contributions only from
components of $\psi_1$ and $\psi_2$ with $h_1$ and $h_2$ such that $h_1+h_2+1 \le h$, as we proved
in section~\ref{ss_lemma}. Acting with $\bb_0$ on the equation of motion we have
\be
\ll_0 \Psi + \bb_0 \left(\Psi * \Psi \right) = 0.
\ee
This equation presents an infinite set of equations, one for each state in the Hilbert space at
ghost number one. Let us truncate the equation (but not the string field ) to the subset of states
up to some maximal $h$. Then due to the above identity, and what we have showed above, this
truncated system will depend on exactly the right number of coefficients of the string field. This
is one of the main advantages of our choice of gauge and basis over traditional Siegel gauge and
Virasoro basis.

Solving our equations to the first two $\ll_0$ levels we find
\bea
\Psi &=& \frac{2}{\pi} \tilde c_1 \ket{0} + \frac{1}{2\pi} \left[\left(\ll_0 + \ll_0^\bpz\right)
\tilde c_1 \ket{0} - \left(\bb_0 + \bb_0^\bpz\right) \tilde c_1  \tilde c_0 \ket{0} \right]+
\\\nonumber\
&& +  \frac{1}{24\pi} \left[\left(\ll_0 + \ll_0^\bpz\right)^2  \tilde c_1 \ket{0} - 2\left(\bb_0 +
\bb_0^\bpz\right)\left(\ll_0 + \ll_0^\bpz\right)  \tilde c_1  \tilde c_0 \ket{0} \right] +
\frac{\pi}{48} \tilde c_{-1} \ket{0} + \cdots.
\eea
Continuing further becomes rather tedious, so we have written Mathematica program to do it for us
and going to higher levels we have discovered that all nonzero coefficients at level 12 have factor
691 in the numerator. This number is famous for being the prime numerator of the twelfth Bernoulli
number, so it did not take long to guess the full form of the solution
\bea\label{gn1solBern}
\Psi &=& \sum_{n=0}^\infty \sum_{p=-1,1,3,5,\ldots}  \frac{\pi^p}{2^{n+2p+1}n!} (-1)^n B_{n+p+1}
\left(\ll_0 + \ll_0^\bpz\right)^n \tilde c_{-p} \ket{0} +
\\\nonumber
&& +\sum_{n=0}^\infty \sum_{\mbox{\scriptsize$\begin{array}{c} p,q = -1\\  p+q\ \mbox{odd}
\end{array}$}}^\infty \frac{\pi^{p+q}}{2^{n+2(p+q)+3} n!} (-1)^{n+q} B_{n+p+q+2} \left(\bb_0 +
\bb_0^\bpz\right) \left(\ll_0 + \ll_0^\bpz\right)^n \tilde c_{-p} \tilde c_{-q} \ket{0}
\eea
which we have verified for the first 508 equations with 357 variables.
Actually only 260 equations with 224 variables played role due to the twist symmetry.
The details are presented in appendix~\ref{a_details}.

Direct proof that (\ref{gn1solBern}) is a solution of the equation of motion does not seem to be
easy. In fact as we have checked the proof requires an infinite number of Euler-like identities
like the one in (\ref{bernoulliId}) proved in appendix~\ref{a_Bernoulli}. Much more convenient
starting point for the proof is a form analogous to (\ref{gn0solfinal}):
\be\label{gn1sol}
\Psi = \lim_{N \to \infty} \left[\psi_N - \sum_{n=0}^N \partial_n \psi_n\right],
\ee
where\footnote{Let us note equivalent but simpler form $\psi_n = \frac{1}{\pi} c_1\ket{0} *
\left(B_1^L-B_1^R\right)\ket{n}*c_1\ket{0}$, for $n \geq 1$. Although it will not play a role in
the subsequent analysis, it is worth mentioning that after taking the derivative with respect to
$n$ and summing over it, the ghost number zero solution appears naturally. This fact can possibly
explain the quasi-pattern found in \cite{GRSZpatterns} discussed further in \cite{GR}. This new
form might be also useful for bringing the solution to the partial isometry or pure-gauge like form
advocated in \cite{BfieldSFT}. }
\be\label{gn1sol2}
\psi_n = \frac{2}{\pi^2} U_{n+2}^\bpz U_{n+2}\left[ \left(\bb_0 + \bb_0^\bpz\right) \tilde
c\left(-\frac{\pi}{4} n\right)  \tilde c\left(\frac{\pi}{4} n\right) + \frac{\pi}{2} \left(\tilde
c\left(-\frac{\pi}{4} n\right) + \tilde c\left(\frac{\pi}{4} n\right)\right) \right] \ket{0}.
\ee
One could derive that from (\ref{gn1solBern}) by similar manipulations as in (\ref{gn0manip}), or
by explicit Borel summation. The easiest way to show the equivalence however, is to realize that
just as in the ghost number zero toy model, the expressions (\ref{gn1solBern}) and (\ref{gn1sol})
are related via the Euler--Maclaurin series
\be\label{EMcseries}
\sum_{n=0}^\infty \frac{B_n}{n!} \left[ f^{(n)}(b) - f^{(n)}(a)\right] = \sum_{k=a}^{b-1} f'(k)
\ee
with $a=0$, $b=N+1 \to \infty$ and $f(k)=-\psi_k$. To see that one has to perform the derivatives
with the help of formula (\ref{eLL}). Before we move on to the somewhat involved proof of the
equation of motion, we invite the reader to check that (\ref{gn1sol}) is actually in the $\bb_0$
gauge. To see that, one needs only the anticommutator $\{\bb_0,\tilde c(\tilde z)\} = \tilde z$ and
\be\label{BUU}
\bb_0  U_{n+2}^\bpz U_{n+2} =   U_{n+2}^\bpz U_{n+2} \left[ \bb_0 -  \frac{n}{2} \left(\bb_0 +
\bb_0^\bpz\right) \right]
\ee
which follows readily from the formulas in appendix~\ref{a_Bgauge}.

%\newpage
%%%%%%%%%%%%%%%%%%%%%%%%%%%%%%%%%%%%%%%%%%%%%%%%%%%%%%%%%%%%%%%%%%%%%%%%%%%%%%
\subsection{Proof of the equation of motion}
\label{ss_eom}
%%%%%%%%%%%%%%%%%%%%%%%%%%%%%%%%%%%%%%%%%%%%%%%%%%%%%%%%%%%%%%%%%%%%%%%%%%%%%%

We shall now give a proof that (\ref{gn1sol}) is indeed a solution to the equation of motion
$Q_B\Psi+\Psi*\Psi=0$. Let us start by ignoring the first term in (\ref{gn1sol}) which, as one can
readily verify by an explicit calculation, is effectively zero. By that we mean that all its
contractions with Fock space states are zero. It can be also shown using (\ref{psi_n_psi_m}) that
it is irrelevant when star multiplied with itself or the other term in (\ref{gn1sol}).

The action of $Q_B$ on $\Psi$ is quite simple
\bea\label{QPsi}
Q_B \Psi &=& -\frac{2}{\pi^2} \sum_{n=0}^\infty \frac{d}{dn} \left\{ U_{n+2}^\bpz  U_{n+2}
\left[\left(\ll_0 + \ll_0^\bpz\right) \tilde c\left(-\frac{\pi}{4} n\right)  \tilde
c\left(\frac{\pi}{4} n\right) + \frac{\pi}{2}\left( \tilde c\partial \tilde c\left(-\frac{\pi}{4}
n\right) + \tilde c\partial \tilde c\left(\frac{\pi}{4} n\right) \right) \right.\right.
\nonumber\\
&& \qquad\qquad\qquad \left. \left. - \left(\bb_0 + \bb_0^\bpz\right) \left( \tilde c\partial
\tilde c\left(-\frac{\pi}{4} n\right)  \tilde c\left(\frac{\pi}{4} n\right)-
 \tilde c\left(-\frac{\pi}{4} n\right) \tilde c\partial \tilde c\left(\frac{\pi}{4} n\right)
\right) \right] \right\} \ket{0}.
\eea
To calculate the star product $\Psi*\Psi$ it is convenient to rewrite (\ref{gn1sol2}) as
\bea\label{gn1sol2L}
\psi_n &=& \frac{2}{\pi} U_{n+2}^\bpz U_{n+2}\left[B_1^L \tilde c\left(-\frac{\pi}{4} n\right)
\tilde c\left(\frac{\pi}{4} n\right) + \tilde c\left(-\frac{\pi}{4} n\right) \right] \ket{0}
\\\label{gn1sol2R}
&=& \frac{2}{\pi} U_{n+2}^\bpz U_{n+2}\left[-B_1^R \tilde c\left(-\frac{\pi}{4} n\right)  \tilde
c\left(\frac{\pi}{4} n\right) + \tilde c\left(+\frac{\pi}{4} n\right) \right] \ket{0}.
\eea
Using the forms (\ref{gn1sol2L}), (\ref{gn1sol2R}) and the general rules of star multiplication
from section~\ref{s_star} one finds
\bea\label{psi_n_psi_m}
\psi_n * \psi_m &=& \left( \frac{2}{\pi} \right)^2\! U_{q+2}^\bpz U_{q+2} \left[ \frac{1}{\pi}
\left(\bb_0 + \bb_0^\bpz\right) \tilde c\left(\frac{\pi}{4}q\right) \tilde
c\left(-\frac{\pi}{4}q\right) - \frac{1}{2} \left( \tilde c\left(\frac{\pi}{4}q\right)+  \tilde
c\left(-\frac{\pi}{4}q\right)\right) \right]
\nonumber\\
&&\qquad \qquad\qquad\times \left(\tilde c\left(\frac{\pi}{4} (r+1)\right) - \tilde
c\left(\frac{\pi}{4} (r-1)\right) \right) \ket{0},
\eea
where $q=n+m+1$ and $r=m-n$. It is important to note that the $q$ and $r$ dependent parts are
factorized. Moreover, when we re-express the double sum over $n$ and $m$ as
\bdm
\sum_{n,m=0}^\infty \,= \, \sum_{q=1}^\infty \! \sum_{\mbox{\scriptsize$\begin{array}{c}r\!=-q+1\\ {\rm step}\, 2\end{array}$}}^{q-1}
\edm
we see that summation of (\ref{psi_n_psi_m}) over $r$ becomes trivial and is given by the first and
last terms. Before the summation we have to of course act with $\partial_m \partial_n =
\partial_q^2 - \partial_r^2$, but that does not spoil this property. Using the identity $(\partial^2
A) B - A \partial^2 B = \partial\left(A(\leftpartial - \rightpartial) B\right)$ we find
\bea\label{PsiPsi}
\Psi*\Psi &=&  \left( \frac{2}{\pi} \right)^2 \sum_{q=1}^\infty \frac{d}{dq} \left\{ U_{q+2}^\bpz
U_{q+2} \left[ \frac{1}{\pi} \left(\bb_0 + \bb_0^\bpz\right) \tilde c\left(\frac{\pi}{4}q\right)
\tilde c\left(-\frac{\pi}{4}q\right) - \frac{1}{2} \left( \tilde c\left(\frac{\pi}{4}q\right)+
\tilde c\left(-\frac{\pi}{4}q\right)\right) \right] \right.
\nonumber\\
&& \qquad \qquad\qquad\times \left. \left(\leftpartial_q - \rightpartial_q \right) \left( \tilde
c\left(\frac{\pi}{4}q\right)-  \tilde c\left(-\frac{\pi}{4}q\right)\right) \right\} \ket{0}.
\eea
After a little manipulation one can bring (\ref{PsiPsi}) to the form of (\ref{QPsi}) with a minus
sign, and noting that the $n=0$ term of (\ref{QPsi}) actually vanishes, the equation of motion
$Q_B\Psi+\Psi*\Psi =0$ is proven.

%\newpage
%%%%%%%%%%%%%%%%%%%%%%%%%%%%%%%%%%%%%%%%%%%%%%%%%%%%%%%%%%%%%%%%%%%%%%%%%%%%%%
\subsection{Proof of Sen's first conjecture}
\label{ss_sen}
%%%%%%%%%%%%%%%%%%%%%%%%%%%%%%%%%%%%%%%%%%%%%%%%%%%%%%%%%%%%%%%%%%%%%%%%%%%%%%

Now we are going to prove Sen's first conjecture  using the explicit form of the solution
(\ref{gn1sol}). Sen's first conjecture states \cite{senconj1,senconj2}, that the energy density of
the true vacuum found by solving the open \sft equations of motion should be equal to minus the
tension of the D25 brane i.e. $-1/(2\pi^2 g_o^2)$. The energy density of a static configuration is
minus the action\footnote{Since we are interested in translationally invariant solutions, we
normalize our correlators such that they do not depend on the volume of the space. In other words
we are setting $V_{25}=1$ and we are not distinguishing between the energy and its density.} so we
are going to prove
\be
V(\Psi) = \frac{1}{g_o^2} \left[ \frac{1}{2} \aver{\Psi, Q_B \Psi} + \frac{1}{3} \aver{\Psi,
\Psi*\Psi} \right] = - \frac{1}{2 \pi^2 g_o^2} \, .
\ee
Since $\Psi$ is a solution of the equations of motion, all we have to show is that
\be
\aver{\Psi, Q_B \Psi} = - \frac{3}{\pi^2} \, .
\ee
Using the correlators from appendix~\ref{a_correlators} we find
\bea\label{anm}
\aver{ \psi_n, Q_B \psi_m} &=& \frac{1}{\pi^2}\left(1+\cos\left( \frac{\pi r}{p}\right)\right)
\left(-1 + \frac{p}{\pi} \sin\left(\frac{2\pi}{p}\right) \right)+
\\\nonumber
&& + 2\sin^2\left(\frac{\pi}{p}\right)\left[- \frac{p-1}{\pi^2} + \frac{(p-2)^2-r^2}{4\pi^2}
\cos\left( \frac{\pi r}{p}\right) + \frac{p r}{2\pi^3}\sin\left( \frac{\pi r}{p}\right)\right],
\eea
where we have introduced $p=m+n+2$ and $r=m-n$.
We also find
\be\label{ddanm}
\aver{ \partial_n \psi_n, Q_B  \partial_m \psi_m}= - \frac{4(p-1)}{p^4}
\cos\left(\frac{2\pi}{p}\right) +\frac{1}{8p^4} \left[ f_p(r+2)-f_p(r) + f_p(-r+2)-f_p(-r) \right],
\ee
where we denote
\be\label{deffp}
f_p(r) = - \left( (p-2)^2 - (r-2)^2 \right) (p^2-r^2) \cos\left( \frac{\pi r}{p}\right).
\ee
Let us re-express the double sum as
\bdm
\sum_{n,m=0}^\infty \,= \, \sum_{p=2}^\infty \! \sum_{\mbox{\scriptsize$\begin{array}{c}r\!=-p+2\\
{\rm step}\, 2\end{array}$}}^{p-2} \, ,
\edm
and observe that the special structure of (\ref{ddanm}) with the help of (\ref{deffp}) gives
readily
\be\label{zerorsum}
\sum_{\mbox{\scriptsize$\begin{array}{c}r\!=-p+2\\ {\rm step}\, 2\end{array}$}}^{p-2}
\aver{ \partial_n \psi_n, Q_B  \partial_m \psi_m}=0.
\ee
This is of course welcome since it shows that the energy of pure gauge solutions $\Psi_{\lambda} =
-\sum_{n=0}^\infty \lambda^{n+1} \partial_n \psi_n$ is manifestly zero. But it also shows that if
one carelessly interpreted (\ref{gn1sol}) as $-\sum_{n=0}^\infty \partial_n \psi_n$ one would find
zero energy, at least with the above order of summation. In fact one could find arbitrary result
since the double sum $\sum_{n,m=0}^\infty \aver{ \partial_n \psi_n, Q_B  \partial_m \psi_m}$ is not
absolutely convergent.

In our case, however, the sum is properly regularized (\ref{gn1sol}) and there is thus no ambiguity
left. We would like to stress that the regularization (\ref{gn1sol}) is in no way ad-hoc, but was
imposed on us by the use of the Euler--Maclaurin formula and confirmed by the analogy with ghost
number zero toy model. In the next two subsections we shall provide two other rather orthogonal
numerical verifications, which give the same energy with high precision.

With our regularization we thus have
\be\label{E-split}
\aver{ \Psi, Q_B \Psi} = \lim_{N \to \infty} \left[ \aver{ \psi_N, Q_B \psi_N} - 2 \sum_{m=0}^N
\aver{ \psi_N, Q_B \partial_m \psi_m} +  \sum_{n=0}^N \sum_{m=0}^N \aver{ \partial_n \psi_n, Q_B
\partial_m \psi_m}\right].
\ee
For the first term one readily finds from (\ref{anm})
\be
\lim_{N \to \infty} \aver{ \psi_N, Q_B \psi_N} = \frac{1}{2} + \frac{2}{\pi^2}.
\ee
For the third term we rewrite the sum over the square $(n,m) \in [0,N]\times [0,N]$
as a sum over the lower left and upper right triangles, i.e.
\bdm
 \sum_{n=0}^N \sum_{m=0}^N \,= \,
\sum_{p=2}^{N+2} \! \sum_{\mbox{\scriptsize$\begin{array}{c}r\!=-p+2\\ {\rm step}\,
2\end{array}$}}^{p-2} + \sum_{p=N+3}^{2N+2} \!
\sum_{\mbox{\scriptsize$\begin{array}{c}r\!=-2N+p-2\\ {\rm step}\, 2\end{array}$}}^{2N-p+2}.
\edm
The first double sum does not contribute by (\ref{zerorsum}), the second one gives
\bea
&& \sum_{p=N+3}^{2N+2} \sum_{\mbox{\scriptsize$\begin{array}{c} r\!=\!-2N\!+\!p\!-\!2\\ {\rm
step}\, 2\end{array}$}}^{2N-p+2} \!\!\!\!\!\! \aver{ \partial_n \psi_n, Q_B  \partial_m \psi_m} =
\sum_{j=1}^N \frac{4}{(2+j+N)^4} \left[ \left(j^2 - (N+1)^2 \right)
\cos\left(\frac{2\pi}{2+j+N}\right) + \right.
\nonumber\\
&& \qquad\qquad\qquad \left. + (j^2-1)(N+1)^2 \cos \left(\frac{\pi(j-N)}{2+j+N}\right) +j^2 N(N+2)
\cos\left(\frac{2j \pi}{2+j+N}\right) \right]\,.
\eea
Note that for every fixed $j$ the summand on the right hand side goes as $16\pi^2 (j^3-j)/N^4$
for large $N$. The dominant contribution comes therefore from large $j$'s.
Let us introduce $x=j/N$ and expand the summand in $1/N$ keeping $x \in (0,1]$ fixed
\be
\frac{8\pi x^2}{(1+x)^5} \sin\left(\pi\frac{1-x}{1+x}\right) \frac{1}{N} +
O\left(\frac{1}{N^2}\right).
\ee
Since the sum involves $N$ bounded terms we can safely ignore the $ O(1/N^2)$ part and the sum of
the first term is in the limit nothing but the Riemann definition of an integral. Therefore
\be
\lim_{N \to \infty} \left[ \sum_{n=0}^N \sum_{m=0}^N \aver{ \partial_n \psi_n, Q_B \partial_m
\psi_m}\right] = \int_0^1 dx \frac{8\pi x^2}{(1+x)^5} \sin\left(\pi\frac{1-x}{1+x}\right) =
\frac{1}{2} - \frac{1}{\pi^2} \,.
\ee

Similarly for the middle term in (\ref{E-split}) we find using (\ref{anm}) and
$\partial_m = \partial_p + \partial_r$ expression where we set $p=N(1+x)+2$, $r=(x-1)N$.
Expanding in $N$ keeping $x=m/N$ fixed we find again Riemann integral
\be
\lim_{N \to \infty} \left[ \sum_{m=0}^N \aver{ \psi_N, Q_B \partial_m \psi_m} \right] = \int_0^1 dx
\frac{4\pi x}{(1+x)^4} \sin\left(\pi\frac{1-x}{1+x}\right) = \frac{1}{2} + \frac{2}{\pi^2} \, .
\ee
Altogether
\be
\aver{ \Psi, Q_B \Psi} =  \frac{1}{2} + \frac{2}{\pi^2} - 2\left( \frac{1}{2} + \frac{2}{\pi^2}
\right) + \frac{1}{2} - \frac{1}{\pi^2} = - \frac{3}{\pi^2} \, ,
\ee
which completes our proof of Sen's first conjecture.

%\newpage
%%%%%%%%%%%%%%%%%%%%%%%%%%%%%%%%%%%%%%%%%%%%%%%%%%%%%%%%%%%%%%%%%%%%%%%%%%%%%%
\subsection{Transforming to the Virasoro basis}
\label{ss_L0basis}
%%%%%%%%%%%%%%%%%%%%%%%%%%%%%%%%%%%%%%%%%%%%%%%%%%%%%%%%%%%%%%%%%%%%%%%%%%%%%%

In this section we would like to demonstrate that our solution (\ref{gn1sol}, \ref{gn1sol2})
is a well behaved element of string field theory Hilbert space, just like the
Siegel gauge solution found by Sen and Zwiebach in their seminal paper \cite{SZ}.
We shall not delve here into the issue of regularity in string field theory,
instead we shall give first few coefficients of the solution in the standard Virasoro
basis, which is the one used in level truncation.

As we have already mentioned, the first term in (\ref{gn1sol}) does not contribute in level
truncation, and thus with a little manipulation we arrive to a convenient, almost normal ordered
form
\be\label{gn1solVir}
\Psi = - \frac{1}{\pi} \sum_{n=2}^\infty
 \frac{d}{dn} \left\{ U_n^\bpz
\left[ \frac{n}{\pi} \bb_0^\bpz \tilde c\left(-\frac{\pi}{2} \frac{n-2}{n} \right) \tilde
c\left(\frac{\pi}{2} \frac{n-2}{n} \right) + \tilde c\left(-\frac{\pi}{2} \frac{n-2}{n} \right) +
\tilde c\left(\frac{\pi}{2} \frac{n-2}{n} \right) \right]\right\}.
\ee
Using the explicit form of the canonically ordered wedge state (\ref{nowedge}), see also
appendix~\ref{a_surface}, and the definitions of $\bb_0^\bpz$ and $\tilde c$ we easily derive all
the coefficients at low levels. Just for illustration let us write the exact solution up to level 4
following the notation of Sen and Zwiebach:
\bea
\Psi &=& t c_1 \ket{0} + u c_{-1} \ket{0} + v  L_{-2}\, c_1 \ket{0} + w  b_{-2} c_{0} c_{1} \ket{0} +\nonumber\\
&& + A  L_{-4}\, c_1 \ket{0} + B L_{-2} L_{-2}\, c_1 \ket{0} + C c_{-3} \ket{0} + D  b_{-3} c_{-1} c_{1} \ket{0}+ \nonumber\\
&& + E  b_{-2} c_{-2} c_{1} \ket{0} + F  L_{-2} c_{-1} \ket{0} + \nonumber\\
&& + w_1 L_{-3} c_0 \ket{0} + w_2 b_{-2} c_{-1} c_{0} \ket{0} + w_3 b_{-4} c_{0} c_{1} \ket{0} + w_4  L_{-2} b_{-2} c_{0} c_{1} \ket{0}
\eea
For the first four coefficients (level 0 and level 2) we find
\bea\label{tuvwsums}
t &=& \sum_{n=2}^\infty  \frac{d}{dn} \left[ \frac{n}{\pi} \sin^2\left( \frac{\pi}{n}\right)
\left(-1 + \frac{n}{2\pi}\sin\left( \frac{2\pi}{n}\right) \right) \right]
\nonumber\\
u &=& \sum_{n=2}^\infty  \frac{d}{dn} \left[\left(\frac{4}{n\pi}- \frac{n}{\pi}\sin^2\left(
\frac{\pi}{n}\right) \right) \left(-1 + \frac{n}{2\pi}\sin\left( \frac{2\pi}{n}\right) \right)
\right]
\nonumber\\
v &=& \sum_{n=2}^\infty  \frac{d}{dn} \left[\left(\frac{4}{3 n\pi}- \frac{n}{3\pi}
\right)\sin^2\left( \frac{\pi}{n}\right) \left(-1 + \frac{n}{2\pi}\sin\left( \frac{2\pi}{n}\right)
\right) \right]
\nonumber\\
w &=& \sum_{n=2}^\infty  \frac{d}{dn} \left[ \sin^2\left( \frac{\pi}{n}\right) \left(\frac{8}{3
n\pi}- \frac{2n}{3\pi} + \frac{n^2}{3\pi^2} \sin\left( \frac{2\pi}{n}\right) \right) \right].
\eea
These sums do not appear to have simple analytic expressions, although they can be rewritten in an
interesting way using the Bernoulli numbers. We can simply expand the trigonometric functions into
their Taylor series and exchange the two infinite sums to find fast converging sums such as
\bea\label{tzeta}
t &=& \sum_{k=2}^\infty (-1)^{k} (2\pi)^{2k-1}  \frac{(2k-1)\left(2^{2k}-2k-2 \right)}{(2k+1)!}
\zeta(2k) =
\\
&=&  \sum_{k=2}^\infty (2\pi)^{4k-1} \frac{(2k-1)\left(k+1-2^{2k-1}\right)}{(2k)!(2k+1)!} B_{2k}.
\eea
For practical purposes one can keep the sums (\ref{tuvwsums}) as they are, since all summands
behave as $1/n^4$ for large $n$, and can be easily evaluated numerically with arbitrary
precision.\footnote{
To speed up the convergence it is convenient to sum first explicitly given number of terms
(e.g. first one hundred), expand the remaining terms in powers of $1/n$ keeping only first
few orders and sum them exactly using the Riemann zeta function. Note that
in the $1/n$ expansion only terms $1/n^4$, $1/n^6$, $1/n^8 \ldots$ appear. }

We have computed the exact coefficients with nine digit precision up to level 10, some numerical
results are given in appendix~\ref{a_Virasoro}.\footnote{To be completely honest, at level 10 due
to inefficiency of our computer program, we left our solution expressed in terms of $L_n^{tot}$ and
the ghosts, instead of $L_n^{matter}$ and the ghosts. This does not affect the check on the D-brane
energy as given below, which we were primarily interested in.} Let us present here the complete
list of the exact coefficients up to level 4:
\bdm\small
\begin{array}{|cclcclcclccl|}
\hline
t &=&   0.55346558  & \quad A &=&  -0.030277583   & \quad E &=&   0.17942652 & \quad w_1 &=& 0
\nonumber \\
u &=&   0.45661043  & \quad B &=&   0.0045805832  & \quad F &=&   0.022748278 & \quad w_2 &=& 0.020943544
\nonumber \\
v &=&   0.13764616  & \quad C &=&  -0.16494614    &         & &               & \quad w_3 &=& 0.088982260
\nonumber \\
w &=&  -0.14421001  & \quad D &=&   0.16039444    &         & &               & \quad w_4 &=& -0.0084696519
\nonumber \\
\hline
\end{array}
\edm One thing one can do with these coefficients is to check whether the solution obeys the
expected symmetries. From the explicit form  (\ref{gn1sol}, \ref{gn1sol2}) or  (\ref{gn1solVir})
one sees that $K_1^{matter} \Psi=0$, since $K_1^{matter}$ commutes with all the terms, including
the common factor $U_n^\bpz U_n$. It is a general rule, that if a solution of $Q_B\Psi+\Psi*\Psi=0$
is annihilated by a star algebra derivative~$D$, it must be annihilated also by $[Q_B,D]_{\pm}$.
For the case at hand, it is easy to check
\bea\label{LTsym}
5 A + 3 B + v &=& 0 \\
w_1 &=& 0 \\
20 A + 12 B + 4 D - 4 F - 8 w_1 &=& 0 \\
15 A + 9 B + v + w - 10 w_1 + 5 w_3 + 3 w_4 &=& 0
\eea
as dictated by $K_1^{matter} \Psi = [K_1^{matter},Q_B] \Psi = 0$. Actually one can see those
identities to be true also from the explicit expressions (\ref{tuvwsums}) before the sums are
carried out.

In Siegel gauge there is somewhat unexpected $SU(1,1)$ symmetry
\cite{SiegelZw,HS,Trimming,GR}, which implies $(c\partial c)_0 \Psi =0$.
Note that $(c\partial c)_0$ is a star algebra derivative, whose commutator with $Q_B$ is zero.
This invariance enforces the Siegel gauge $b_0 \Psi =0$. In addition it implies a constraint $C+3D=0$
and even more constraints at higher levels. It is definitely of some interest to see whether our $\bb_0$
gauge solution possesses similar symmetries. Given the fact that we have expressions like (\ref{tuvwsums})
we can look for such symmetries systematically. Surprisingly we have found one more independent identity
\be
2A+4D-3E+2F-3w_2+3w_4=0.
\ee
We were not able to find any simple origin, it might be just an accidental symmetry. Apart of
$K_1^{matter} \Psi = [K_1^{matter},Q_B] \Psi = 0$ there is one more obvious symmetry $K_1 B_1
\bb_0^\bpz \Psi =0$ which gives some exact constraints manifest in level truncation, but they
become nontrivial only at level~$6$. To complete the discussion of symmetries we remind the reader
at this point of the obvious twist symmetry
\be
(-1)^{L_0-1} \Psi = \Psi,
\ee
which we in fact imposed when solving the equations of motion in the $\ll_0$ basis, and which our
solution shares with the Siegel gauge solution. Finally, there is yet another symmetry, which as
far as we can see, is obvious only from the solution (\ref{gn1solBern}) in the $\ll_0$ basis. All
the terms with $\ll_0$ eigenvalue equal to $m$, are multiplied by the Bernoulli number $B_{m+1}$.
Now, all odd index Bernoulli numbers vanish except $B_1$ and it turns out that the term multiplied
by $B_1$ is annihilated by all three operators $Q_B$, $\bb_0$ and $\ll_0$. We can thus write the
symmetry as
\be\label{Lsym}
(-1)^{\ll_0-1} \ll_0 \Psi = \ll_0 \Psi.
\ee
It would be interesting to see if it can be translated to the Virasoro basis.

One slight disadvantage of the $\bb_0$ gauge is that the gauge fixing condition is broken by level
truncation. As we noted earlier, by virtue of (\ref{BUU}) the solution indeed obeys $\bb_0\Psi =0$
exactly, but after truncating it to level $4$ the gauge conditions become
\bea
w_i &=& 0, \qquad i=1,2,3,4 \\
\frac{2}{3} E + w &=& 0,
\eea
out of which only $w_1=0$ is true exactly. The last condition is true within $83 \%$ and for
$w_{2,3,4}$ we can only say that they are two to three times smaller than similar coefficients
without $c_0$. We hope that this level dependent gauge fixing would not pose problems and that the
numerical high level computations of Moeller and Taylor \cite{MT} and Gaiotto and Rastelli
\cite{GR} would converge to our solution. On the other hand, we do not expect convergence
properties superior to the Siegel gauge, because of our experience with the ghost number zero
equation discussed in section~\ref{s_gn0}.

Finally we would like to demonstrate that our solution yields the correct D25-brane energy density
also in level truncation. Note that this is the main problem with the identity based solutions
\cite{TT}. To check the energy we have evaluated the kinetic term $\aver{\Psi, Q_B \Psi}$ up to
level 10. The values are summarized in the following table
\begin{table}[h]
\begin{center}
\begin{tabular}{|l|l|l|l|l|l|}
\hline
$L=0$  & $L=2$ & $L=4$ & $L=6$  & $L=8$ & $L=10$ \\
\hline
-1.007766 & -1.007815 & -1.004499 & -1.003217  & -1.002556 & -1.002130 \\
\hline
\end{tabular}
\caption[Energy in level truncation]{\small Energy density normalized by the D-brane tension at various levels of truncation
of the exact solution. The numbers which appear are $\aver{\Psi, Q_B \Psi}$ divided by $3/\pi^2$. }
\end{center}
\end{table}

%\newpage
%%%%%%%%%%%%%%%%%%%%%%%%%%%%%%%%%%%%%%%%%%%%%%%%%%%%%%%%%%%%%%%%%%%%%%%%%%%%%%
\subsection{Pad\'e approximants and Borel summation}
\label{ss_pade}
%%%%%%%%%%%%%%%%%%%%%%%%%%%%%%%%%%%%%%%%%%%%%%%%%%%%%%%%%%%%%%%%%%%%%%%%%%%%%%

Instead of passing through the representation in terms of wedge states or using level
truncation one could attempt to compute the energy density or coefficients in the Virasoro basis
directly from the tachyon solution
(\ref{gn1solBern}) written in terms of Bernoulli numbers.
As we shall see both tasks lead to divergent series, but ones which can be handled with.
Let us start by `regularizing' our solution by replacing $\Psi$ with $z^{\ll_0} \Psi$.
In the $\ll_0$ level expansion different levels will acquire different integer powers of
$z$. The 'regularization' is then removed in the limit $z \to 1$. We have put regularization in
quotation marks, since as we shall see $z$ does not quite regularize the energy nor the Virasoro coefficients
but merely provides an expansion parameter for an asymptotic series.

%\subsubsection*{Energy}
\bigskip\noindent
{\it\large  Energy}
\medskip

\noindent Let us start with the computation of the energy as a formal expansion in $z$. Using the
explicit solution (\ref{gn1solBern}) and few correlators from appendix~\ref{a_correlators} we
arrive to \footnote{ The closed form expression we found for the series contains six fold sum of a
product of two Bernoulli numbers, six factorials (five of them in the denominator) and some powers
of $2$ and $\pi$. We didn't dare to simplify it, however we noticed that at given order of $z^{n}$
the term with the highest power of $\pi$ simplifies to $-\frac{1}{2(n+2)n!}|B_{n+2}|
\left(\frac{\pi}{2}\right)^n$ for $n \ge 1$. There is an easy proof which uses the Euler identity
(\ref{Euler}). }
\bea\label{z_expansion_Energy}
\aver{ \Psi,z^{\ll_0^\bpz} Q_B z^{\ll_0} \Psi} &=& -\frac{4}{\pi^2 z^2} + \left( \frac{1}{12} +
\frac{1}{3\pi^2} \right) - \left(\frac{1}{90} + \frac{\pi^2}{1920} \right) z^2 + \left(
\frac{17}{5040} - \frac{11\pi^2}{17920}- \frac{\pi^4}{193536}\right) z^4
\nonumber\\
&& + \left(-\frac{113}{60480}+\frac{2413\pi^2}{1935360} - \frac{137 \pi^4}{5806080}
-\frac{\pi^6}{22118400} \right) z^6 +\cdots.
\eea
Trying to evaluate the series numerically for $z=1$ one immediately finds that the series is
divergent. The most common method for dealing numerically with divergent series is the Pad\'e
approximation. This, a bit mysterious, but often very successful method approximates a series
outside its radius of convergence by a rational function. Given a formal power series $f(z) \sim
\sum a_n z^n$, Pad\'e approximant $P_M^N(z)$ is a ratio of two polynomials of degree $N$ and $M$,
such that its power series matches the one of $f(z)$ up to $z^{M+N}$.
\begin{table}[ht]
\begin{center}
\begin{tabular}{|l|l|l|}
\hline
  & $P_2^{2n}$ & $P_{n+2}^n$ \\
\hline
$n=0$  & -1.3333                            & -1.33333333 \\
\hline
$n=2$  & -1.0015                            & -0.99501646 \\
\hline
$n=4$  & -0.98539                           & -1.00100097 \\
\hline
$n=6$  & -1.0327                            & -1.00032831 \\
\hline
$n=8$  & -1.3054                            & -1.00042520 \\
\hline
$n=10$ & \, 6.7582                          & -1.00003423 \\
\hline
$n=12$ & \, 256.34                          & -0.99999846 \\
\hline
$n=14$ & -21575.                            & -0.99999945 \\
\hline
$n=16$ & -3.6391$\times 10^6$               & -0.99999819 \\
\hline
$n=18$ & \, 6.5671$\times 10^7$             & -1.00000064 \\
\hline
\end{tabular}
\caption[Energy via Pad\'e approximation]{\small The Pad\'e approximation for the normalized energy
$\frac{\pi^2}{3}\aver{ \Psi,z^{\ll_0^\bpz} Q_B z^{\ll_0} \Psi}$ evaluated at \mbox{$z=1$}. The
first column is in fact a trivial approximation, a naively summed series with behavior typical for
asymptotic series. The second column nicely confirms Sen's first conjecture despite somewhat
irregular convergence at higher orders. } \label{t_Pade}
\end{center}
\end{table}
In table \ref{t_Pade} we give a Pad\'e approximation $P_{n+2}^n$ for even $n=0,\ldots, 18$ and compare it with
the naive evaluation which can be viewed as $P_2^{2n}$. Note that both $P_{n+2}^n$ and $P_2^{2n}$ match
(\ref{z_expansion_Energy}) to the same order.

The first column with $P_2^{2n}$ is also nothing but the definition of the energy in the
$\ll_0$-level truncation. Interestingly we see that with level $2$ we get very close to the exact
value and up to level $6$ we are still within few percent. At higher levels the divergent character
of the series starts to show up.\footnote{ This raises the unwelcome possibility that ordinary
$L_0$-level truncation in Siegel gauge would show up a similar behavior, perhaps at some higher
level $\gtrsim 20$. The overshooting of the correct energy at level $14$ found by Gaiotto and
Rastelli \cite{GR} could be attributed to it. It is not clear to us whether the high level
extrapolations of \cite{wati-tc,GR} resolve the issue.}

Given the relative ease of evaluating (\ref{z_expansion_Energy}) we carried out the expansion up to
$z^{50}$ and to our surprise we found that to this order the Pad\'e approximations do not improve
much beyond the $10^{-6}$ accuracy. Looking separately at the contribution of the first term in
(\ref{gn1solBern}) only, we found that the convergence is rather irregular with a rough pattern of
plateaux of constant accuracy and occasional bigger jumps towards better accuracy. It seems that to
reach accuracy of $10^{-9}$ one would need at least a Pad\'e approximant $P_{52}^{50}$. The
somewhat irregular convergence is in sharp contrast with the behavior of other series such as the
celebrated Euler series $\sum (-1)^n n! z^n$ or $\sum B_n z^n$, (for which we know the exact answer
by Borel summation), and where we checked that the Pad\'e approximants converge to the exact answer
monotonically.

%\subsubsection*{Tachyon coefficient}
\bigskip\noindent
{\it\large Tachyon coefficient}
\medskip

Let us now see how one can get the tachyon coefficient $t$ directly from (\ref{gn1solBern}). As in
the case of the energy we 'regularize' our solution by considering $z^{\ll_0} \Psi$ and the tachyon
coefficient will become $z$ dependent. Let us write $t=t_1+t_2$ for the two contributions coming
from the two terms in (\ref{gn1solBern}). Using (\ref{tczmp}) from appendix~\ref{a_correlators} one
finds
\bea\label{t1z}
t_1(z) &=& \sum_{n=0}^\infty \sum_{\mbox{\scriptsize$\begin{array}{c}p=-1 \\ p \, {\rm
odd}\end{array}$}}^\infty z^{n+p} \frac{(-1)^n}{n!} \left(\frac{\pi}{2}\right)^p
\frac{B_{n+p+1}}{2^{n+p+1}} (p-1)_n \left[\frac{2^p}{(p+1)!}(-1)^{\frac{p+1}{2}} + \frac{1}{2}
\delta_{p,-1} \right], \qquad
\\\label{t2z}
t_2(z) &=& \sum_{n=0}^\infty \sum_{\mbox{\scriptsize$\begin{array}{c}p=-1 \\ p \, {\rm
odd}\end{array}$}}^\infty \sum_{\mbox{\scriptsize$\begin{array}{c}q=0 \\ q \, {\rm
even}\end{array}$}}^\infty z^{n+p+q+1} \frac{(-1)^{n+q}}{n!} \left(\frac{\pi}{2}\right)^{p+q}
\frac{B_{n+p+q+2}}{2^{n+p+q+2}} (p+q)_n
\\\nonumber
&& \qquad \qquad \qquad \qquad \times \left[\frac{2^p}{(p+1)!}(-1)^{\frac{p+1}{2}} + \frac{1}{2}
\delta_{p,-1} \right] \left[\frac{2^q}{(q+1)!}(-1)^{\frac{q+2}{2}} - \delta_{q,0} \right],
\eea
which as one can easily check are again divergent series due to the presence of Bernoulli numbers
and the Pochhammer symbol $(x)_n=x(x+1)\ldots(x+n-1)$. Before attempting the Borel summation which
may not always work and often requires some labor we propose as a rule of thumb to check the series
first with Pad\'e approximants.\footnote{There is actually a theorem that under certain conditions
both Borel summation and Pad\'e approximation lead to the same result. I thank J.~Fischer for a
discussion on this issue.} We have found that the Pad\'e approximants $P_N^N$ to $t_{1,2}(z)$
evaluated at $z=1$ approach the expected values
\bea\label{t1}
t_1 &=& \frac{\pi}{2} + \sum_{k=1}^\infty (-1)^k (2\pi)^{2k+1} \frac{k}{(2k+2)!} \zeta(2k+1) \,=\, 0.277658977\ldots
\\\label{t2}
t_2 &=& -t_1 + \sum_{k=1}^\infty (-1)^{k} (2\pi)^{2k-1}  \frac{(2k-1)\left(2^{2k}-2k-2
\right)}{(2k+1)!} \zeta(2k) \, =\, 0.275806609\ldots
\eea
in a similar manner as the energy, or perhaps a bit faster. Originally we found these exact
expressions by performing the Borel summation, but they can be most easily derived from
(\ref{gn1sol}, \ref{gn1sol2}). Note that the first term in (\ref{gn1sol}) contributes $\pi/2$ in
(\ref{t1}) and $-\pi/2$ in (\ref{t2}) inside $-t_1$. Although it is not true in general that the
Pad\'e approximation to a sum of functions is a sum of their Pad\'e approximations, we see that the
sum of Pad\'e  approximants to $t_1$ and $t_2$ at $z=1$ approaches the correct value
$t=0.553465587\ldots$. One could perform the Pad\'e approximation directly for the sum $t_1+t_2$
with the same results, although for finite $N$ the results differ.

Let us now sketch how one can perform Borel summation for the series (\ref{t1z}, \ref{t2z}). This
was actually our first computation of $t$ before we discovered the simple representation in terms
of wedge states. First observe that both expressions  (\ref{t1z}, \ref{t2z}) contain a common part
which can be summed separately for $r \ge 2$ and $\re z> 0$
\be
\sum_{n=0}^\infty  (r-2)_n \frac{B_{n+r}}{n!} \left(\frac{z}{2}\right)^{n+r-1} = (-1)^r \left[
\frac{(r-1)(r-2)}{z} + \frac{r-2}{2}+\frac{z}{12} -
\sum_{j=0}^{r-3}\left(\frac{2}{z}\right)^{j+2}\!\!\! \frac{r!\,\,
\psi_{j+1}\left(\frac{2}{z}\right) }{j!(j+3)!(r-j-3)!} \right].
\ee
In this formula $\psi_{n}(z) = (-1)^{n+1} n! \sum_{k=0}^\infty (z+k)^{-n-1}$
denotes the polygamma function.
For lower values $r=0$ and $r=1$ the sum on the left hand side terminates and one finds
$2z^{-1} + 1 + z/12$ and $\, -(z+6)/12$ respectively.
Using this result one can readily derive
\bea\label{t1zfin}
t_1(z) &=& \frac{\pi}{2z} + \sum_{k=1}^\infty (-1)^k \left(\frac{2\pi}{z}\right)^{2k+1}
\frac{k}{(2k+2)!} \zeta\left(2k+1,\frac{2}{z}\right)
\\\label{t2zfin}
t_2(z) &=& -t_1(z) + \frac{1}{z} \sum_{k=1}^\infty (-1)^{k} \left(\frac{2\pi}{z}\right)^{2k-1}
\frac{(2k-1)\left(2^{2k}-2k-2 \right)}{(2k+1)!} \zeta\left(2k,\frac{2}{z}\right),
\eea
where $\zeta(n,z) = \sum_{k=0}^\infty (k+z)^{-n}$ is the Hurwitz zeta function.
For the sum $t(z)=t_1(z)+t_2(z)$ we find finally
\bea\label{tzfin1}
t(z) &=&  + \frac{1}{z} \sum_{k=1}^\infty (-1)^{k} \left(\frac{2\pi}{z}\right)^{2k-1}
\frac{(2k-1)\left(2^{2k}-2k-2 \right)}{(2k+1)!} \zeta\left(2k,\frac{2}{z}\right)
\\\label{tzfin2}
&=& \sum_{n=0}^\infty  \frac{d}{d(n z)} \left[ \frac{2+nz}{\pi} \sin^2\left(
\frac{\pi}{2+nz}\right) \left(-1 + \frac{2+nz}{2\pi}\sin\left( \frac{2\pi}{2+nz}\right) \right)
\right].
\eea
Let us make few comments. The easiest way to obtain (\ref{tzfin2}) is by using the wedge state
representation (\ref{gn1sol}, \ref{gn1sol2}). Note that the action of $z^{\ll_0}$ effectively
replaces all factors of $n$ with $nz$. From (\ref{tzfin1}) for $z=1$ follows immediately
(\ref{tzeta}), one has to use the identity $\zeta(n,2)=\zeta(n)-1$ and observe that the term $-1$
does not contribute. Finally observe that the function $t(z)$ is holomorphic at $z=\infty$ (the
functions $t_1(z)$ and $t_2(z)$ have first order poles there which cancel each other). On the
contrary it has an essential singularity at $z=0$ as can be seen from (\ref{tzfin2}), since $z=0$
is a cumulation point of essential singularities at $z=-2/n$. This explains why the series
(\ref{t1z}) and (\ref{t2z}) have zero radius of convergence.

%%%%%%%%%%%%%%%%%%%%%%%%%%%%%%%%%%%%%%%%%%%%%%%%%%%%%%%%%%%%%%%%%%%%%%%%%%%%%%
\sectiono{Conclusions and outlook}
\label{s_conclusions}
%%%%%%%%%%%%%%%%%%%%%%%%%%%%%%%%%%%%%%%%%%%%%%%%%%%%%%%%%%%%%%%%%%%%%%%%%%%%%%

We have found the first exact and fully explicit nonsingular solution describing the
non-perturbative tachyon vacuum in Witten's cubic open bosonic \sf theory. We also definitely
proved Sen's first conjecture, which relates the value of the tachyon potential at the minimum to
the D-brane tension known from the annulus computation. Good evidence was presented that our
solution is quite regular from the point of view of level truncation. It would be interesting to
confirm it by direct numerical computations.

We have presented our solution in two different forms. In the first form, the solution is written
in the basis of $\ll_0$ eigenstates, and is given in terms of Bernoulli numbers. In this basis it
was rather straightforward to find the solution, although it was not easy to prove that it actually
{\em is} a solution. Another advantage of this basis is that it is rather easy to study exactly a
large sector of the full infinite-dimensional gauge symmetry and thus clearly discriminate the
tachyon solution from pure gauge solutions.

The second form can be most elegantly obtained from the first one by noticing that it is an
Euler--Maclaurin series of certain sum over wedge states with ghost insertions. For this form of
the solution it was fairly easy to prove that it solves the equations of motion, and that it obeys
Sen's first conjecture.

Clearly now we are at a stage where many new exciting things can be done. There are still two other
Sen's conjectures that remain to be proved. We believe that with the tools developed in this paper
the cohomology of the kinetic operator at the vacuum can be studied rather easily, and hopefully
shown to be empty. To study space-time-dependent solutions, such as higher codimension D-branes or
rolling tachyon backgrounds could also be possible with the presented methods, although the
presence of nontrivial contractions among matter operators makes their study a challenge. We hope
that one could also study the question of how closed strings emerge at the tachyon vacuum.  It
seems very likely that our techniques could be used for efficient computation of off-shell string
amplitudes in the $\bb_0$ gauge, which would be much simpler than those in the Siegel gauge.

In this paper we focused solely on the open bosonic \sf theory. It seems quite possible that our
methods extend to the Berkovits super\sf theory, since it is based on Witten's associative star
product. On the other hand, for closed string field theory \cite{closed}, we are much less
optimistic because of the multitude of higher order vertices and especially because of the level
matching condition $b_0^{-} \Psi =0$, which does not fit well into our algebraic framework. So far
all attempts to eliminate the level matching condition, or put it on the same footing as a gauge
choice have been unsuccessful.

\section*{Acknowledgments}

I am grateful to Ian Ellwood for collaboration at the initial stages of this work. I am indebted to
Barton Zwiebach for useful discussions and especially for detailed reading of the manuscript which
helped clarify many issues and improve the presentation. I would also like to thank organizers of
Benasque Center for Science workshop on string theory for providing lovely environment while this
work was in progress.

\newpage
\appendix

%%%%%%%%%%%%%%%%%%%%%%%%%%%%%%%%%%%%%%%%%%%%%%%%%%%%%%%%%%%%%%%%%%%%%%%%%%%%%%
\sectiono{Comments on surface states}
\label{a_surface}
%%%%%%%%%%%%%%%%%%%%%%%%%%%%%%%%%%%%%%%%%%%%%%%%%%%%%%%%%%%%%%%%%%%%%%%%%%%%%%

The wedge states discussed extensively in section~\ref{s_star} are a prime example of more general
surface states \cite{RZ, RSZ-BCFT}. The surface states are in one-to-one correspondence with
conformal maps $f(z)$ holomorphic inside the unit disk $|z|<1$. They are defined by the relation
\be\label{surfacedef}
\aver{f | \phi} = \aver{f\circ \phi}, \qquad \forall \phi.
\ee
In the operator formalism they can be expressed as $\bra{f} = \bra{0} U_{f}$, where $U_f =
\exp(\sum v_n L_n )$ is an exponential of non-negatively moded Virasoro generators. The
coefficients $v_n$ can be thought of as Laurent coefficients of a vector field $v(z) = \sum v_n
z^{n+1}$ which is related to the map $f(z)$ by the Julia equation $v(z)\partial_z f(z) = v(f(z))$.
In practise, given $v(z)$ the equation is fairly easy to integrate to find $f(z)$
\cite{GRSZ-projectors}, the inverse problem is much harder and usually one has to resort to an
iterative procedure to determine the coefficients of the vector field. One class of solutions
\cite{LPP2} is particularly useful however, the maps
\be
f_{n,t}(z) = \frac{z}{\left(1-t n z^{n}\right)^{1/n}}
\ee
are generated by a vector field $v(z)= t z^{n+1}$, so that $U_{f_{n,t}} = e^{t L_n}$. These maps
played pivotal role recently in the study of butterfly projectors
\cite{GRSZ-ghost,butt,GRSZ-projectors} within the context of vacuum string field theory.

Apart of their importance for the butterfly projectors these maps can be taken as some kind of a
basis for holomorphic maps. Any map $f(z)$ holomorphic at the origin $z=0$ and vanishing there can
be uniquely decomposed as
\be\label{fdecomp}
f(z) = f_{0,t_0} \circ f_{1,t_1} \circ f_{2,t_2} \circ \ldots \, .
\ee
In a sense this is a complete parametrization of the space of conformal maps holomorphic at the
origin.\footnote{There is also another, as far as we can see unrelated, parameterization of this
space using harmonic moments which can be thought of as times of dispersionless Toda hierarchy.
This has been applied to the study of wedge states \cite{BR} and of the three-vertex \cite{BS}.}
Given a power series expansion around the origin, this decomposition is unique. It can be easily
implemented on a computer since $f_{n,t} = z + t z^{n+1} + O(z^{2n+1})$. The decomposition
(\ref{fdecomp}) is useful because using the composition rule $U_{f \circ g} = U_f U_g$ (reflecting
the fact that $U_f$ form a representation of the conformal group) the operator $U_f$ can be written
as
\be\label{Ufdecomp}
U_f =  e^{t_0 L_0} \, e^{t_1 L_1} \, e^{t_2 L_2} \ldots \, .
\ee
For the surface states $\bra{f} = \bra{0}U_f$ the first two exponentials are of course
irrelevant. Expanding the other exponentials in powers of $L_n$ yields automatically
canonically ordered form
\be
\bra{f} = \sum_{k_2,k_3,k_4,\ldots} \frac{t_2^{k_2} t_3^{k_3} t_4^{k_4}\ldots}{k_2! k_3! k_4! \ldots}
 \bra{0} L_{2}^{k_2} L_{3}^{k_3}L_{4}^{k_4} \ldots,
\ee
which is very useful in level truncation. This decomposition was found to take very simple form for
the identity state \cite{EFHM} and also for the 'nothing state' projector in \cite{butt,
GRSZ-projectors}.

For the purposes of the present paper, where we use at several occasions the level truncation to
check and illustrate certain exact computations, we need a decomposition (\ref{Ufdecomp}) for the
wedge states. With the help of a computer we easily find for $f_r = \tan\left(\frac{2}{r}\arctan
z\right)$
\be\label{UrNO}
U_r \equiv U_{f_r} = \left(\frac{2}{r}\right)^{L_0} e^{u_2 L_2} e^{u_4 L_4}  e^{u_6 L_6} e^{u_8
L_8} e^{u_{10} L_{10}} \ldots,
\ee
where the coefficients $u_n$ are given by
\bea\label{ucoeff}
u_2 &=& -\frac{r^2-4}{3r^2} \nonumber\\
u_4 &=& \frac{r^4-16}{30r^4} \nonumber\\
u_6 &=& -\frac{16 (r^2-4)(r^2-1)(r^2+5)}{945 r^6} \nonumber\\
u_8 &=& \frac{(r^2-4)(109 r^6 + 436 r^4 - 944r^2+1344)}{11340 r^8} \nonumber\\
u_{10} &=& -\frac{16 (r^2-4)(r^2-1)(9r^6+45 r^4-64r^2 + 160)}{22275 r^{10}} \, .
\eea
Note that all the coefficients vanish for $r=2$, i.e. the vacuum, but also $u_6=u_{10}=0$ for $r=1$
in accord with the observation of Ellwood et al.~\cite{EFHM}.

%\subsubsection*{Conservation laws}
\bigskip\noindent
{\it\large Conservation laws}
\medskip

Conservation laws in string field theory are quite a useful tool. We use them to tell us what is
the action of a given mode of an arbitrary operator on a surface state, or any kind of $n$-vertex.
They were first studied systematically by Rastelli and Zwiebach in \cite{RZ}, although some of them
appeared in the literature much earlier. In what follows we will be mainly interested in the so
called Virasoro conservation laws associated with the energy-momentum tensor and for simplicity we
shall assume zero central charge.

The basic conservation laws for arbitrary surface state $\bra{f}=\bra{0} U_f$ can be
written trivially as
\be\label{conslawtriv}
\bra{f} f^{-1} \circ L_{-n} = 0
\ee
since $f^{-1} \circ L_{-n} = U_f^{-1} L_{-n} U_f$ and $L_{-n}$ annihilates the vacuum $\bra{0}$. In
the language of \cite{RZ} one would write
\be\label{vTglob}
\bra{f} \oint v^{w}(w) T_{ww}(w) dw = 0
\ee
for any vector field in the global coordinate $w$ that is holomorphic everywhere including infinity
except possibly the puncture. Transforming to the local coordinate $z$ gives
\be
\bra{f} \oint v^{z}(z) T_{zz}(z) dz = 0.
\ee
Using the transformation law for the vector field $v^{z}(z) = (f'(z))^{-1} v^{w}(w)$ one finds
conservation laws
\be
\bra{f} \res{w} \frac{f(w)^{-n+1}}{f'(w)} \sum_{m=-n}^\infty \frac{L_m}{w^{m+2}} = 0,
\ee
which are identical to (\ref{conslawtriv}). An example of such conservation laws for the sliver was
given in (\ref{sliverconslaw}).

The disadvantage of this form of conservation laws is that it expresses an action of an operator
$L_n$ for $n \ge -1$ not as a sum of operators $L_k$ with $k \le -2$ acting on $\ket{f}$ but it
involves also operators $L_k$ with $-2 < k < n$, for which one would like to use the conservation
laws again. It is clearly desirable to have more direct way of writing the conservation laws one
needs.

There is actually a trick to do that. The key observation is that the $b(z)$ ghost has the same conservation
laws as the energy-momentum tensor $T(z)$ with zero central charge. For the $b$ ghost the conservation
laws in the right form follow readily from the Neumann matrix representation. Thus the task is reduced to
find the Neumann matrix such that in the ghost sector
\be
\bra{f}_{ghost} = \bra{0} e^{\sum c_p S_{pq} b_q} .
\ee
There is a simple way to do it.\footnote{I thank Barton Zwiebach for suggesting the method.} We
can simply evaluate the correlator
\be
\bra{f} b_{-n} c_{-m} c\partial c \partial^2 c(0) \ket{0}
\ee
in two different ways and match the results.
Using the fact that $\bra{f} = \bra{0} U_f$ and performing the conformal
transformation on the ghosts we find
\be
\bra{f} b(z) c(w) c\partial c \partial^2 c(0) \ket{0} = \frac{\left(f'(z)\right)^2}{f'(w)}
\left(\frac{f(w)}{f(z)}\right)^3 \frac{2}{f(w)-f(z)},
\ee
and therefore
\be
\bra{f} b_{-n} c_{-m} c\partial c \partial^2 c(0) \ket{0} = \oint \frac{dz}{2\pi i} \oint
\frac{dw}{2\pi i} \frac{1}{z^{n-1}} \frac{1}{w^{m+2}} \frac{\left(f'(z)\right)^2}{f'(w)}
\left(\frac{f(w)}{f(z)}\right)^3 \frac{2}{f(w)-f(z)}.
\ee
On the other hand using the normalization $\aver{c\partial c \partial^2 c(0)}=-2$
(i.e. $\aver{c_{-1} c_0 c_1}=1$) we get
\be
\bra{0} e^{\sum c_p S_{pq} b_q} b_{-n} c_{-m} c\partial c \partial^2 c(0) \ket{0} = 2 S_{nm}
\ee
and hence
\be
S_{nm} = \oint \frac{dz}{2\pi i} \oint \frac{dw}{2\pi i} \frac{1}{z^{n-1}} \frac{1}{w^{m+2}}
\frac{\left(f'(z)\right)^2}{f'(w)} \left(\frac{f(w)}{f(z)}\right)^3 \frac{1}{f(w)-f(z)}.
\ee
The conservation laws then read
\bea
\bra{f} b_n  &=& - \sum_{m=2}^\infty S_{nm} \bra{f} b_{-m},
\\\label{smartlaws}
\bra{f} L_n  &=& - \sum_{m=2}^\infty S_{nm} \bra{f} L_{-m}.
\eea

One application where we used the conservation laws (\ref{smartlaws}) was to test the frequently
used conservation law for the wedge states
\be\label{fucl}
\ll_0 \ket{r} = \frac{2-r}{r} \ll_0^\dagger \ket{r},
\ee
which follows directly from (\ref{ULU}). This conservation law can be also derived following
Rastelli and Zwiebach using a vector field
\be
v^{w}(w) = 2(1+w^2) \arccot w.
\ee
This vector field is not globally defined, but is holomorphic everywhere outside the unit circle
including the infinity, so that (\ref{vTglob}) still holds for a contour encircling the infinity.
By deforming the contour onto the unit circle and passing to the local coordinate one finds
\be
v^{z}(z) = (r-2) (1+z^2) \arctan z + r (1+z^2) \arccot z
\ee
from which
\be
\bra{r} \left( r \ll_0^\dagger + (r-2) \ll_0 \right) = 0
\ee
follows, and hence also (\ref{fucl}). Although we have derived or proved (\ref{fucl}) in many ways
we wanted to see whether it really works in level truncation. Using (\ref{smartlaws}) we calculated
the $L_{-2}$ coefficient of $\ll_0 \ket{r}$ and compared with the expected result
$\frac{8(2-r)}{3r^3} L_{-2} \ket{0}+\cdots$ from the right hand side of (\ref{fucl}). The numerical
agreement turned out to be quite good for finite $r$, but for $r$ set to infinity $\ll_0
\ket{\infty}$ did not seem to converge, although formally it can be set to zero. This only stresses
the importance of the observation made in section~\ref{s_gn0} and further discussed in
appendix~\ref{a_sumsliver}, that the sliver and the sum part of (\ref{gn0solfinal}) cancel each
other to a large extent.

\newpage
%%%%%%%%%%%%%%%%%%%%%%%%%%%%%%%%%%%%%%%%%%%%%%%%%%%%%%%%%%%%%%%%%%%%%%%%%%%%%%
\sectiono{Bernoulli numbers}
\label{a_Bernoulli}
%%%%%%%%%%%%%%%%%%%%%%%%%%%%%%%%%%%%%%%%%%%%%%%%%%%%%%%%%%%%%%%%%%%%%%%%%%%%%%

The Bernoulli numbers are among the most important number sequences in number theory.
They are defined through
\be\label{BernDef}
\frac{x}{e^x-1} = \sum_{n=0}^\infty \frac{B_n x^n}{n!}.
\ee
The first few nontrivial numbers are
$B_0 =1, B_1 = -\frac{1}{2}, B_2 = \frac{1}{6}, B_4 = -\frac{1}{30}, B_6 = \frac{1}{42},
B_8 = -\frac{1}{30}, B_{10} = \frac{5}{66}, B_{12} = -\frac{691}{2730}, \dots $.
The obey number of remarkable properties, the most basic ones are
\bea
B_{2k+1} &=& 0, \qquad \forall k \ge 1 \\
\label{BernIdLin}
\sum_{k=0}^n \frac{B_k}{k!} \frac{1}{(n+1-k)!} &=& 0, \qquad \forall n \ge 1.
\eea
Well known is also the Euler identity
\be\label{Euler}
(n+1) B_n = -\sum_{k=2}^{n-2} \frac{n!}{k! (n-k)!} B_k B_{n-k}, \qquad \forall n \ge 3.
\ee
There are number of other linear, quadratic or higher order identities \cite{Weisstein}. It appears
however that the ones we have discovered by solving the string field theory equations of motion
were previously unknown. The first one is quite similar to the Euler identity
\be\label{BernIdNew}
(n-1) B_n = - \sum_{\mbox{\scriptsize$\begin{array}{c} 0 \le p,q \le n \\ p+q \le n \end{array}$}}
\frac{n!}{p! q! (n-p-q)!} B_p B_q, \qquad \forall n \ge 0.
\ee
A simple proof using (\ref{BernIdLin}) goes as follows. Let us write the right hand side
of (\ref{BernIdNew}) as
\bea
&& -B_n - \sum_{q=0}^{n-1} \sum_{p=0}^{n-q} \frac{n! (n-p-q+1)}{p! q! (n-p-q+1)!} B_p B_q =
 -B_n + \sum_{q=0}^{n-1} \sum_{p=1}^{n-q} \frac{n!}{(p-1)! q! (n-p-q+1)!} B_p B_q = \nonumber\\
&&  -B_n + \sum_{p=1}^{n} \sum_{q=0}^{n-p} \frac{n!}{(p-1)! q! (n-p-q+1)!} B_p B_q = (n-1) B_n.
\nonumber
\eea
In the first sum only the $-p$ term from the factor $(n-p-q+1)$  in the numerator was contributing
thanks to (\ref{BernIdLin}). In the last sum only $p=n, q=0$ term was contributing thanks to the
same identity.

Another important fact about Bernoulli numbers we need, is their asymptotics
\be
B_{2k} = 2(-1)^{k-1} \frac{(2k)!}{(2\pi)^{2k}} \zeta(2k) =  2(-1)^{k-1} \frac{(2k)!}{(2\pi)^{2k}}
\left(1+ O\left(2^{-2k}\right)\right).
\ee

\newpage
%%%%%%%%%%%%%%%%%%%%%%%%%%%%%%%%%%%%%%%%%%%%%%%%%%%%%%%%%%%%%%%%%%%%%%%%%%%%%%
\sectiono{Proof of the sum-sliver cancellation}
\label{a_sumsliver}
%%%%%%%%%%%%%%%%%%%%%%%%%%%%%%%%%%%%%%%%%%%%%%%%%%%%%%%%%%%%%%%%%%%%%%%%%%%%%%
In this appendix we shall apply the Euler--Maclaurin formula to establish rigorous lower bound on
positive constant $A_p$ such that
\be
\sum_{k=2}^\infty \left. \frac{d}{d\alpha} \left[1-\left(\frac{2}{k+\alpha}\right)^p \right]^M
\right|_{\alpha=0} = 1 + O\left(e^{-A_p M^{1/(p+1)}}\right).
\ee
This in turn will imply the estimates (\ref{bound2}, \ref{bound4}) proving thus the cancellation
between the two terms in $\ket{\infty}-\sum_{n=2}^\infty  \partial_n \ket{n}$ for two classes of
high level coefficients, namely $\left(L_{-2}\right)^M \ket{0}$ and $\left(L_{-4}\right)^M
\ket{0}$.

The Euler--Maclaurin formula states that (see e.g. \cite{WW,Edwards})
\be\label{EMc}
\sum_{k=a}^{b-1} f(k) = \sum_{n=0}^N \frac{B_n}{n!} \left[ f^{(n-1)}(b) - f^{(n-1)}(a)\right] +
R_N,
\ee
where $B_n$ are the Bernoulli numbers (see appendix~\ref{a_Bernoulli}), and by $f^{(-1)}$ we denote
the primitive function $\int^t f(t) dt$.  The remnant $R_N$ for arbitrary $N$ is given by
\be\label{EMcerror}
R_N = \frac{1}{N!} \int_a^b B_N(t-[t]) f^{(N)}(t) dt,
\ee
where $B_n(x)$ are the Bernoulli polynomials and $[t]$ denotes the integer part of $t$. For a given
function, the Euler--Maclaurin formula is typically useful only up to certain maximal $N$ which
minimizes the error. This is because of the eventual factorial growth of the Bernoulli numbers and
polynomials.\footnote{The Euler--Maclaurin formula written for infinite $N$ without a remnant in
most cases presents an asymptotic series which can be summed via Borel summation technique
\cite{Hardy}. The cases when the series converges by itself are rare and the most prominent
examples are polynomials and exponentials.}

Applying the formula (\ref{EMc}) to our sum and taking the harmless limit $b \to \infty$ we see
that for $0 < n \le N < M$ the $(n-1)$-th derivatives of our function $f(t)= \partial_t
\left[1-\left(\frac{2}{t}\right)^p \right]^M$ all vanish at $t=2$ and $t=\infty$. Thus it is only
the first term in (\ref{EMc}) with $n=0$ and the remnant $R_N$ which contribute. For the upper
bound on the remnant one can use $|B_N(x)| \leq |B_N|$ for $x \in [0,1]$ for $N$ even and hence
\be
|R_N| \le \frac{|B_N|}{N!} \int_2^\infty |f^{(N)}(t)| dt.
\ee
The strategy is now to find a value of $N, 0<N<M$ such that $|R_N|$ is minimized. For that we need
accurate estimate of $ \int_2^\infty |f^{(N)}(t)| dt$ which is actually the hardest part of the
proof.

Naive expansion of $\frac{d^{N+1}}{dt^{N+1}}\left[1-\left(\frac{2}{t}\right)^p \right]^M$ into the
binomial series, taking the absolute value of each term and integrating it, wouldn't work. The
estimate would be too crude and useless, since it would not take into account that at $t=2$ the
integrand vanishes.

Let us start with the formula for derivative of a composite function
\be
\frac{d^n}{dx^n} F(\phi(x)) = \sum_{m=0}^n
\sum_{p_j\left|\!\!\mbox{\scriptsize$\begin{array}{c}\sum p_j =m \\ \sum j p_j =n
\end{array}$}\right.} \frac{n!}{p_1! p_2! \ldots p_l!} \frac{d^m F}{dy^m} \prod_{j=1}^l
\left(\frac{\phi^{(j)}(x)}{j!}\right)^{p_j}.
\ee
Inserting the identity in the forms of $1=\oint \frac{dz}{2\pi i} \frac{1}{z^{n+1}}\prod_j z^{j p_j}$ and
$1=\oint \frac{dw}{2\pi i} \frac{1}{w^{m+1}}\prod_j w^{p_j}$, and performing the sum over all $p_j$'s we find
\be
\frac{d^n}{dx^n} F(\phi(x)) = n! \sum_{m=0}^n \frac{d^m F}{dy^m} \oint \frac{dz}{2\pi i}
\frac{1}{z^{n+1}} \frac{1}{m!} \left[\phi(x+z)-\phi(x)\right]^m.
\ee
Let us set $F(y)= y^M, \phi(x) = 1 - \left(\frac{2}{x}\right)^p $ and assume $n>0$. Then by using
$\phi(x+z)-\phi(x) =  \left(\frac{2}{x}\right)^p \left[1-\left(1+\frac{z}{x}\right)^{-p}\right]$
and with the help of binomial expansion for the $m$-th power, we find easily by direct integration
\be
\int_2^\infty \left|\frac{d^n}{dx^n} \left[1-\left(\frac{2}{x}\right)^p \right]^M \right| \, \leq
\, \frac{M!\, 2^{-n+1} p^{-1}}{\Gamma\left(M+\frac{n-1}{p}+1\right)} \sum_{m=1}^n \sum_{k=1}^m
\frac{\Gamma\left(m+\frac{n-1}{p}\right) (pk+n-1)!}{k!(m-k)!(pk-1)!}.
\ee
The double sum on the right hand side can be replaced by the maximal term times a factor of $n^2$
which is not going to affect the leading behavior. The maximum is achieved for $m=n$ and $k =c_p n$
where $c_p$ is a solution to \bdm \left(1+\frac{1}{p c_p}\right)^p = \frac{c_p}{1-c_p}, \edm i.e.
$c_2 = 0.738, c_4 = 0.758$  for the cases of interest. Now setting $n=N+1$, using $B_N/N! \sim
(2\pi)^{-N}$ we can minimize the remnant. The minimum is attained for $n \propto M^{1/(p+1)}$ and
by calculating the exact coefficient we find
\be
|R_N| \le K_p  e^{-A_p M^{1/(p+1)}},
\ee
where $K_p$ is some finite constant and $A_p = \left(4\pi \frac{1-c_p}{1+p c_p}\right)^{p/(p+1)}$.
Again for the cases of interest we find $A_2 = 1.210$ and $A_4=0.799$ which seem to be smaller by a
factor of four from what numerical fits would suggest. To obtain more precise estimate of $A_p$ and
not just upper bound would be more challenging, since $B_N(t-[t])$ is a periodic function and large
cancellations in (\ref{EMcerror}) are taking place. Anyway it is nice, that apart of proving an
upper bound we were able to capture the qualitative behavior, i.e. the power of $M^{1/(p+1)}$ in
the exponent.

\newpage
%%%%%%%%%%%%%%%%%%%%%%%%%%%%%%%%%%%%%%%%%%%%%%%%%%%%%%%%%%%%%%%%%%%%%%%%%%%%%%
\sectiono{Collection of useful formulas}
\label{a_collection}
%%%%%%%%%%%%%%%%%%%%%%%%%%%%%%%%%%%%%%%%%%%%%%%%%%%%%%%%%%%%%%%%%%%%%%%%%%%%%%

\subsection{$\bb_0$-gauge formulas}
\label{a_Bgauge}

We define
\bea
\bb_0 &=& \oint \frac{d\tilde z}{2\pi i} \tilde z b_{\tilde z \tilde z}(\tilde z) \, = \,\tan \circ \, b_0 \,
= \, b_0 + \sum_{k=1}^\infty \frac{2(-1)^{k+1}}{4k^2-1} b_{2k} \, =
\, b_0 +  \frac{2}{3} b_2 - \frac{2}{15} b_4 + \cdots
\\
\bb_0^\dagger &=& \oint \frac{d\tilde z}{2\pi i} \tilde z b_{\tilde z \tilde z}\left(\tilde
z-\frac{\pi}{2}\right) \, = \, b_0 + \sum_{k=1}^\infty \frac{2(-1)^{k+1}}{4k^2-1} b_{-2k}\, = \,
b_0 + \frac{2}{3} b_{-2} - \frac{2}{15} b_{-4} + \cdots
\\
B_1 &=& \oint \frac{d\tilde z}{2\pi i}  b_{\tilde z \tilde z}(\tilde z)
= b_1 + b_{-1}
\\
B_1^L &=& \oint_{C_L} \frac{d\tilde z}{2\pi i}  b_{\tilde z \tilde z}(\tilde z) =  \frac{1}{2} B_1
+ \frac{1}{\pi} \left( \bb_0 + \bb_0^\dagger \right)
\\
B_1^R &=& \oint_{C_R} \frac{d\tilde z}{2\pi i}  b_{\tilde z \tilde z}(\tilde z) =  \frac{1}{2} B_1
- \frac{1}{\pi} \left( \bb_0 + \bb_0^\dagger \right),
\eea
where the open contours $C_L$ and $C_R$ are the left half ($\re \tilde z>0$) and the right half
($\re \tilde z<0$) of the unit circle. These objects clearly satisfy $B_1^L+B_1^R = B_1$ and
further
\bea
B_1^L \left( \phi_1 * \phi_2 \right) &=&  \left(B_1^L \phi_1 \right) * \phi_2, \\
B_1^R \left( \phi_1 * \phi_2 \right) &=&  (-1)^{\mathrm{gn}(\phi_1)} \phi_1 * \left(B_1^R \phi_2\right), \\
B_1   \left( \phi_1 * \phi_2 \right) &=&  \left(B_1 \phi_1 \right) * \phi_2 +
(-1)^{\mathrm{gn}(\phi_1)} \phi_1 * \left(B_1 \phi_2\right).
\eea
The first two equations are manifestations of the fact that Witten's star product glues together
right part of the first string with the left part of the second string, so that $B_1^L$ acting on
the first string can be pulled out of the product. Same is true for $B_1^R$ on the second string
with appropriate Grassman sign. The last equation tells us, that $B_1$ is a graded derivation of
the star algebra.

We also frequently need to commute $\bb_0^\dagger$ through the operator $U_r=(2/r)^{\ll_0}$,
or  $\bb_0$ through $U_r^\dagger$
\bea
U_{r}\, \bb_0^\dagger \,U_{r}^{-1} &=& \frac{2-r}{r}  \,\bb_0 + \frac{2}{r} \,\bb_0^\dagger,
\nonumber\\
U_{r}^{\dagger\,-1}\, \bb_0\, U_{r}^{\dagger} &=& \frac{2}{r} \, \bb_0 + \frac{2-r}{r} \,
\bb_0^\dagger,
\nonumber\\
U_{r}^{-1}\, \bb_0^\dagger \,U_{r} &=& \frac{r-2}{2}  \,\bb_0 + \frac{r}{2} \,\bb_0^\dagger,
\nonumber\\
U_{r}^{\dagger}\, \bb_0\, U_{r}^{\dagger\,-1} &=& \frac{r}{2} \, \bb_0 + \frac{r-2}{2} \,
\bb_0^\dagger.
\eea
Useful anticommutators are
\bea
\left\lbrace \bb_0, \tilde c(\tilde z) \right\rbrace &=& \tilde z,
\nonumber\\
\left\lbrace B_1, \tilde c(\tilde z) \right\rbrace &=& 1.
\eea

%%%%%%%%%%%%%%%%%%%%%%%%%%%%%%%%%%%%%%%%%%%%%%%%%%%%%%%%%%%%%%%%%%%%%%%%%%%%%%
\subsection{Some correlators}
\label{a_correlators}
%%%%%%%%%%%%%%%%%%%%%%%%%%%%%%%%%%%%%%%%%%%%%%%%%%%%%%%%%%%%%%%%%%%%%%%%%%%%%%

Using the definitions $\tilde c(x) = \cos^2(x) c(\tan x)$ and the fact that inversion acts simply
as a translation $I \circ \tilde c(x) = \tilde c(x-\pi/2) = \tilde c(x+\pi/2)$ we readily
derive\footnote{All the correlators are taken on the upper half-plane. Also, it would be more
consistent with our previous notation if all $x$, $y$ and $z$ had a tilde.}
\bea
\aver{ \tilde c(x)\, \tilde c(y)\, \tilde c(z)} &=& \sin(x-y) \sin(x-z) \sin (y-z),
\nonumber\\
\aver{I \circ \tilde c(x)\, \tilde c(y)\, \tilde c(z)} &=& \cos(x-y) \cos(x-z) \sin (y-z),
\nonumber\\
\aver{I \circ \tilde c(x)\, I \circ \tilde c(y)\, \tilde c(z)} &=& \sin(x-y) \cos(x-z) \cos (y-z),
\nonumber\\
\aver{I \circ \tilde c(x)\, I \circ \tilde c(y)\, I \circ \tilde c(z)} &=& \sin(x-y) \sin(x-z) \sin
(y-z),
\nonumber\\
\aver{\tilde c(x) \, \tilde c \partial\tilde c(y)} &=& - \sin(x-y)^2,
\nonumber\\
\aver{I \circ  \tilde c(x) \, \tilde c \partial\tilde c(y)} &=& - \cos(x-y)^2.
\eea
Useful correlators involving the $\bb_0 +\bb_0^\bpz$ operator are
\bea
\aver{I \circ  \tilde c(x) I \circ  \tilde c(-x) \left(\bb_0 +\bb_0^\bpz \right)  \tilde c(y)
\tilde c(-y)}
&=& 2y \sin(2x) \cos(x+y) \cos(x-y) + \nonumber\\
&& + 2x \sin(2y) \cos(x+y) \cos(x-y)
\\
\aver{I \circ  \tilde c(x) I \circ  \tilde c(-x) \left(\bb_0 +\bb_0^\bpz \right) \tilde c
\partial\tilde c(y)} &=& -x \left( \cos^2(x+y) + \cos^2(x-y) \right) +\\ \nonumber && + \left(y \partial_y -1
\right) \sin(2x) \cos(x-y) \cos(x+y).
\eea

Evaluating the above correlators for particular modes is not necessarily a simple task.
We can use $\aver{c_{-1} c_0 c_1}=1$ and
\bea\label{tczmp}
\tilde c_{-2k} &=& (-1)^k \frac{2^{2k}}{(2k+1)!} c_0 + \cdots,
\nonumber\\
\tilde c_{-(2k-1)} &=& (-1)^k \frac{2^{2k}}{(2k)!} \frac{c_1-c_{-1}}{2} +
\delta_{k,0} \frac{c_1+c_{-1}}{2} + \cdots,
\eea
where the dots indicate modes other than $c_{-1}$, $c_0$ and $c_1$. We then find
\be
\aver{\left(\tilde c_{-p}\right)^\bpz Q_B \tilde c_{-q}} = \frac{2^{p+q+1}}{(p+1)!(q+1)!}
(-1)^{\frac{p+q}{2}} - \frac{1}{2} \delta_{p,-1} \delta_{q,-1}.
\ee
Assuming $p_1$ and $q_1$ to be odd and $p_2$ and $q_2$ to be even we find further:
\be
\aver{\left(\tilde c_{-p_1}\right)^\bpz  \tilde c_{-q_1} \tilde c_{-q_2}} = - (-1)^{\frac{q_2}{2}}
\frac{2^{q_2}}{(q_2+1)!} \left[\delta_{q_1,-1} \frac{2^{p_1}}{(p_1+1)!}(-1)^{\frac{p_1+1}{2}} +
\delta_{p_1,-1} \frac{2^{q_1}}{(q_1+1)!}(-1)^{\frac{q_1+1}{2}} \right],
\ee
\bea
\aver{\left(\tilde c_{-p_1}\right)^\bpz \left(\tilde c_{-p_2}\right)^\bpz \left(\bb_0+\bb_0^\bpz
\right) \tilde c_{-q_1} \tilde c_{-q_2}} &=& -\left[ \delta_{p_2,0} (-1)^{\frac{q_2}{2}}
\frac{2^{q_2}}{(q_2+1)!} + \delta_{q_2,0} (-1)^{\frac{p_2}{2}} \frac{2^{p_2}}{(p_2+1)!}\right]
\\\nonumber
&& \times \left[ \delta_{p_1,-1} \frac{2^{q_1}}{(q_1+1)!}(-1)^{\frac{q_1+1}{2}} + \delta_{q_1,-1}
\frac{2^{p_1}}{(p_1+1)!}(-1)^{\frac{p_1+1}{2}} \right].
\eea

\newpage
%%%%%%%%%%%%%%%%%%%%%%%%%%%%%%%%%%%%%%%%%%%%%%%%%%%%%%%%%%%%%%%%%%%%%%%%%%%%%%
\sectiono{Details for ghost number one equation of motion}
\label{a_details}
%%%%%%%%%%%%%%%%%%%%%%%%%%%%%%%%%%%%%%%%%%%%%%%%%%%%%%%%%%%%%%%%%%%%%%%%%%%%%%

In this appendix we provide few intermediate steps for plugging the ansatz (\ref{gn1ansatz})
\bdm
\Psi = \sum_{n,p} f_{n,p}\, \lll^n \tilde c_p \ket{0} + \sum_{n,p,q} f_{n,p,q}\, \bbb \lll^n \tilde
c_p \tilde c_q\ket{0}
\edm
into the equation of motion $Q_B\Psi+\Psi*\Psi=0$. The action of the BRST charge $Q_B$ is quite
simple, since it annihilates the vacuum, commutes with $\lll^n$ and its anticommutator with $\bbb$
is $\lll$. Least obvious is perhaps the action on the $\tilde c$ ghost $\{Q_B,\tilde c(\tilde z)
\}= \tilde c \tilde\partial \tilde c(\tilde z)$, which takes the same form as in the $z$
coordinate. For the first term in the equation of motion we find easily
\be
Q_B\Psi = \sum_{n,k,l} \left[\frac{k-l}{2} f_{n,k+l} + f_{n-1,k,l} \right] \lll^n \tilde c_k \tilde
c_l\ket{0} - \bbb  \sum_{n,k,l,q} (k-l)f_{n,k+l,q} \lll^n \tilde c_k \tilde c_l \tilde c_q \ket{0}.
\ee
For the second term $\Psi*\Psi$ we use results from section~\ref{ss_lemma}. Denoting the two terms
in (\ref{gn1ansatz}) as $\Psi = \Psi^{(1)}+\Psi^{(2)}$, the second one containing the $\left(\bb_0
+ \bb_0^\bpz\right)$ factor, we find {\small
\bea
\Psi^{(1)}*\Psi^{(1)} &=& \sum_{N,n,m,k,l,p,q} (-1)^k \left(\frac{\pi}{4}\right)^{k+l}D_{n,m,l,k}^N
\binom{k+p-2}{k} \binom{l+q-2}{l} f_{n,k+p} f_{m,l+q}\, \lll^N \tilde c_p \tilde c_q \ket{0}
\nonumber\\
\\
\Psi^{(1)}*\Psi^{(2)} &=& \frac{\pi}{2} \!\!\! \sum_{N,n,m,k,l,p,q} (-1)^k
\left(\frac{\pi}{4}\right)^{k+l} D_{n,m,k+l,0}^N \binom{k+p-2}{k} \binom{l+q-2}{l}  f_{n,1}
f_{m,p+k,q+l}
\, \lll^N  \tilde c_p \tilde c_q \ket{0} +\nonumber\\
&& -\!\!\!\!\!\! \sum_{N,n,m,k_1,k_2,l,p_1,p_2,q}  \left(\frac{\pi}{4}\right)^{k_1+k_2+l}
(-1)^{k_1}
D_{n,m,k_2+l,k_1}^N f_{n,p_1+k_1} f_{m,p_2+k_2,q+l} \nonumber\\
&& \qquad \qquad\qquad \binom{k_1+p_1-2}{k_1} \binom{k_2+p_2-2}{k_2} \binom{l+q-2}{l} \, \bbb \,
\lll^N \tilde c_{p_1} \tilde c_{p_2} \tilde c_q \ket{0}
\nonumber\\
\\
\Psi^{(2)}*\Psi^{(1)} &=& \frac{\pi}{2} \!\!\! \sum_{N,n,m,k,l,p,q} (-1)^k
\left(-\frac{\pi}{4}\right)^{k+l} D_{n,m,0,k+l}^N \binom{k+p-2}{k} \binom{l+q-2}{l}  f_{n,p+k,q+l}
f_{m,1}
\, \lll^N  \tilde c_p \tilde c_q \ket{0} +\nonumber\\
&& +\!\!\!\!\!\! \sum_{N,n,m,k_1,k_2,l,p_1,p_2,q}  \left(\frac{\pi}{4}\right)^{k_1+k_2+l}
(-1)^{k_1+l}
D_{n,m,k_2,k_1+l}^N f_{n,p_1+k_1,q+l} f_{m,p_2+k_2} \nonumber\\
&& \qquad \qquad\qquad \binom{k_1+p_1-2}{k_1} \binom{k_2+p_2-2}{k_2} \binom{l+q-2}{l} \, \bbb \,
\lll^N  \tilde c_{p_1}  \tilde c_q \tilde c_{p_2}\ket{0}
\nonumber\\
\\
\Psi^{(2)}*\Psi^{(2)} &=& \pi^2  \!\!\! \!\!\!\!\!\! \sum_{N,n,m,q_1,q_2,l_1,l_2} \!\!\!\!\!\!
(-1)^{l_1} \left(\frac{\pi}{4}\right)^{l_1+l_2} D_{n,m,l_2,l_1}^N \binom{l_1+q_1-2}{l_1}
\binom{l_2+q_2-2}{l_2} f_{n,1,q_1} f_{m,1,q_2}
\, \lll^N \!\tilde c_{q_1} \tilde c_{q_2} \ket{0} +\nonumber\\
&& -\frac{\pi}{2} \!\!\! \sum_{N,n,m,p_1,p_2,q_1,q_2,k_1,k_2,l_1,l_2}
\left(\frac{\pi}{4}\right)^{k_1+l_1+k_2+l_2} (-1)^{k_1+l_1} D_{n,m,k_2+l_2,k_1+l_1}^N
f_{n,p_1+k_1,q_1+l_1} f_{m,p_2+k_2,q_2+l_2} \nonumber\\
&& \times \binom{k_1+p_1-2}{k_1} \binom{l_1+q_1-2}{l_1}
\binom{k_2+p_2-2}{k_2}\binom{l_2+q_2-2}{l_2} \bbb B_1\, \lll^N  \!\tilde c_{p_1} \tilde c_{q_1}
\tilde c_{p_2} \tilde c_{q_2} \ket{0},
\nonumber\\
\eea
}
where
\bea
D_{n,m,k,l}^N &=& \frac{n! m!}{N!} (-2)^{n+m-N} \oint\frac{dr}{2\pi i} \oint\frac{ds}{2\pi i}
\frac{(r+s-3)^N}{(r-2)^{n+1}(s-2)^{n+1}} (r-1)^k (s-1)^l
\nonumber\\
&=& \frac{n! m!}{N!} (-2)^{n+m-N} \sum_{j=0}^N \binom{N}{j} \binom{k}{n-j} \binom{N-j+l}{m}.
\eea
Although the twelve-fold sum with up to seven binomial factors looks prohibitively complicated, it
is actually quite easy to plug the expressions to the computer. Imposing twist symmetry, i.e.
$f_{n,p}=0$ for $p$ even and $f_{n,p,q}=0$ for $p+q$ even, we find
\bea
0 &=& f_{0,1} + \pi \left[ -\frac{1}{2} f_{0,1}^2 +  f_{0,1} \left( f_{1,1} + 2 f_{0,1,0} \right)
\right]
\\
0 &=& f_{1,1}+ 2f_{0,1,0} + \pi \left[ \frac{1}{4} f_{0,1}^2 - \frac{3}{2} f_{0,1} f_{1,1} -
f_{0,1} f_{0,1,0} + f_{1,1}^2 + 2 f_{1,1} f_{0,1,0} + 2 f_{0,1} \left(f_{2,1}+f_{1,1,0}\right)
\right]
\nonumber
\\
0 &=& f_{2,1} + 2f_{1,1,0} +\pi \left[-\frac{1}{16} f_{0,1}^2 + \frac{5}{8} f_{0,1} f_{1,1}
-f_{1,1}^2 -2 f_{0,1} f_{2,1}
+ 3f_{1,1} f_{2,1} + f_{0,1} \left(3f_{3,1} + 2 f_{2,1,0}\right) + \right. \nonumber\\
&& \qquad \qquad \qquad \left. \frac{1}{4} f_{0,1} f_{0,1,0} - f_{1,1} f_{0,1,0} + 2 f_{2,1}
f_{0,1,0} - f_{0,1} f_{1,1,0} + 2f_{1,1} f_{1,1,0} \right]
\nonumber\\
0 &=& f_{0,-1} + \pi \left[ \frac{1}{2} f_{0,1} f_{0,-1} + f_{0,1} \left( - f_{1,-1} + 2
f_{0,0,-1}\right) \right] + \pi^3 \left[\frac{1}{32} f_{0,1}^2 -\frac{3}{16} f_{0,1} f_{1,1} +
\frac{1}{4} f_{1,1}^2 + \right.
\nonumber\\
&& \quad \left. + \frac{1}{4} f_{0,1} f_{2,1} -\frac{1}{2} f_{1,1} f_{2,1} + \frac{1}{8} f_{0,1}
f_{0,1,0} - \frac{1}{2} f_{1,1} f_{0,1,0} + f_{2,1} f_{0,1,0} -f_{0,1,0}^2 + 2 f_{0,1,0} f_{1,1,0}
\right]
\nonumber\\
0 &=& 3 f_{0,-1} + \pi \left[-\frac{3}{2} f_{0,-1} f_{0,1} + 3 f_{0,-1} f_{1,1} + 2 f_{0,1}
f_{0,1,-2} \right] + \pi^3 \left[ -\frac{1}{32} f_{0,1}^2 + \frac{3}{16} f_{0,1} f_{1,1}
-\frac{3}{4} f_{0,1} f_{2,1} + \right.
\nonumber\\
&& \quad \left. + \frac{3}{2} f_{0,1} f_{3,1} + \frac{3}{8} f_{0,1} f_{0,1,0} - \frac{3}{2} f_{1,1}
f_{0,1,0} + 3 f_{2,1} f_{0,1,0} \right]
\nonumber\\
&\dots& .
\eea
These equations are equations for the coefficients of $\tilde c_1 \tilde c_0\ket{0}$,
$(\ll_0+\ll_0^\bpz) \tilde c_1 \tilde c_0\ket{0}$, $(\ll_0+\ll_0^\bpz)^2 \tilde c_1 \tilde
c_0\ket{0}$, $\tilde c_0 \tilde c_{-1}\ket{0}$, $\tilde c_1 \tilde c_{-2}\ket{0}, \ldots $ in the
equation of motion $Q_B\Psi + \Psi*\Psi=0$. It is interesting to see that imposing the $\bb_0$
gauge condition
\be
f_{n,p,0}+\frac{n+1}{2} f_{n+1,p} = 0
\ee
eliminates all the terms in the round brackets, and therefore the equations become
exactly solvable one after each other. We have proved this general pattern in
section~\ref{s_gn1}. For example the first equation implies
\bdm
f_{0,1} = \frac{2}{\pi}, \qquad \mbox{or} \qquad f_{0,1} = 0.
\edm
In the first case $f_{0,1} = \frac{2}{\pi}$ we readily find
\bdm
f_{1,1} = \frac{1}{2\pi}, \qquad f_{2,1} = \frac{1}{24\pi}, \qquad f_{0,-1} = \frac{\pi}{48},
\qquad f_{3,1} = -\frac{4}{3\pi^2} f_{0,1,-2}
\edm
and so on. Continuing up to level 12, i.e. finding coefficients like $f_{12,1}$, it is easy to guess
the complete form
\bea
f_{n,-p} &=& \frac{\pi^p}{2^{n+2p+1}n!} (-1)^n B_{n+p+1}, \qquad \mbox{$p$ odd}, \\
f_{n,-p,-q} &=&  \frac{\pi^{p+q}}{2^{n+2(p+q)+3} n!} (-1)^{n+q} B_{n+p+q+2}, \qquad \mbox{$p+q$
odd},
\eea
and hence (\ref{gn1solBern}) follows. The only proof that our guess is a true solution is given in
section~\ref{ss_eom} using the wedge state representation. From the mathematical point of view, it
would be interesting to find a direct proof using the form (\ref{gn1solBern}), since this would
presumably lead to an infinite set of Euler--Ramanujan type of identities for Bernoulli numbers.

\bigskip\noindent
{\it\large Pure gauge solutions}
\medskip

In the second case $f_{0,1} = 0$ we find $f_{1,1}=\beta$ to be a free parameter which determines
\bdm
f_{2,1} = -\frac{\pi}{2} \beta^2, \qquad f_{0,-1} = -\frac{\pi^3}{4} \beta^2.
\edm
Going one step further we would find
\bdm
f_{3,1} = \frac{\pi}{8} \beta^2 + \frac{\pi^2}{4} \beta^3, \qquad f_{1,-1} = \frac{3\pi^3}{16}
\beta^2 + \frac{3\pi^4}{8}  \beta^3, \qquad f_{0,1,-2} = -\frac{3\pi^3}{32} \beta^2 -
\frac{3\pi^4}{16} \beta^3.
\edm
This solution clearly corresponds to a pure gauge. One particular value of $\beta$ deserves perhaps
a special attention. For  $\beta= -\frac{1}{2\pi}$ we found that all the terms $f_{3,1}$,
$f_{1,-1}$, $f_{0,1,-2}$ vanish, it seems that the solution shares the symmetry (\ref{Lsym}) with
the tachyon solution. The other low level coefficients for this value of $\beta$ are given by
\bdm
f_{2,1} = -\frac{1}{8\pi}, \qquad f_{0,-1} = -\frac{\pi}{16},
\edm
\bdm
f_{4,1} = \frac{1}{384\pi}, \qquad f_{2,-1}= - f_{1,1,-2}= \frac{\pi}{128}, \qquad f_{0,-3}=
\frac{\pi^3}{256}.
\edm

\newpage
%%%%%%%%%%%%%%%%%%%%%%%%%%%%%%%%%%%%%%%%%%%%%%%%%%%%%%%%%%%%%%%%%%%%%%%%%%%%%%
\sectiono{Coefficients of the tachyon condensate in the Virasoro basis}
\label{a_Virasoro}
%%%%%%%%%%%%%%%%%%%%%%%%%%%%%%%%%%%%%%%%%%%%%%%%%%%%%%%%%%%%%%%%%%%%%%%%%%%%%%

Complete table of the exact coefficients up to level 4 is
\begin{displaymath}
\begin{tabular}{|r|l|}
\hline
$ c_1 \ket{0}$ & 0.55346558 \\
\hline
$ c_{-1} \ket{0}$ & 0.45661043 \\
\hline
$ L_{-2}\, c_1 \ket{0}$ & 0.13764616 \\
\hline
$ b_{-2} c_{0} c_{1} \ket{0}$ & -0.14421001 \\
\hline
$ L_{-4}\, c_1 \ket{0}$ &  -0.030277583 \\
\hline
$ L_{-2} L_{-2}\, c_1 \ket{0}$ &  0.0045805832 \\
\hline
$ c_{-3} \ket{0}$ & -0.16494614 \\
\hline
$ b_{-3} c_{-1} c_{1} \ket{0}$ & 0.16039444 \\
\hline
$ b_{-2} c_{-2} c_{1} \ket{0}$ & 0.17942652 \\
\hline
$ L_{-2} c_{-1} \ket{0}$ & 0.022748278 \\
\hline
$ L_{-3} c_{0} \ket{0}$ & 0 \\
\hline
$ b_{-2} c_{-1} c_{0} \ket{0}$ & 0.020943544\\
\hline
$ b_{-4} c_{0} c_{1} \ket{0}$ & 0.088982260\\
\hline
$ L_{-2} b_{-2} c_{0} c_{1} \ket{0}$ & -0.0084696519\\
\hline
\end{tabular}
\end{displaymath}
Let us also list the coefficients in the matter sector up to level 10
\begin{displaymath}
\begin{tabular}{|r|l|}
\hline
$ c_1 \ket{0}$ & 0.55346558 \\
\hline
$ L_{-2}\, c_1 \ket{0}$ & 0.13764616 \\
\hline
$ L_{-4}\, c_1 \ket{0}$ &  -0.030277583 \\
\hline
$ L_{-2} L_{-2}\, c_1 \ket{0}$ &  0.0045805832 \\
\hline
$ L_{-6}\, c_1 \ket{0}$ &   0.01245732489\\
\hline
$ L_{-4} L_{-2} \, c_1 \ket{0}$ & -0.0015475008 \\
\hline
$ L_{-2} L_{-2} L_{-2}\, c_1 \ket{0}$ & -0.00015818471 \\
\hline
$ L_{-8}\, c_1 \ket{0}$ &  -0.00694735218\\
\hline
$ L_{-6} L_{-2} \, c_1 \ket{0}$ & 0.000722255152 \\
\hline
$ L_{-4} L_{-4} \, c_1 \ket{0}$ & 0.0001290340047 \\
\hline
$ L_{-4} L_{-2}  L_{-2} \, c_1 \ket{0}$ &  0.000085720253 \\
\hline
$ L_{-2} L_{-2} L_{-2} L_{-2}\, c_1 \ket{0}$ &  2.5529377 $10^{-6}$\\
\hline
$ L_{-10}\, c_1 \ket{0}$ &  0.004375158716\\
\hline
$ L_{-8} L_{-2} \, c_1 \ket{0}$ & -0.000396885628 \\
\hline
$ L_{-6} L_{-4} \, c_1 \ket{0}$ & -0.000120886555 \\
\hline
$ L_{-6} L_{-2}  L_{-2} \, c_1 \ket{0}$ &  -0.000039274125\\
\hline
$ L_{-4} L_{-4} L_{-2} \, c_1 \ket{0}$ & -0.000015086291 \\
\hline
$ L_{-4} L_{-2} L_{-2} L_{-2} \, c_1 \ket{0}$ & -2.2863989  $10^{-6}$ \\
\hline
$ L_{-2} L_{-2} L_{-2} L_{-2} L_{-2}\, c_1 \ket{0}$ &  4.0674798 $10^{-8}$\\
\hline
\end{tabular}
\end{displaymath}
It is worth noticing that the coefficients of the states $ L_{-2}^n \, c_1 \ket{0}$ decay quite
rapidly, at a similar rate as in the Siegel gauge. This can be contrasted with identity based
solution, where the decay is much slower, leading eventually to the divergence of the energy.

\end{document}

%% file: two_vertex.pstex_t
\begin{picture}(0,0)%
\includegraphics{two_vertex.pstex}%
\end{picture}%
\setlength{\unitlength}{3552sp}%
\begingroup\makeatletter\ifx\SetFigFont\undefined%
\gdef\SetFigFont#1#2#3#4#5{%
  \reset@font\fontsize{#1}{#2pt}%
  \fontfamily{#3}\fontseries{#4}\fontshape{#5}%
  \selectfont}%
\fi\endgroup%
\begin{picture}(8700,2403)(1051,-3031)
\put(2551,-961){\makebox(0,0)[lb]{\smash{\SetFigFont{10}{12.0}{\familydefault}{\mddefault}{\updefault}$w$}}}
\put(9751,-961){\makebox(0,0)[lb]{\smash{\SetFigFont{10}{12.0}{\familydefault}{\mddefault}{\updefault}$\tilde z$}}}
\put(9001,-2986){\makebox(0,0)[lb]{\smash{\SetFigFont{9}{10.8}{\familydefault}{\mddefault}{\updefault}$\frac{\pi}{4}$}}}
\put(9601,-2986){\makebox(0,0)[lb]{\smash{\SetFigFont{9}{10.8}{\familydefault}{\mddefault}{\updefault}$\frac{\pi}{2}$}}}
\put(7126,-2986){\makebox(0,0)[lb]{\smash{\SetFigFont{9}{10.8}{\familydefault}{\mddefault}{\updefault}-$\frac{\pi}{2}$}}}
\put(7726,-2986){\makebox(0,0)[lb]{\smash{\SetFigFont{9}{10.8}{\familydefault}{\mddefault}{\updefault}-$\frac{\pi}{4}$}}}
\put(3526,-2986){\makebox(0,0)[lb]{\smash{\SetFigFont{9}{10.8}{\familydefault}{\mddefault}{\updefault}-$1$}}}
\put(4801,-2986){\makebox(0,0)[lb]{\smash{\SetFigFont{9}{10.8}{\familydefault}{\mddefault}{\updefault}$0$}}}
\put(5926,-2986){\makebox(0,0)[lb]{\smash{\SetFigFont{9}{10.8}{\familydefault}{\mddefault}{\updefault}$1$}}}
\put(8401,-2986){\makebox(0,0)[lb]{\smash{\SetFigFont{9}{10.8}{\familydefault}{\mddefault}{\updefault}$0$}}}
\put(2476,-1561){\makebox(0,0)[lb]{\smash{\SetFigFont{9}{10.8}{\familydefault}{\mddefault}{\updefault}$\pi$}}}
\put(1051,-1561){\makebox(0,0)[lb]{\smash{\SetFigFont{9}{10.8}{\familydefault}{\mddefault}{\updefault}$0$}}}
\put(6301,-961){\makebox(0,0)[lb]{\smash{\SetFigFont{10}{12.0}{\familydefault}{\mddefault}{\updefault}$z$}}}
\put(1726,-1786){\makebox(0,0)[lb]{\smash{\SetFigFont{10}{12.0}{\familydefault}{\mddefault}{\updefault}$M$}}}
\put(4726,-1486){\makebox(0,0)[lb]{\smash{\SetFigFont{10}{12.0}{\familydefault}{\mddefault}{\updefault}$M$}}}
\put(8326,-736){\makebox(0,0)[lb]{\smash{\SetFigFont{10}{12.0}{\familydefault}{\mddefault}{\updefault}$M$}}}
\put(1726,-2986){\makebox(0,0)[lb]{\smash{\SetFigFont{10}{12.0}{\familydefault}{\mddefault}{\updefault}$P$}}}
\put(4726,-2686){\makebox(0,0)[lb]{\smash{\SetFigFont{10}{12.0}{\familydefault}{\mddefault}{\updefault}$P$}}}
\put(8326,-2686){\makebox(0,0)[lb]{\smash{\SetFigFont{10}{12.0}{\familydefault}{\mddefault}{\updefault}$P$}}}
\put(1426,-1486){\makebox(0,0)[lb]{\smash{\SetFigFont{10}{12.0}{\familydefault}{\mddefault}{\updefault}$R$}}}
\put(2026,-1486){\makebox(0,0)[lb]{\smash{\SetFigFont{10}{12.0}{\familydefault}{\mddefault}{\updefault}$L$}}}
\put(3526,-2011){\makebox(0,0)[lb]{\smash{\SetFigFont{10}{12.0}{\familydefault}{\mddefault}{\updefault}$R$}}}
\put(5926,-2011){\makebox(0,0)[lb]{\smash{\SetFigFont{10}{12.0}{\familydefault}{\mddefault}{\updefault}$L$}}}
\put(7501,-1486){\makebox(0,0)[lb]{\smash{\SetFigFont{10}{12.0}{\familydefault}{\mddefault}{\updefault}$R$}}}
\put(9076,-1486){\makebox(0,0)[lb]{\smash{\SetFigFont{10}{12.0}{\familydefault}{\mddefault}{\updefault}$L$}}}
\end{picture}

%% file: three_vertex.pstex_t
\begin{picture}(0,0)%
\includegraphics{three_vertex.pstex}%
\end{picture}%
\setlength{\unitlength}{3552sp}%
\begingroup\makeatletter\ifx\SetFigFont\undefined%
\gdef\SetFigFont#1#2#3#4#5{%
  \reset@font\fontsize{#1}{#2pt}%
  \fontfamily{#3}\fontseries{#4}\fontshape{#5}%
  \selectfont}%
\fi\endgroup%
\begin{picture}(3792,2403)(451,-3031)
\put(3001,-2986){\makebox(0,0)[lb]{\smash{\SetFigFont{9}{10.8}{\familydefault}{\mddefault}{\updefault}$\frac{\pi}{4}$}}}
\put(3601,-2986){\makebox(0,0)[lb]{\smash{\SetFigFont{9}{10.8}{\familydefault}{\mddefault}{\updefault}$\frac{\pi}{2}$}}}
\put(1126,-2986){\makebox(0,0)[lb]{\smash{\SetFigFont{9}{10.8}{\familydefault}{\mddefault}{\updefault}-$\frac{\pi}{2}$}}}
\put(1726,-2986){\makebox(0,0)[lb]{\smash{\SetFigFont{9}{10.8}{\familydefault}{\mddefault}{\updefault}-$\frac{\pi}{4}$}}}
\put(2401,-2986){\makebox(0,0)[lb]{\smash{\SetFigFont{9}{10.8}{\familydefault}{\mddefault}{\updefault}$0$}}}
\put(451,-2986){\makebox(0,0)[lb]{\smash{\SetFigFont{9}{10.8}{\familydefault}{\mddefault}{\updefault}-$\frac{3\pi}{4}$}}}
\put(4126,-2986){\makebox(0,0)[lb]{\smash{\SetFigFont{9}{10.8}{\familydefault}{\mddefault}{\updefault}$\frac{3\pi}{4}$}}}
\put(3976,-1036){\makebox(0,0)[lb]{\smash{\SetFigFont{10}{12.0}{\familydefault}{\mddefault}{\updefault}$\tilde z$}}}
\put(2326,-736){\makebox(0,0)[lb]{\smash{\SetFigFont{10}{12.0}{\familydefault}{\mddefault}{\updefault}$M$}}}
\put(2326,-2686){\makebox(0,0)[lb]{\smash{\SetFigFont{10}{12.0}{\familydefault}{\mddefault}{\updefault}$P_2$}}}
\put(1126,-2686){\makebox(0,0)[lb]{\smash{\SetFigFont{10}{12.0}{\familydefault}{\mddefault}{\updefault}$P_3$}}}
\put(3526,-2686){\makebox(0,0)[lb]{\smash{\SetFigFont{10}{12.0}{\familydefault}{\mddefault}{\updefault}$P_1$}}}
\put(1876,-1486){\makebox(0,0)[lb]{\smash{\SetFigFont{10}{12.0}{\familydefault}{\mddefault}{\updefault}$R$}}}
\put(3076,-1486){\makebox(0,0)[lb]{\smash{\SetFigFont{10}{12.0}{\familydefault}{\mddefault}{\updefault}$R$}}}
\put(751,-1486){\makebox(0,0)[lb]{\smash{\SetFigFont{10}{12.0}{\familydefault}{\mddefault}{\updefault}$R$}}}
\put(3976,-1486){\makebox(0,0)[lb]{\smash{\SetFigFont{10}{12.0}{\familydefault}{\mddefault}{\updefault}$L$}}}
\put(2776,-1486){\makebox(0,0)[lb]{\smash{\SetFigFont{10}{12.0}{\familydefault}{\mddefault}{\updefault}$L$}}}
\put(1576,-1486){\makebox(0,0)[lb]{\smash{\SetFigFont{10}{12.0}{\familydefault}{\mddefault}{\updefault}$L$}}}
\end{picture}

%% file: star_product.pstex_t
\begin{picture}(0,0)%
\includegraphics{star_product.pstex}%
\end{picture}%
\setlength{\unitlength}{3552sp}%
\begingroup\makeatletter\ifx\SetFigFont\undefined%
\gdef\SetFigFont#1#2#3#4#5{%
  \reset@font\fontsize{#1}{#2pt}%
  \fontfamily{#3}\fontseries{#4}\fontshape{#5}%
  \selectfont}%
\fi\endgroup%
\begin{picture}(3717,2403)(1126,-3031)
\put(4726,-2986){\makebox(0,0)[lb]{\smash{\SetFigFont{9}{10.8}{\familydefault}{\mddefault}{\updefault}$\frac{3\pi}{4}$}}}
\put(4126,-2986){\makebox(0,0)[lb]{\smash{\SetFigFont{9}{10.8}{\familydefault}{\mddefault}{\updefault}$\frac{\pi}{2}$}}}
\put(3526,-2986){\makebox(0,0)[lb]{\smash{\SetFigFont{9}{10.8}{\familydefault}{\mddefault}{\updefault}$\frac{\pi}{4}$}}}
\put(2926,-2986){\makebox(0,0)[lb]{\smash{\SetFigFont{9}{10.8}{\familydefault}{\mddefault}{\updefault}$0$}}}
\put(2326,-2986){\makebox(0,0)[lb]{\smash{\SetFigFont{9}{10.8}{\familydefault}{\mddefault}{\updefault}-$\frac{\pi}{4}$}}}
\put(1726,-2986){\makebox(0,0)[lb]{\smash{\SetFigFont{9}{10.8}{\familydefault}{\mddefault}{\updefault}-$\frac{\pi}{2}$}}}
\put(1126,-2986){\makebox(0,0)[lb]{\smash{\SetFigFont{9}{10.8}{\familydefault}{\mddefault}{\updefault}-$\frac{3\pi}{4}$}}}
\put(4576,-1036){\makebox(0,0)[lb]{\smash{\SetFigFont{10}{12.0}{\familydefault}{\mddefault}{\updefault}$\tilde z$}}}
\put(2326,-2686){\makebox(0,0)[lb]{\smash{\SetFigFont{10}{12.0}{\familydefault}{\mddefault}{\updefault}$P_2$}}}
\put(3526,-2686){\makebox(0,0)[lb]{\smash{\SetFigFont{10}{12.0}{\familydefault}{\mddefault}{\updefault}$P_1$}}}
\put(1876,-1486){\makebox(0,0)[lb]{\smash{\SetFigFont{10}{12.0}{\familydefault}{\mddefault}{\updefault}$R$}}}
\put(3076,-1486){\makebox(0,0)[lb]{\smash{\SetFigFont{10}{12.0}{\familydefault}{\mddefault}{\updefault}$R$}}}
\put(3976,-1486){\makebox(0,0)[lb]{\smash{\SetFigFont{10}{12.0}{\familydefault}{\mddefault}{\updefault}$L$}}}
\put(2776,-1486){\makebox(0,0)[lb]{\smash{\SetFigFont{10}{12.0}{\familydefault}{\mddefault}{\updefault}$L$}}}
\put(1576,-1486){\makebox(0,0)[lb]{\smash{\SetFigFont{10}{12.0}{\familydefault}{\mddefault}{\updefault}$L$}}}
\put(1276,-2686){\makebox(0,0)[lb]{\smash{\SetFigFont{10}{12.0}{\familydefault}{\mddefault}{\updefault}$P_3$}}}
\put(2926,-736){\makebox(0,0)[lb]{\smash{\SetFigFont{10}{12.0}{\familydefault}{\mddefault}{\updefault}$M$}}}
\put(4276,-1486){\makebox(0,0)[lb]{\smash{\SetFigFont{10}{12.0}{\familydefault}{\mddefault}{\updefault}$R$}}}
\end{picture}

%% file: wedge.pstex_t
\begin{picture}(0,0)%
\includegraphics{wedge.pstex}%
\end{picture}%
\setlength{\unitlength}{3552sp}%
\begingroup\makeatletter\ifx\SetFigFont\undefined%
\gdef\SetFigFont#1#2#3#4#5{%
  \reset@font\fontsize{#1}{#2pt}%
  \fontfamily{#3}\fontseries{#4}\fontshape{#5}%
  \selectfont}%
\fi\endgroup%
\begin{picture}(6342,2403)(901,-3031)
\put(6976,-1036){\makebox(0,0)[lb]{\smash{\SetFigFont{10}{12.0}{\familydefault}{\mddefault}{\updefault}$\tilde z$}}}
\put(6526,-2986){\makebox(0,0)[lb]{\smash{\SetFigFont{9}{10.8}{\familydefault}{\mddefault}{\updefault}$\frac{n\pi}{4}$}}}
\put(1726,-2986){\makebox(0,0)[lb]{\smash{\SetFigFont{9}{10.8}{\familydefault}{\mddefault}{\updefault}-$\frac{n\pi}{4}$}}}
\put(2326,-2986){\makebox(0,0)[lb]{\smash{\SetFigFont{9}{10.8}{\familydefault}{\mddefault}{\updefault}-$\frac{(n-1)\pi}{4}$}}}
\put(901,-2986){\makebox(0,0)[lb]{\smash{\SetFigFont{9}{10.8}{\familydefault}{\mddefault}{\updefault}-$\frac{(n+1)\pi}{4}$}}}
\put(4576,-2986){\makebox(0,0)[lb]{\smash{\SetFigFont{9}{10.8}{\familydefault}{\mddefault}{\updefault}$\frac{(n-3)\pi}{4}$}}}
\put(5251,-2986){\makebox(0,0)[lb]{\smash{\SetFigFont{9}{10.8}{\familydefault}{\mddefault}{\updefault}$\frac{(n-2)\pi}{4}$}}}
\put(5851,-2986){\makebox(0,0)[lb]{\smash{\SetFigFont{9}{10.8}{\familydefault}{\mddefault}{\updefault}$\frac{(n-1)\pi}{4}$}}}
\put(7051,-2986){\makebox(0,0)[lb]{\smash{\SetFigFont{9}{10.8}{\familydefault}{\mddefault}{\updefault}$\frac{(n+1)\pi}{4}$}}}
\put(1876,-1486){\makebox(0,0)[lb]{\smash{\SetFigFont{10}{12.0}{\familydefault}{\mddefault}{\updefault}$R$}}}
\put(3076,-1486){\makebox(0,0)[lb]{\smash{\SetFigFont{10}{12.0}{\familydefault}{\mddefault}{\updefault}$R$}}}
\put(3976,-1486){\makebox(0,0)[lb]{\smash{\SetFigFont{10}{12.0}{\familydefault}{\mddefault}{\updefault}$L$}}}
\put(2776,-1486){\makebox(0,0)[lb]{\smash{\SetFigFont{10}{12.0}{\familydefault}{\mddefault}{\updefault}$L$}}}
\put(1576,-1486){\makebox(0,0)[lb]{\smash{\SetFigFont{10}{12.0}{\familydefault}{\mddefault}{\updefault}$L$}}}
\put(4276,-1486){\makebox(0,0)[lb]{\smash{\SetFigFont{10}{12.0}{\familydefault}{\mddefault}{\updefault}$R$}}}
\put(5926,-2686){\makebox(0,0)[lb]{\smash{\SetFigFont{10}{12.0}{\familydefault}{\mddefault}{\updefault}$P_1$}}}
\put(4726,-2686){\makebox(0,0)[lb]{\smash{\SetFigFont{10}{12.0}{\familydefault}{\mddefault}{\updefault}$P_2$}}}
\put(5176,-1486){\makebox(0,0)[lb]{\smash{\SetFigFont{10}{12.0}{\familydefault}{\mddefault}{\updefault}$L$}}}
\put(5476,-1486){\makebox(0,0)[lb]{\smash{\SetFigFont{10}{12.0}{\familydefault}{\mddefault}{\updefault}$R$}}}
\put(6676,-1486){\makebox(0,0)[lb]{\smash{\SetFigFont{10}{12.0}{\familydefault}{\mddefault}{\updefault}$R$}}}
\put(6376,-1486){\makebox(0,0)[lb]{\smash{\SetFigFont{10}{12.0}{\familydefault}{\mddefault}{\updefault}$L$}}}
\put(4126,-736){\makebox(0,0)[lb]{\smash{\SetFigFont{10}{12.0}{\familydefault}{\mddefault}{\updefault}$M$}}}
\put(2326,-2686){\makebox(0,0)[lb]{\smash{\SetFigFont{10}{12.0}{\familydefault}{\mddefault}{\updefault}$P_n$}}}
\end{picture}